\documentclass[a4paper,12pt]{article}
\pdfoutput=1
\usepackage{amsmath}
\usepackage{latexsym}
\usepackage{amsfonts}
\usepackage[normalem]{ulem}
\usepackage{array}
\usepackage{amssymb}
\usepackage{graphicx}
\usepackage{url}
\usepackage{bm}
\usepackage[
backend=bibtex,
style=authoryear,
citestyle=authoryear-comp,
sorting=ynt, 
maxcitenames=2, 
isbn=false,
doi=false,
url=false,
]{biblatex}

\usepackage{amsmath}
\usepackage{amsthm}
\usepackage{unicode}
\usepackage[english]{babel}
\usepackage[utf8]{inputenc}
\usepackage[T1]{fontenc}
\usepackage{csquotes}
\usepackage{caption}
\usepackage{float}
\usepackage{lscape} 
\usepackage{afterpage}
\usepackage{color} 
\usepackage[export]{adjustbox}

\newcommand{\puna}[1]{{\color{red} #1}}

\def\ind{\mathrel{\hbox{\rlap{\hbox to 7.5pt{\hrulefill}}\raise6.6pt\hbox{\eka\char'167}}}}

\numberwithin{equation}{section}

\usepackage{booktabs,array}

\newcount\rowc

\makeatletter
\def\ttabular{%
\hbox\bgroup
\let\\\cr
\def\rulea{\ifnum\rowc=\@ne \hrule height 1.3pt \fi}
\def\ruleb{
\ifnum\rowc=1\hrule height 1.3pt \else
\ifnum\rowc=6\hrule height \heavyrulewidth 
   \else \hrule height \lightrulewidth\fi\fi}
\valign\bgroup
\global\rowc\@ne
\rulea
\hbox to 10em{\strut \hfill##\hfill}%
\ruleb
&&%
\global\advance\rowc\@ne
\hbox to 10em{\strut\hfill##\hfill}%
\ruleb
\cr}
\def\endttabular{\crcr\egroup\egroup}

\usepackage[ruled,vlined]{algorithm2e}
\usepackage[noend]{algpseudocode}

\usepackage{caption}
\usepackage{subcaption}
\usepackage{wrapfig}
\usepackage{wasysym}
\usepackage{enumitem}
\usepackage{adjustbox}
\usepackage{ragged2e}
\usepackage[svgnames,table]{xcolor}
\usepackage{tikz}
\usepackage{longtable}
\usepackage{changepage}
\usepackage{setspace}
\usepackage{hhline}
\usepackage{multicol}
\usepackage{tabto}
\usepackage{float}
\usepackage{multirow}
\usepackage{makecell}
\usepackage{fancyhdr}
\usepackage[toc,page]{appendix}
\usetikzlibrary{shapes.symbols,shapes.geometric,shadows,arrows.meta}
\tikzset{>={Latex[width=1.5mm,length=2mm]}}
\usepackage{flowchart}\usepackage[paperheight=11.69in,paperwidth=8.27in,left=0.79in,right=0.79in,top=0.98in,bottom=0.98in,headheight=1in]{geometry}
\TabPositions{0.91in,1.82in,2.73in,3.64in,4.55in,5.46in,6.37in,}
\usepackage{xr-hyper} 
\usepackage{hyperref}
\hypersetup{
    colorlinks=false,
}
\usepackage{titling}

\usepackage{placeins}
\newcommand{\beginsupplement}{%
        \setcounter{table}{0}
        \renewcommand{\thetable}{S\arabic{table}}%
        \setcounter{figure}{0}
        \renewcommand{\thefigure}{S\arabic{figure}}%
     }

\setlistdepth{9}
\renewlist{enumerate}{enumerate}{9}
		\setlist[enumerate,1]{label=\arabic*)}
		\setlist[enumerate,2]{label=\alph*)}
		\setlist[enumerate,3]{label=(\roman*)}
		\setlist[enumerate,4]{label=(\arabic*)}
		\setlist[enumerate,5]{label=(\Alph*)}
		\setlist[enumerate,6]{label=(\Roman*)}
		\setlist[enumerate,7]{label=\arabic*}
		\setlist[enumerate,8]{label=\alph*}
		\setlist[enumerate,9]{label=\roman*}

\renewlist{itemize}{itemize}{9}
		\setlist[itemize]{label=$\cdot$}
		\setlist[itemize,1]{label=\textbullet}
		\setlist[itemize,2]{label=$\circ$}
		\setlist[itemize,3]{label=$\ast$}
		\setlist[itemize,4]{label=$\dagger$}
		\setlist[itemize,5]{label=$\triangleright$}
		\setlist[itemize,6]{label=$\bigstar$}
		\setlist[itemize,7]{label=$\blacklozenge$}
		\setlist[itemize,8]{label=$\prime$}
\def\macroheight{0.60\textwidth}

\def\Q{\mathbb{Q}}
\def\N{\mathbb{N}}
\def\E{\mathbb{E}}

\def\P{\mathbb{P}}

\def\T{\mathbb{T}}
\def\Lik{\mbox{Lik}}

\DeclareBoldMathCommand\bfalpha{\alpha}
\DeclareBoldMathCommand\bftheta{\theta}
\DeclareBoldMathCommand\bfrho{\rho}
\DeclareBoldMathCommand\bfeta{\eta}

\usepackage{authblk}
\begin{document}
\author[1]{Elja Arjas} 
\author[2]{Dario Gasbarra}
\affil[1]{University of Helsinki and University of Oslo}
\affil[2]{University of Helsinki}
\title{Adaptive treatment allocation and selection in multi-arm  clinical trials: a Bayesian perspective}
\date{\today}
\maketitle
\begin{abstract}
Clinical trials are an instrument for making informed decisions based on evidence from well-designed experiments. Here we consider adaptive designs mainly from the perspective of multi-arm Phase II clinical trials, in which one or more experimental treatments are compared to a control. Treatment allocation of individual trial participants is assumed to take place according to a fixed block randomization, albeit with an important twist: The performance of each treatment arm is assessed after every measured outcome, in terms of the posterior distribution of a corresponding model parameter. Different treatments arms are then compared to each other, according to pre-defined criteria and using the joint posterior as the basis for such assessment. If a treatment is found to be sufficiently clearly inferior to the currently best candidate, it can be closed off either temporarily or permanently from further participant accrual. The latter possibility provides a method for adaptive treatment selection, including early stopping of the trial. The main development in the paper is in terms of binary outcomes, but some extensions, notably for handling time-to-event data, are discussed as well. The presentation is to a large extent comparative and expository.      
\end{abstract}
  

\section{Introduction}
 
 From the earliest contributions to the present day, the statistical methodology for designing and executing clinical trials has been dominated by frequentist ideas, most notably, on  testing a precise hypothesis of "no effect difference" against an alternative, using a fixed sample size, and applying a pre-specified significance level to control for Type 1 error, as a means to guard against false positives in long term. An important drawback of this basic form of the standard methodology is that the design does not include the possibility of interim analyses during the trial. Particularly in exploratory studies during Phase II aimed at finding effective treatments from among a  number of experimental candidates it is natural look for extended designs that allow the execution of the trial to be modified based on the results from interim analyses. For example, such results could provide reasons for terminating the accrual of additional patients to some treatments for lack of efficacy or, if the opposite is true, for allocating more patients to the treatments that turned out more successful. Allowing for earlier dissemination of such findings may then also benefit the patient population at large.  

These motivations have led to the development of a whole spectrum of adaptive trial designs, and of corresponding methods for the  statistical analysis of such data.   An authoritative presentation of group sequential methods is provided in the monograph \textcite{d9aeaab36017460aa28e3461d3310ffe}. More general reviews of adaptive clinical trial designs, from the perspective of classical inference, can be found in, e.g., \textcite{chow2008adaptive},  \textcite{mahajan2010adaptive}, \textcite{chow2014adaptive}, \textcite{chang2016adaptive}, \textcite{Pallmann2018} and  \textcite{atkinson2019randomised}. While such adaptive designs allow for greater flexibility in the running of actual trials, their assessment is usually based on selected frequentist performance measures. In the standard version, interim analyses are planned before the trial is started, and need then to be accounted for, due to the consequent multiple testing, in computing the probability of Type 1 error. Although such rigid form of planning can be relaxed when employing the so-called alpha spending functions (e.g., \textcite{pocock1977group}, \textcite{o1979multiple}, \textcite{demets1994interim}), 
looking into the data before reaching the pre-planned end of the trial carries a cost either in terms of an inflated probability of Type 1 error or, if that is fixed, in a reduced power of the test to detect meaningful differences between the considered treatments.   

These classical approaches in the design and execution of clinical trials have been challenged  from both foundational and practical perspectives.  Important early contributions include, e.g., \textcite{thompson1933likelihood}, \textcite{Flhler1983BayesianAT}, \textcite{berry1985interim},   \textcite{SPIEGELHALTER19868}, \textcite{ berger1988statistical}, \textcite{spiegelhalter1994bayesian} and  \textcite{Thall1994}; for a brief historical account and a large number of references, see \textcite{Grieve2016IdleTO}. Comprehensive expositions of the topic are provided in the monographs \textcite{spiegelhalter2004bayesian}, \textcite{berry} and \textcite{yuan2017bayesian}.

The key argument here is the change of focus: instead of guarding against false positives in a series of trials in long term, the main aim is to utilize the full information potential in the observed data from the ongoing trial itself. Then, looking into the data in interim analyses is not viewed as something incurring a cost, but rather, as providing an opportunity to act more wisely. The foundational arguments enabling this change are provided by the adoption of the likelihood principle, e.g., \textcite{berger}. 

In practice, this also implies a change of the inferential paradigm, from frequentist into Bayesian. In Bayesian inference, the conditional (posterior) distribution for unknown model parameters is being updated based on the available data, via updates of the corresponding likelihood. In a clinical trial, it is even possible to continuously monitor the outcome data as they are observed, and thereby utilize such data in a fully adaptive fashion during the execution of the trial. The advantages of this approach are summarized neatly in the short review paper \textcite{Berry2006}, in \textcite{doi:10.1200/JCO.2010.32.2685}, \textcite{Lee2012BayesianCT}, and more recently, in \textcite{yin2017bayesian}, \textcite{doi:10.1080/00031305.2019.1566091} and \textcite{ijerph18020530}. The paper \textcite{Villar2015} contains a useful review of the theoretical background, connecting the theory of the optimal design of clinical trials with that of \textit{multi-armed bandit} problems. Unfortunately, general results on optimal strategies are largely lacking and their application in practice often infeasible because of computational complexity; however, see  \cite{Press}. Recently,  simulation based approximations have been used for applying Bayesian decision theory in the clinical trials context (e.g., \textcite{Mller2017ClinicalTD}, \textcite{yuan2017bayesian},   \textcite{Alban2018EXTENDINGAB}). 

Importantly, the posterior probabilities provide intuitively meaningful and directly interpretable answers to questions concerning the mutual comparison of different treatments, given the available evidence, and do so without needing reference to concepts such as sampling distribution of a test statistic under given hypothetical circumstances.

Here we consider adaptive designs mainly from the perspective of multi-arm Phase II clinical trials, in which one or more experimental treatments are compared to a control. However, the same ideas can be applied, essentially without change, in confirmatory Phase III trials, where only a single experimental treatment is compared to a control, but the planned size of the trial is larger. In both situations, treatment allocation of individual trial participants is assumed to take place according to a fixed block randomization, albeit with an important twist: The performance of each treatment arm is assessed after every measured outcome in terms of the posterior distribution of a corresponding model parameter. Different treatments arms are then compared to each other according to pre-defined criteria. If a treatment arm is found to be inferior in such a comparison to the others, it can be closed off either temporarily or permanently from further accrual.

Of the recent clinical trials literature, the papers by \textcite{Villar2015} and \textcite{Jacob2016} seem most closely related to our approach, although in different ways. In the latter part of \textcite{Villar2015}, the authors discuss and compare several adaptive strategies according to which patients can be allocated to different treatments in a multi-arm trial. Although the paper uses Bayesian inferential methods in parameter estimation, the final comparison between alternative methods is based on frequentist ideas and measures: testing of hypotheses, using fixed sample size and given significance level. In contrast to this,  \textcite{Jacob2016}  introduces three dynamic rules for dropping inferior treatment arms during the trial; these rules are closely similar to our Rules 1 and 2 below. On the other hand, and unlike \textcite{Villar2015}, \textcite{Jacob2016} does not explicitly consider the possibility of adaptive treatment allocation. 

We consider first, in Section \ref{section:no:2}, the simple  situation in which the outcomes are binary, and they can be observed soon after the treatment has been delivered. Section \ref{section:simulation} reports results from corresponding simulation experiments, following closely the settings of two examples in \textcite{Villar2015} but applying the adaptive methods presented in Section \ref{section:no:2}. In Section \ref{section:no:3}, the approach is extended to cover situations in which either binary outcomes are measured after a fixed time lag from the treatment, or the data consist of time-to-event measurements, with the possibility of right censoring. This section includes also some notes on vaccine efficacy trials.
The paper concludes with a discussion in Section \ref{section:no:4}. The presentation is to a large extent comparative and expository, particularly in Sections \ref{section:simulation} and \ref{section:no:4}. As a companion to this paper, we provide an implementation of the proposed method in the form of a freely available R package \textcite{barts} that facilitates the simulation of clinical trials with adaptive treatment allocation. 

\section{ The case of Bernoulli outcomes}
\label{section:no:2}
\subsection{An adaptive method for treatment allocation: Rule 1 }
\label{subsection:no:2.1}

 As in the papers \textcite{Villar2015} and
 \textcite{Jacob2016}, consider the 
 ‘prototype’ example of a trial with binary outcomes and two types of treatments,
 one type representing a \textit{control} or \textit{reference} treatment indexed by 0, and  $ K $  \textit{experimental} treatments indexed by  $ k, 1 \leq k \leq K $.
 Motivated by a \textit{conditional exchangeability} postulate between trial participants (with conditioning corresponding to their assignment to the different treatment arms), independent 
 Bernoulli outcomes can in this case be assumed for all treatments,
 with respective response rates  $\theta _{0}  $ and  $  \theta_{1}, \theta_{2}, \ldots , \theta_{K} $
 considered as model parameters.  

We index the participants in their order of recruitment to the trial by  $ i, 1 \leq i \leq N_{\max}, $  where  $ N_{\max}$ is an assumed maximal size of the trial. If no such maximal size is specified, we choose $ N_{\max} $ to be infinite.  
In this prototype version it is assumed that, for each  $ i, $  the outcome  $ Y_{i}  $ from the treatment of patient  $ i $ 
is observed soon after the treatment has been delivered. 
This assumption simplifies the consideration of adaptive designs, as the rule applied for deciding the treatment given to each participant can then directly account for information on such earlier outcomes. The meaning of ‘soon’ here should be understood in a relative sense to the accrual of participants to the trial. If the considered medical condition is rare in the background population, accrual will usually be slow with relatively long times between the arrivals. Then this requirement of outcome information being available when the next participant arrives may apply even if ‘soon’ is not literally true in chronological time. Extensions of this simple situation  are considered in Section \ref{section:no:3}.

We assume that, before starting the trial, a sequential block randomization to the treatment arms $0,1,...,K$ has been performed. We index by $n \geq 1 $  the positions on that list, calling $n$ \textit{list index}, and denote by $r(n)$ the corresponding treatment arm. Thus, we have a fixed sequence $((r(1),r(2),...r(K+1)),(r(K+2),r(K+3),...r(2(K+1)),...)$ of randomized blocks of length $K + 1$, where the blocks are independent random permutations of the treatment arm indexes ${0,1,...,K}$.

Assignment of the participants to the different treatment arms is now assumed to follow this list, but with the possibility of skipping a treatment arm in case it has been determined to be in the \textit{dormant} state for the considered value of $n$. This leads to a balanced design in the sense that, as long as no treatment arms have been skipped by the time of considering list index $n$, the numbers of participants assigned to different treatments can differ from each other by at most $1$, and they are equal when $n$ is a multiple of $K+1$.

Denote by $I_{k,n} $ the binary indicator variable of arm $k$ being in \textit{active} state at list index value $n$,  $n\geq 0$, $0 \leq k\leq K $, and let $I_{n}=(I_{0,n},I_{1,n},...,I_{K,n})$ be the corresponding activity state vector. The values of these vectors are determined in an inductive manner to be specified later.

By inspection we find that, at the time   a value $n \geq 1$ of the list index is considered, altogether 
\begin{align}  \label{eq:no:1} 
N(n) = \sum _{m=1}^{n}I_{r(m),m-1} \end{align}
trial participants  have so far arrived and been assigned to some treatment. Clearly $N(n) \leq n$. Let now  the sequence  $\left\{N^{-1}(i);i \geq 1\right \} $  be defined recursively by 
\begin{align}  \label{eq:no:2}
N^{-1}(1) = 1;\; N^{-1}(i) = \inf \left\{ n > N^{-1}(i-1): I_{r(n),n-1} = 1  \right
\}, i > 1.
\end{align} 
Then $N^{-1}(i)$ is the value of the list index $n$ at which participant  $i$ is assigned to a treatment, while $A_{i} = r(N^{-1}(i))$ is the index of the corresponding treatment arm. Having postulated independent Bernoulli outcomes with treatment arm specific parameters $\theta_{k}, 0 \leq k \leq K$, we then get that $Y_{i}$ is distributed according to \textit{Bernoulli}$(\theta_{r(N^{-1}(i))}).$

The distinction between active and dormant states is that no trial participants are assigned, at a value $n$ of the list index, to a treatment arm $r(n)$ if it is in the dormant state. Generally speaking, treatments whose performance in the trial has been poor, in a relative sense to the others, are more likely to be transferred into the dormant sate. However, with more data, there may later turn out to be sufficient evidence for such a trial arm to be returned back to the active state. 

The data $D_{n}$ that have accrued from the trial when it has proceeded up to list index value $n$ consist of the values of the state indicators  $I_{k,m-1}, \;  0 \leq k \leq K,  1 \leq m \leq n$, and of treatments $A_{i}$ and  outcomes $Y_{i}$ for $i \leq N(n)$. 

Next, we outline the inductive rule  by which  the values of state vectors $I_{n}=(I_{0,n},I_{1,n},...,I_{K,n})$ in a data sequence $\{ D_n; n \geq 1 \}$ are updated when the value of $n$ is increased by $1$. We write  $  \theta = \left(  \theta_{0}, \theta_{1}, \ldots  \theta_{K} \right)  $ and use, for clarity, boldface notation $\bftheta _{k}$ when the parameters are unknown and considered as random variables. Denote also $\bftheta_{\vee}=\max\{\bftheta_0, \bftheta_1,\dots, \bftheta_K\}$. 

According to this rule, called Rule 1, for $n \geq 1$ and if $r(n)=k $ is an experimental treatment arm, we let $I_{k,n}=0 $ if  $\P_{\pi}(\bftheta _{k} =  \bftheta_{\vee} \big\vert  D_{n} ) < \varepsilon$, and otherwise $I_{k,n}=1 $. Similarly, for the control arm $r(n)=0$ we let $I_{0,n}=0 $ if $\P_{\pi}(\bftheta _{0}+\delta \geq  \bftheta_{\vee} \big\vert  D_{n}) < \varepsilon$, and otherwise $I_{0,n}=1 $. Here the threshold values $\varepsilon > 0$ and $\delta \geq 0$ are selected \textit{operating characteristics} of the algorithm. A smaller value of $\varepsilon$ reflects then a more conservative attitude towards moving a treatment into the dormant state. The value of $\delta$ can be viewed as specifying the \textit{
minimal important difference} (MID) or\textit{ minimal clinically important difference} (MCID) in the trial; if positive, it provides
some extra protection to the control arm  from being moved into the dormant state. 

At the beginning, for $n=0$, the coordinates of $I_{0} =  (I_{0,0},I_{1,0},...,I_{K,0})$  are   determined in a similar fashion directly from the prior. In practice, the prior is never so strong that we would not have $I_0 = (1,1,...,1)$.

\begin{enumerate}
\item[\textbf{Rule 1}] \textit{Adaptive method for treatment allocation}.  

\begin{algorithm}[H]
\SetAlgoLined
\uIf{$\pi \left( \bftheta _{0} + \delta \geq 
\bftheta_{\vee}  \right)<\varepsilon$}
{ $I_{0,0} \leftarrow 0$\;}
\Else
{$I_{0,0} \leftarrow 1$\;}

\For{$k\leftarrow 1$ \KwTo $K$ (\it{experimental treatment arms}) }{
\uIf {${\pi} \left( \bftheta _{k} = 
\bftheta_{\vee}  \right)<\varepsilon $}
{           
$I_{k,0} \leftarrow  0$\;}
\Else
{$I_{k,0} \leftarrow  1$\;
} 
}

$I_{0} \leftarrow  (I_{0,0},I_{1,0},...,I_{K,0})$\; 
$\Lik_{0}(\theta)\leftarrow 1$   \;
$n\leftarrow 0$ \;
$N(0) \leftarrow  0$ \;

\While{$N(n)<N_{\max}$}{
$n \leftarrow n+1$\;
\uIf {$I_{r(n),n-1} = 0$}
{$N(n) \leftarrow N(n-1) $\; 
$I_{n} \leftarrow I_{n-1}$\;
$\Lik_{n}(\theta) \leftarrow  \Lik_{n-1}(\theta)$\; }
 \Else
 {(\it {in this case} $I_{r(n),n-1}=1)$\;
 $N(n) \leftarrow   N(n-1) + 1$\;
 $A_{N(n)}\leftarrow r(n)$\; 
$\Lik_{n}(\theta) \leftarrow \Lik_{n-1}(\theta)\times\theta_{r(n)}^{Y_{N(n)}} \left( 1- \theta_{r(n)} \right)^{1 - Y_{N(n)} } $\;

\For{$k\leftarrow 1$ \KwTo $K$ (\it{experimental treatment arms}) }{
\uIf {$\P_{\pi} \left( \bftheta _{k} = 
\bftheta_{\vee} \big\vert  D_{n} \right)<\varepsilon $}
{           
$I_{k,n} \leftarrow  0$\;}
\Else
{$I_{k,n} \leftarrow  1$\;
} 
}
\uIf{$\P_{\pi} \left( \bftheta _{0} + \delta \geq 
\bftheta_{\vee} \big\vert  D_{n} \right)<\varepsilon $} 
{ $I_{0,n} \leftarrow 0$\;}
\Else
{$I_{0,n} \leftarrow 1$\;}
}
 }
\end{algorithm}
\end{enumerate}

As a byproduct, successive applications of Rule 1 give us an explicit expression for the likelihood  $ L \left(  \theta  \vert D_{n} \right) = \Lik_{n}, n \geq 1,   $ arising from observing data  $ D_{n} $ as specified above. According to this rule, the likelihood expression $ L \left(  \theta  \vert D_{n} \right)$ is updated only at values of $n $ at which $I_{r(n),n} = 1, $ and then this is done by multiplying the previous value $ L \left(  \theta  \vert D_{n-1} \right)$ by the factor $\theta _{r(n)}^{Y_{N(n)}} \left( 1- \theta _{r(n)} \right)^{1 - Y_{N(n)} }$. By repeatedly applying the chain multiplication rule for conditional probabilities, we get that
\begin{align} \label{eq:no:2.3}  
L \left(  \theta  \vert D_{n} \right) =  \prod_{m=1}^{n}\theta _{r(m)}^{I_{r(m),m}Y_{N(m)}} \left( 1- \theta _{r(m)} \right)^{I_{r(m),m}(1 - Y_{N(m)}) } = 
 \prod_{k=0}^{K} \theta _{k}^{N_{k,1} \left( n \right) } \left( 1- \theta _{k} \right) ^{N_{k,0} \left( n \right) }.  
\end{align}
The right hand side expression is obtained by re-arranging the terms and denoting by
 \begin{align}  \label{eq:no:2.4}
N_{k,1}(n) =
\sum _{m=1}^{n}I_{k,m}{ 1}_{
\{ Y_{N(m)}=1 \}}, \;
N_{k0}(n) = \sum _{m=1}^{n}I_{k,m}
{ 1}_{\{ Y_{N(m)}=0 \}}, \; 0 \leq k \leq K, \; n \geq 1 ,
\end{align}
respectively, the number of successful and failed outcomes from treatment $k$ when considering list index values up to $n$. Of intrinsic importance in this derivation is that, when conditioning sequentially at $n$ on the data $D_{n}$, the criteria according to which the values of the indicators $I_{k,n}$ are updated to $I_{k,n+1}$ do not depend on the parameter $\theta$. As a consequence, these updates do not contribute to the likelihood terms that would depend on $\theta$.  Different formulations of this result can be found in many places, e.g., \textcite{Villar2015}.
 
As a consequence we can change the focus from the full data $\{D_n, n \geq 1 \},$ indexed according to the original list indexes used for randomization, to  "condensed" data  $\{D_i^{*}, i \geq 1 \}$ indexed according to the order in which the  participants were treated. We denote by
 \begin{align}  \label{eq:no:3}
S_{k}(i) =
 \sup \left\{ N_{k,1}(n): N(n) \leq i  \right
\}, \;F_{k}(i) =
 \sup \left\{ N_{k,0}(n): N(n) \leq i  \right
\}, \;  0 \leq k \leq K,  
\end{align}
respectively, the number of successful and failed outcomes from treatment $k$ when considering the first $i$  participants. Let \begin{align}   
S(i) =
\sum _{k=0}^{K}S_{k}(i),\;F(i) =
\sum _{k=0}^{K}F_{k}(i)
\end{align}
be the corresponding total number of successes and of failures, across all treatment arms. 

Following the usual practice in similar contexts, we assume that the unknown parameter values  $  \bftheta _{0}, \bftheta _{1}, \ldots , \bftheta _{K} $  have been assigned independent \textit{Beta}-priors, with
\textit{Beta}$\left( \theta _{k} \vert  \alpha _{k},\beta _{k} \right)$ for treatment arm  $ k$, where $\alpha _{k}$ and $\beta _{k}$ are separately chosen hyperparameters. The choice of appropriate values of these hyperparameters (e.g., \textcite{Thall1994}) is always context specific, and is not discussed here further. Then, due to the well-known conjugacy property of the
\textit{Beta}-priors and the Bernoulli-type likelihood \eqref{eq:no:2.3}, the  posterior  $ p \left(  \theta _{k} \vert  D_{k,i}^{*} \right)$  for $\bftheta _{k}$,   corresponding to data $D_i^{*}$, has the form of Beta-distribution with its parameters updated directly from the data:
\begin{align}
\label{eq:no:3a} p \left(  \theta _{k} \vert  D_{i,k}^{*} \right) =  \mbox{\textit{Beta}}   \left(  \theta _{k} \vert   \alpha _{k}+S_{k} \left(i \right) , \beta _{k}+F_{k} \left(i \right)  \right) ,~ i \geq 1, k=0, 1, \ldots ,K. \end{align}

This, together with the product form of the likelihood \eqref{eq:no:2.3} and the assumed independence of the priors $\pi$, allows then for an easy computation of the joint posterior distribution for  $  \left(  \bftheta _{0}, \bftheta _{1}, \ldots , \bftheta _{K} \right)   $ for any  $i. $ The density  $ p_{\pi}\left(  \theta _{0}, \theta _{1}, \ldots , \theta _{K} \vert D_{i,k}^{*}  \right)   $ becomes the product of  $ K+1 $  \textit{Beta}-densities. For example, posterior probabilities of the form $\P_{\pi} \left( \bftheta _{k} = \bftheta_{\vee} \big\vert  D_{n} \right)$, or posterior distributions for pairwise differences of the type  $  \bftheta _{k}- \bftheta _{0} $  or $  \bftheta _{k}- \bftheta _{l},$ can be computed numerically, in practice either by numerical integration as in \textcite{Jacob2016}, or by performing Monte Carlo sampling from this distribution; see also \textcite{Zaslavsky2012}. In our numerical examples in Section 3 we have applied this latter possibility.  

While Rule 1 may at least temporarily inactivate some less successful treatment arms and thereby close them off from further accrual, this closure need not be final. As long as a treatment arm is in the dormant state, the posterior for the corresponding parameter $\bftheta _{k}$ remains fixed. In contrast, with the accrual of  participants to active treatment arms still continuing, the posteriors for their parameters can be expected to become less and less dispersed. As a consequence, returns from dormant to active state tend to become increasingly rare. 

\textbf{Thompson's rule.} Rule 1 has much similarity with Thompson's rule (\textcite{thompson1933likelihood},  see also, e.g.,  \textcite{Thall2015StatisticalCI}, \textcite{Villar2015}), and both can be viewed as particular versions of \textit{response-adaptive randomization} (RAR) designs (\textcite{chow2008adaptive}). In its standard version, this Thompson's rule randomizes new patients to different treatment arms $k,0 \leq k \leq K,$ directly according to the posterior probabilities $\P_{\pi} \left( \bftheta _{k} = \bftheta_{\vee} \big\vert  D_{n} \right)$, updating the values of these probabilities as described above. Fractional versions of Thompson's rule use probability weights for this purpose, based on powers $\left(\P_{\pi} \left( \bftheta _{k} = \bftheta_{\vee} \big\vert  D_{n}  \right)\right)^{\kappa},$ with $0 \leq \kappa \leq 1$, normalized into probabilities by dividing such terms by their sum over different values of $k$. Thus, for $\kappa=0$, the randomization is symmetric to all $K+1$ treatments, and its adaptive control mechanism becomes stronger with increasing $\kappa$. We return to considering Thompson's rule in Section \ref{section:simulation}, in the context of the simulation experiments described there.

\subsection{An adaptive method for treatment selection: Rule 2 }
\label{subsection:no:2.2}

 While an open end recipe such as Rule 1 or Thompson's algorithm may seem attractive, for example, from the  perspective of drawing increasingly accurate inferences on the response parameters, practical considerations will often justify incorporation of rules for more definitive selection of some treatments and elimination of others. This is the case if the continued availability of more than one experimental treatment alternative at a later point in time is judged to be impracticable, as when entering the study into Phase III. Another reason is that incorporation of such decision rules enables us to make more direct comparisons to trial designs utilizing classical hypothesis testing ideas.
  
With this in mind, we complement Rule 1 with an optional possibility to conclusively terminate the accrual of additional  participants to the less successful treatment arms. Rule 2 below is an adaptation and extension of the corresponding definitions in, e.g., \textcite{Thall2007}, \textcite{berry}, \textcite{xie2012bayesian} and \textcite{Jacob2016}. In the commonly adopted terminology of adaptive designs, Rule 2 can be said to be a combination of versions of \textit{response-adaptive randomization} (RAR) and \textit{drop-the-losers} designs (\textcite{chow2008adaptive}).
 
In the definition of the algorithm, the letter $\T$ is used as a generic notation for the set of treatment arms still left in the trial at the considered value of $n$. Each elimination of a treatment reduces its size by one. Rule 2 contains, as part, Rule 1 for moving treatments to the dormant state. It then involves, in addition to the operating characteristics $\varepsilon$ and $\delta$ for Rule 1, three new parameters, viz. $\theta_{low}$, $\varepsilon_1$ and 
$\varepsilon_2$. Specifying a value for  $\theta_{low}$ means setting up a level of \textit {minimum required treatment response rate} (MRT), e.g., \textcite{xie2012bayesian}. A  treatment $k \in \T$ is eliminated from the trial if the posterior probability for $\{\bftheta_k > \theta_{low}\}$ falls below $\varepsilon_1$. The criteria for eliminating treatments are formally identical to those for moving them into the dormant state except that the bounds for the posterior probabilities then need to be tighter, $\varepsilon_1 \leq \varepsilon$.  

\begin{enumerate}
 \item[\textbf{Rule 2}]\textit{Adaptive rule for treatment allocation and selection}.  
 
 \begin{algorithm}[H]
\footnotesize
\SetAlgoLined

\uIf{$\pi \left( \bftheta _{0} + \delta \geq 
\bftheta_{\vee}  \right)<\varepsilon$}
{ $I_{0,0} \leftarrow 0$\;}
\Else
{$I_{0,0} \leftarrow 1$\;}

\For{$k\leftarrow 1$ \KwTo $K$ (\it{experimental treatment arms}) }{
\uIf {${\pi} \left( \bftheta _{k} = 
\bftheta_{\vee}  \right)<\varepsilon $}
{           
$I_{k,0} \leftarrow  0$\;}
\Else
{$I_{k,0} \leftarrow  1$\;
} 
}
$I_{0} \leftarrow  (I_{0,0},I_{1,0},...,I_{K,0})$\; 
$\T\leftarrow\{0,1,...,K\}$\;
$N(0)\leftarrow 0$\; 
$\Lik_{0}(\theta)\leftarrow 1$\;
$n\leftarrow 0$ \;
\While{$N(n)<N_{\max}$}
{
$n\leftarrow n+1$\;
\uIf{ $I_{r(n),n-1} = 0$} 
{$N(n)\leftarrow N(n-1)$\; 
$I_{n}\leftarrow I_{n-1}$\;
$\Lik_{n}(\theta)\leftarrow\Lik_{n-1}(\theta)$
\;}
 \Else{$\mbox{\it in this case }(r(n)\in\T)\;\mbox{\it and}\;(I_{r(n),n-1}=1)$\;
 $N(n)\leftarrow N(n-1)+1$\;
 $A_{N(n)}\leftarrow r(n)$\; 
$\Lik_{n}(\theta)\leftarrow\Lik_{n-1}(\theta)\times\theta_{r(n)}^{Y_{N(n)}}\left( 1- \theta_{r(n)}\right)^{1-Y_{N(n)}}$\;
\For{$k\in\T\setminus\{0\} \mbox{(experimental treatment arms)}$}{
\uIf{$\P_{\pi}\left(\bftheta _{k}  \geq 
\theta_{low} \big\vert  D_{n} \right)<\varepsilon_1\; \mbox{\bf or } \P_{\pi} \left( \bftheta _{k}=\max\limits_{\ell\in \T}\bftheta_{\ell}   \big\vert  D_{n} \right)<\varepsilon_2 $}
{$\T \leftarrow\T \setminus\{k\}$\;
$I_{k,n}\leftarrow 0$ \;
$n_{k, last}\leftarrow n$\;
} 
\ElseIf {$\P_{\pi} \left( \bftheta _{k} = \max\limits_{\ell\in \T} 
\bftheta_{\ell}
\big\vert  D_{n} \right)<\varepsilon $}
{           
$I_{k,n} \leftarrow  0$\;}
\Else{$I_{k,n} \leftarrow  1$\;
}
}
\uIf {$0\in \T$}
{
\uIf  
{$\P_{\pi} \left( \bftheta _{0} + \delta \geq \theta_{low} \big\vert  D_{n} \right)<\varepsilon_1\; 
\mbox{\bf or }\P_{\pi} \left( \bftheta _{0} + \delta \geq \max\limits_{\ell\in\T} \bftheta_{\ell}  \big\vert  D_{n} \right)<\varepsilon_2 $} {$\T \leftarrow \T \setminus \{0\}$\; 
$I_{0,n} \leftarrow 0$\;
$n_{0, last} \leftarrow  n$\;
}
\ElseIf{$\P_{\pi} \left( \bftheta _{0} + \delta \geq
\max\limits_{\ell\in \T} 
\bftheta_{\ell} \vert  D_{n} \right)<\varepsilon $}   
{ $I_{0,n} \leftarrow 0$\;}
\Else{$I_{0,n} \leftarrow 1$\;}
}
}
}
 \end{algorithm}
\end{enumerate}

\textbf{Notes}.  The state indicator $I_{r(n),n}$ at list index value $n$  depends on the recorded past trial history  $\{ D_{m}; 1 \leq m \leq n-1 \}$. However, given this history, it is conditionally independent of the model parameters  $  \theta = \left(  \theta_{0}, \theta_{1}, \ldots,  \theta_{K} \right)$. As was the case in   Rule 1, for a given original block randomization, the likelihood expression arising from applying Rule 2 depends only on the outcome data. 

This property is crucially important from the perspective of being able to draw correct statistical inferences from the trial. But it is also important from the perspective of practical implementation. Having assumed the initial randomization $\{r(n): n \geq 1\}$ to be fixed, 
no further randomization is needed when the trial is run since, at any point in time, the next move to be made will be fully determined by the  observed past data. 

After every new observed outcome, the algorithm of Rule 2 determines the current state of each treatment arm, choosing between the three possible options: active, dormant, or dropped. All moves between these states are possible except that the dropped state is absorbing: once a treatment arm has been dropped, it will stay. If an arm is in dormant state, it is at least momentarily closed from further patient accrual.

Consider then the different actions based on Rule 2 in more detail. The  posterior probabilities $\P_{\pi} \left( \bftheta _{k}  \geq \theta_{low} \big\vert  D_{n} \right)$ for the experimental arms, and $\P_{\pi} \left( \bftheta _{0}+\delta  \geq \theta_{low} \big\vert  D_{n} \right)$ for the control arm, express how likely it is, given the currently available data, that their response rate exceeds the pre-specified MRT $\theta_{low}$. The first criterion in Rule 2 then says that if this probability is below a selected threshold value $\varepsilon_{1},$ the treatment arm is dropped from the trial. The value of $\varepsilon_{1}$ can then be said to represent an acceptable risk level of error when concluding that $\{\bftheta _{k}  \geq \theta_{low}\}$, or $\{\bftheta _{0} + \delta  \geq \theta_{low}\}$, would not be true. This part of Rule 2 will obviously not be active if either $\theta_{low}=0$ or $\varepsilon_1=0$. 

The second criterion in Rule 2  makes a comparison of the response rate of a treatment and that of the best treatment in the trial. Both values are unknown, and the comparison is made in terms of the posterior probabilities $\P_{\pi}(\bftheta _{k} = \max\limits_{\ell\in \T}\bftheta_{\ell} \big\vert  D_{n} )$ for the experimental arms and $\P_{\pi}(\bftheta _{0}+\delta \geq \max\limits_{\ell\in \T}\bftheta_{\ell} \big\vert  D_{n})$ for the control. Here $\T \subset \{0, 1, ..., K\} $ is the set of treatment arms left in the trial at time $n$. The composition of $\T$ is determined in an inductive manner, starting from $\T = \{0, 1, ..., K\} $ at $n=1$. A treatment is dropped from the trial if the corresponding posterior probability  falls below the selected threshold level $\varepsilon_{2}$. Thus, for small $\varepsilon_{2}$, the decision to drop an experimental treatment $k$ is made if, in view of the currently available  data $D_{n}$, the event $ \{\bftheta _{k} = \max\limits_{\ell\in \T}\bftheta_{\ell} \}$ is true only with probability close to 0, with $\varepsilon_2$ representing the selected risk level.   The control arm is protected even more strongly from  inadvertent removal from the trial if a positive safety margin $\delta$ is employed; the comparison to experimental arms becomes symmetric if $\delta = 0.$ This entire mechanism of eliminating treatments based on mutual comparisons is inactivated by letting $\varepsilon_2 = 0$.

One should note that, while Rule 1 is compatible with the likelihood principle, Rule 2 has an element which violates it. This is because, in multi-arm trials with $K > 1$, when considered at times $n$ at which some treatment arms have already been dropped, the definition of the maximal response parameter value  $ \theta _{V} = \max\limits_{\ell\in \T}\theta_{\ell} $ ignores those indexed in $  \{0, 1, ..., K\} \setminus \T$. Sequential elimination of treatments, as embodied in Rule 2, while it has an obvious practical appeal in running a clinical trial,  also renders properties such as standard Bayesian consistency inapplicable.

In the third criterion of Rule 2 copies Rule 1: A n experimental treatment arm $k \in \T $  is made dormant if  $\P_{\pi}(\bftheta _{k} = \max\limits_{\ell\in \T}\bftheta_{\ell} \big\vert  D_{n} ) < \varepsilon$, and the control arm if $\P_{\pi}(\bftheta _{0}+\delta \geq \max\limits_{\ell\in \T}\bftheta_{\ell} \big\vert  D_{n}) < \varepsilon$, where $\varepsilon$ is a selected threshold. For this part of Rule 2 to function in a nontrivial way, we need to choose $\varepsilon >  \varepsilon_{1}$ and $\varepsilon >  \varepsilon_{2}$. If either $\varepsilon =  \varepsilon_{1}$ or $\varepsilon =  \varepsilon_{2},$ then the possibility of a treatment arm being moved into the dormant state is ruled out, and if $\varepsilon_1 = \varepsilon_2 = 0,$ then Rule 2 is easily seen to collapse into the simpler Rule 1. Finally, if also $\varepsilon = 0,$ then treatment allocation will follow directly the original block randomization, which was assumed to be symmetric between all treatment arms, and no treatments are dropped before reaching $N_{max}$.

The selection of  appropriate threshold values $\delta $ and $\theta_{low}$ in Rule 1 and Rule 2 should be based on substantive contextual arguments in the trial. If a positive value for  $\delta$ is specified, then, as already mentioned in the context of Rule 1, this is commonly viewed as the   \textit{ minimal clinically important difference} (MCID) in the trial. Employing such a positive threshold value when comparing the response rate of the control arm to that of an experimental arm, and not doing so when comparing two experimental arms to each other, reflects the idea that the design should be more conservative towards moving the control arm to the dormant state, let alone dropping it conclusively from the trial, than when contemplating about a similar move for an experimental treatment. 

Once selected, the design parameters  $\varepsilon,  \varepsilon _{1} $ and $\varepsilon _{2}$ in applying Rule 2, and then  deciding to either drop the treatment or putting it into the dormant state, can be interpreted directly as upper bounds for the risk that this decision was in fact unwarranted. By \textit{risk} is here meant the posterior probability of error, each time conditioned on the current data actually observed. Suppose, for example, that a finite value for $n_{k,last}$  has been established due to $\P_{\pi} \left( \bftheta _{k}  \geq \theta_{low} \big\vert  D_{n_{k,last}} \right) < \varepsilon _{1}$. Further accrual of trial  participants to treatment arm $k$ is then stopped after the patient indexed by $N_{k,last}$ because the response rate $\theta _{k}$ from that arm is judged, with only a small  probability $\leq \varepsilon_{1}$, given the data, to be above the MRT level $\theta _{low}$.  

If all experimental treatments have been dropped as a result of applying Rule 2, the trial ends with a negative result,   \textit{futility}, e.g. \textcite{Thall2007}. On the other hand, if the control arm has been dropped, at least one of the experimental arms was deemed better than the control, which is a positive finding. In case more than two experimental arms were left at that time, the trial design may allow for a continued application of Rule 2, with the goal of ultimately identifying the one with the highest response rate.

As remarked earlier, the application of Rule 2 is optional. If it is not enforced, Rule 1 is open ended and will only control the assignment of new  participants to the different treatments.   Then, if the trial size $N_{\max}$ has been specified and fixed in advance, and regardless of whether Rule 1 was previously employed or not, the posterior probabilities $\P_{\pi}(  \bftheta _{k} \geq \theta _{low} \vert D_{N_{\max}}^{*} )$, $\P_{\pi} ( \bftheta _{0}+\delta \geq \bftheta _{\vee}   \vert D_{N_{\max}}^{*} )$ and $\P_{\pi} (  \bftheta _{k} = \bftheta _{\vee} \vert D_{N_{\max}}^{*} )$ can be computed routinely  after all outcome data $D_{N_{\max}}^{*}$ have been observed, to provide the final assessment of the results from the trial.

\textbf{A frequentist perspective}. A different perspective to the application of Rule 2 is offered by the classical frequentist theory of statistical hypothesis testing. While the main point of this paper is to argue in favor of reasoning directly based on posterior inferences, this may not be sufficient to  satisfy stake holders external to the study itself, including the relevant regulatory authorities in question, which may be concerned about frequentist measures such as the overall Type 1 error rate at a pre-specified significance level (\textcite{chow2008adaptive}). 

From a frequentist point of view, the posterior probabilities $\P_{\pi}(  \bftheta _{0}+\delta \geq \bftheta_{\vee}
\vert  D_{n} )$ and $\P_{\pi}(  \bftheta _{k} = \bftheta_{\vee}
\vert  D_{n})$, via their dependence on the data $D_{n}$, can be viewed as test statistics in respective sequential testing problems, with Rule 2 defining the stopping boundaries. In the case $K=1$, they  correspond to considering two overlapping  hypotheses (e.g., \textcite{lewis1994group}), null  hypothesis $H_0: \theta_1 \leq \theta_0  + \delta$  and its  alternative    $H_1:\theta_1 \geq \theta_0  $. For $K \geq 1$,  the  null  hypothesis becomes  $H_0: \theta_{\vee} \leq \theta_0  + \delta$,  and the   alternative  $H_1:\theta_{\vee} \geq \theta_0 $. The posterior probabilities $\P_{\pi}(  \bftheta _{0}+\delta \geq \bftheta_{\vee}
\vert  D_{n} )$ can then be used as test statistics in testing  $H_{0}$, and $\P_{\pi}(  \bftheta _{\vee}  \geq \bftheta_{0}\vert  D_{n}  ) $ for testing $H_1.$

The size of the test  depends on the hypothesized "true" values of the response parameters $\theta =  (  \theta_{0}, \theta_{1}, \ldots, 
\theta_{K}  )$, on the selected threshold values  $\delta, \theta_{low}, \varepsilon,  \varepsilon _{1} $, $\varepsilon _{2}$ and, if specified in advance, on the maximal size $N_{\max} $ of the trial. For clarity, we denote such a hypothesized distribution generating the data by $\Q$, 
distinct from the mixture distribution $\P_{\pi}$ used, after being conditioned on current data, in applying Rule 1 and Rule 2. 

Frequentist measures such as true and false positive and negative rates, characterizing the performance of a test, can be computed numerically to a good approximation by performing a sufficiently large number of forward simulations from the selected $\Q$ and then averaging the sampled values. Such a consideration is, however, essentially only needed at the design stage when the trial design needs to be approved and no outcome data are yet available. When the trial is then run, it is natural to utilize, at each time $n$, the currently available data $D_n$ and the consequent posterior probabilities such as $\P_{\pi} (  \bftheta _{0}+\delta \geq \bftheta _{\vee}  \vert  D_{n}  )$, $\P_{\pi}  (  \bftheta _{k}= \bftheta _{\vee} \vert  D_{n}  )$ and $\P_{\pi}  (  \bftheta _{k} \geq \theta _{low} \vert  D_{n}  )$. In this context it may be useful to recall the well known result from  general decision theory: for any prior, the smallest Bayes risk is achieved by minimizing "pointwise" the expected loss with respect to the posterior.  We return to considering the frequentist measures  in the next section, in connection of Experiments 1 and 2.

\section{Illustrations of the methods by using simulation experiments}  \label{section:simulation}

Here we illustrate the application of the methods for treatment allocation (Rule 1) and for treatment selection (Rule 2) by performing a number of simulation experiments from hypothesized probability distributions $\Q$. For this, we consider different choices for the "true" parameter values
$ \theta = \left(  \theta_{0}, \theta_{1}, \ldots,  \theta_{K} \right)$, varying also the values of the threshold parameters  $\delta, \varepsilon$ and $\varepsilon _{2}$, and of the maximal trial size $N_{\max} $. For simplicity, and since we do not aim at modeling any contextual real data, we let $\theta_{low}=0$ and $\varepsilon _{1}=0$ in all simulations. 

Before entering the more detailed discussion of the simulation experiments, we consider briefly the choice of the prior distribution for the parameter  $  \bftheta = \left(  \bftheta _{0}, \bftheta _{1},\ldots ,\bftheta _{K} \right)$. In these experiments we are using systematically independent \textit{Uniform}(0,1)-priors for all coordinates $\bftheta _{k}$, corresponding to the hyperparameter values $\alpha_k = \beta_k = 1$ of the \textit{Beta}-distributions. This choice is not intended as a practical guideline, nor to be representative of choices that would be commonly made in real data situations. Instead, it is thought to be appropriate to be used as an illustration of the workings of Rule 1 and Rule 2, as all essential information for running the trial then comes from the registered outcome data from the trial itself. Particularly on the treatment used as  the control arm, there is usually a fair amount of background information from earlier experiments for specifying a more informative prior; for the relevant literature on this topic see, e.g., \textcite{spiegelhalter1994bayesian}, \textcite{Thall1994}, \textcite{spiegelhalter2004bayesian} and \textcite{doi:10.1177/1740774509356002}.

A pair $(\alpha_k, \beta_k)$ of hyperparameters of the \textit{Beta}-distribution is commonly thought to represent prior information equivalent to $\alpha_k$ successes and $\beta_k$ failures from a treatment arm before initiating the trial. If the selected values $\alpha_k$ and $\beta_k$ for some particular treatment arm $k$ are such that their sum $\alpha_k + \beta_k$ is larger than that of the others, say $\alpha_l + \beta_l$, it may be a good idea to postpone the application of Rule 1 on arm $k$ from the start of the trial, and use it to assign the first participants to those other arms $l$ until the sum $\alpha_l + \beta_l + S_l(i) + F_l(i)$ reaches the level of $\alpha_k + \beta_k$. Intuitively speaking, treatments are then compared to each other only after the joint posterior is based in the same number of (pseudo)observations from all arms.
 
\subsection{Simulation studies with a 2-arm trial:  Experiment 1}
\label{subsection:no:3.1}

Our first simulation experiment mimics the setting of the two-arm trial described in \textcite{Villar2015}, Section 5.1. In this comparison of a single experimental treatment to a control, the hypothesis $ H_{0}: \theta_{1} \leq  \theta_{0}$ was tested against the alternative $ H_{1}: \theta_{1} >  \theta_{0}$ by using Fisher's exact test at the significance level of $\alpha=0.05$ for Type 1 error. Two alternative parameter settings were considered in the simulations leading to the numerical results shown in Table 5 of \textcite{Villar2015}, with Type 1 error rate computed at parameter values $\theta _{0} = \theta _{1} = 0.3, $ henceforth denoted by $\Q_{null}$, and the power of rejecting $ H_{0}$ computed at $\theta _{0} = 0.3, \theta _{1} =  0.5$, denoted by $\Q_{alt}$. The simulations and the tests were based on fixed trial size $ N_{\max}=148$.
 
In the present approach, instead of first collecting all planned outcome data and then performing  a test at a given  level of significance, the trial would be run in an adaptive manner, continuously updating the posterior probabilities specified in Rule 1 and/or Rule 2, and then proceeding in an inductive manner according to these rules. 
\subsubsection{Monitoring the operation of Rule 1 in Experiment 1}

We considered three different settings of the design parameters for Rule 1: (a) $\delta = 0.1, \varepsilon =  0.1$, (b) $\delta = 0.1,   \varepsilon =  0.05$, and (c) $\delta = 0.05, \varepsilon = 0.2 $. Note that larger values for  $\delta$ and smaller values for $\varepsilon$  correspond to a higher degree of conservatism towards moving a treatment arm from active to dormant state, and conversely. The choice (b) is therefore more conservative than (a), while (c) is more liberal.

\begin{landscape} 
 \begin{figure}[!htbp] 
 \includegraphics[max size={1.6\textwidth}{1.6\textheight},left]{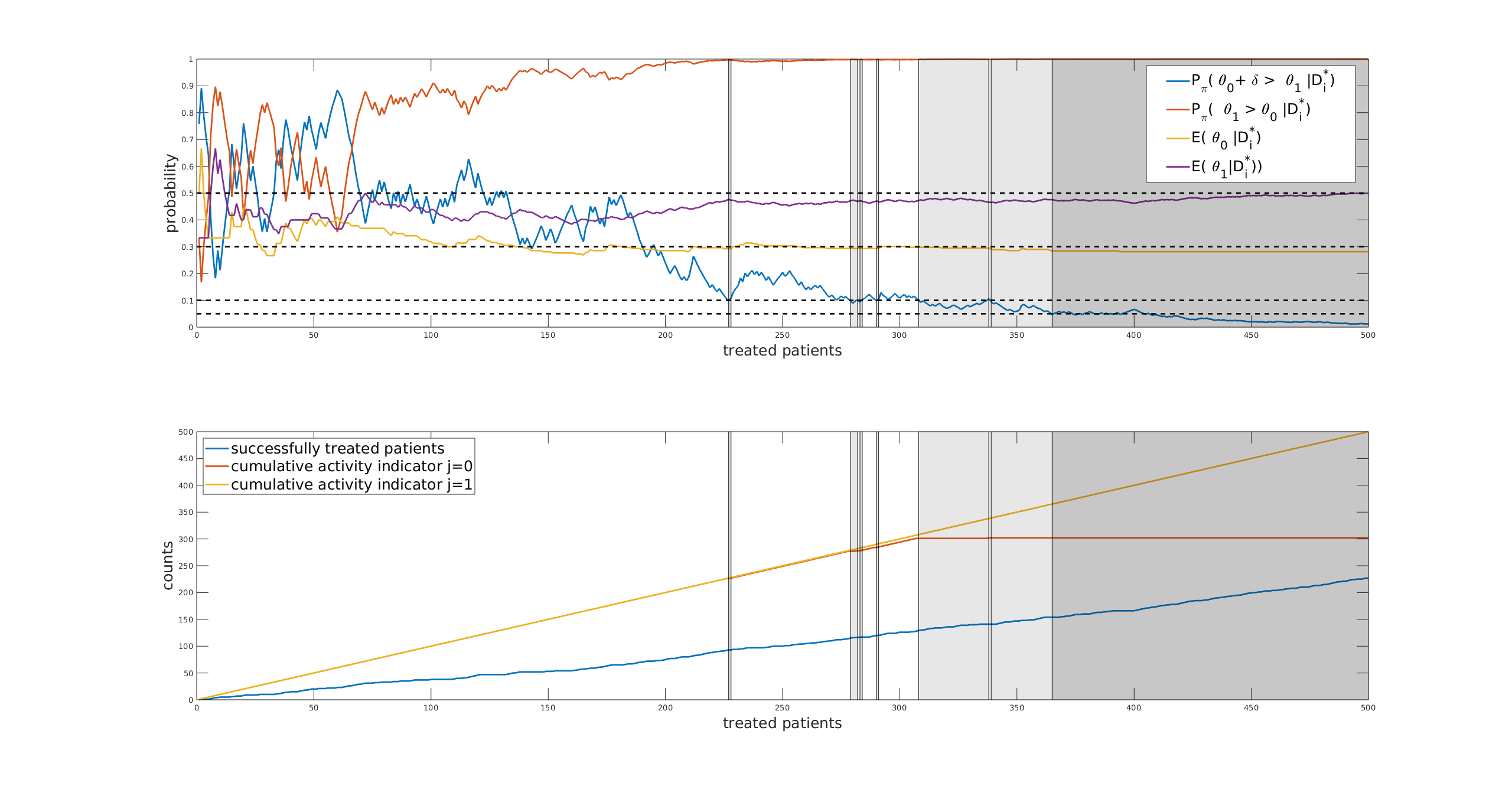}
 \caption{
 \small An example of monitoring  the execution of a 2-arm trial of size $N_{\max}=500,$ simulated from $\Q_{alt}$ with parameter values
$\theta_0=0.3$ and $\theta_1=0.5$. Rule 1 with design parameter values $\varepsilon=0.1$ and  $\delta=0.1$ was applied for treatment allocation, with the control arm in dormant state indicated in grey color. The effect of also involving Rule 2 for treatment selection, with $\varepsilon_1=0$ and $\varepsilon_2=0.05$, is indicated in darker grey. Top: Time-evolution of the posterior probabilities
$\P_{\pi}\bigl( \bftheta _{0}+\delta \geq \bftheta _{\vee}  \big\vert D_i^{*}\bigr) = \P_{\pi}\bigl( \bftheta _{0}+\delta \geq \bftheta _{1}  \big\vert D_i^{*}\bigr) $ and $\P_{\pi}\bigl( \theta_1 = \bftheta_{\vee}\big\vert D_i^{*}\bigr) = \P_{\pi}\bigl( \theta_1 > \bftheta_{0}\big\vert D_i^{*}\bigr),$ 
and of the posterior expectations
$\E_{\pi}\bigl( \bftheta_j\vert D_i^{*}\bigr),
j=0,1; 1 \leq i \leq 500$. Bottom: Cumulative sums of activity indicators of the treatment arms and the cumulative number of treatment successes. For more details, see text. }
 \label{FIGURE:yksi}
 \end{figure}
\end{landscape}

As an illustration of the workings of Rule 1, we performed an experiment emulating a real trial with maximal size $ N_{\max} = 500$, using a single realization generated from $\Q_{alt}$ and applying Rule 1 for treatment allocation with thresholds (a). For this, we considered  the values of the list index $n$ at which a new patient was assigned to either of the two treatments, i.e., $N(n) - N(n-1) = 1$, thereby skipping an index if it corresponded to an arm in the dormant state. For such values of $n$ and until $N(n) = 500$,  we monitored in Figure \ref{FIGURE:yksi}  the development of the posterior probabilities $\P_{\pi}(  \bftheta _{0} + \delta \geq \bftheta _{\vee} \vert  D_{n}) = \P_{\pi}  (  \bftheta _{0} + \delta \geq \bftheta _{1} \vert  D_{n}  )$ and $\P_{\pi}  (  \bftheta _{1} = \bftheta _{\vee} \vert  D_{n}  ) = \P_{\pi}  (  \bftheta _{1} \geq \bftheta _{0} \vert  D_{n}  )$, of the posterior expectations $\E_{\pi}  ( \bftheta _{0} \vert  D_{n}  )$ and $\E_{\pi}  ( \bftheta _{1} \vert  D_{n}  ),$ and of the activity indicators $I_{0,n}$ and $I_{1,n}$. Note that these functions  depend only on the  corresponding "condensed" simulated data  $\{D_i^{*}, 1 \leq i \leq 500 \}.$ 

According to Rule 1 (a), the control arm is in dormant state for  patient $i$ if
$\P_{\pi} ( \bftheta_0+ 0.1 \geq \bftheta_{1} \vert D_i^{*} ) < 0.1$. The values of $i$ for which this was the case in the considered simulation are shown in Figure \ref{FIGURE:yksi}  in grey color. For such $i,$ no new  patients were assigned to the control treatment, and therefore the corresponding cumulative sum of activity indicators $I_{0,n}$ and the posterior expectation $\E_{\pi}( \bftheta_0\vert D_i^{*} )$ remained constant. In contrast, the experimental arm was active during the entire follow-up due to all posterior probabilities $\P_{\pi} ( \bftheta_1 \geq \bftheta_{0} \vert D_i^{*} )$ staying above the threshold $\varepsilon = 0.1.$ Had also Rule 2 been applied, say, with threshold values $\varepsilon_1 = 0$ and $\varepsilon_2 = 0.05$, the control arm would have been dropped from the trial at the first $i$ for which $\P_{\pi} ( \bftheta_0+ 0.1 \geq \bftheta_{1} \vert D_i^{*} ) < 0.05$. In the considered simulation this happened at $i = 365$. In Figure \ref{FIGURE:yksi} this is indicated in dark grey. 

\subsubsection{Effect of the design parameters on treatment allocation}
\label{sub:3.1.2}

\begin{figure}[!htbp]
     \begin{subfigure}[b]{0.5\textwidth}
     \includegraphics[width=\textwidth]{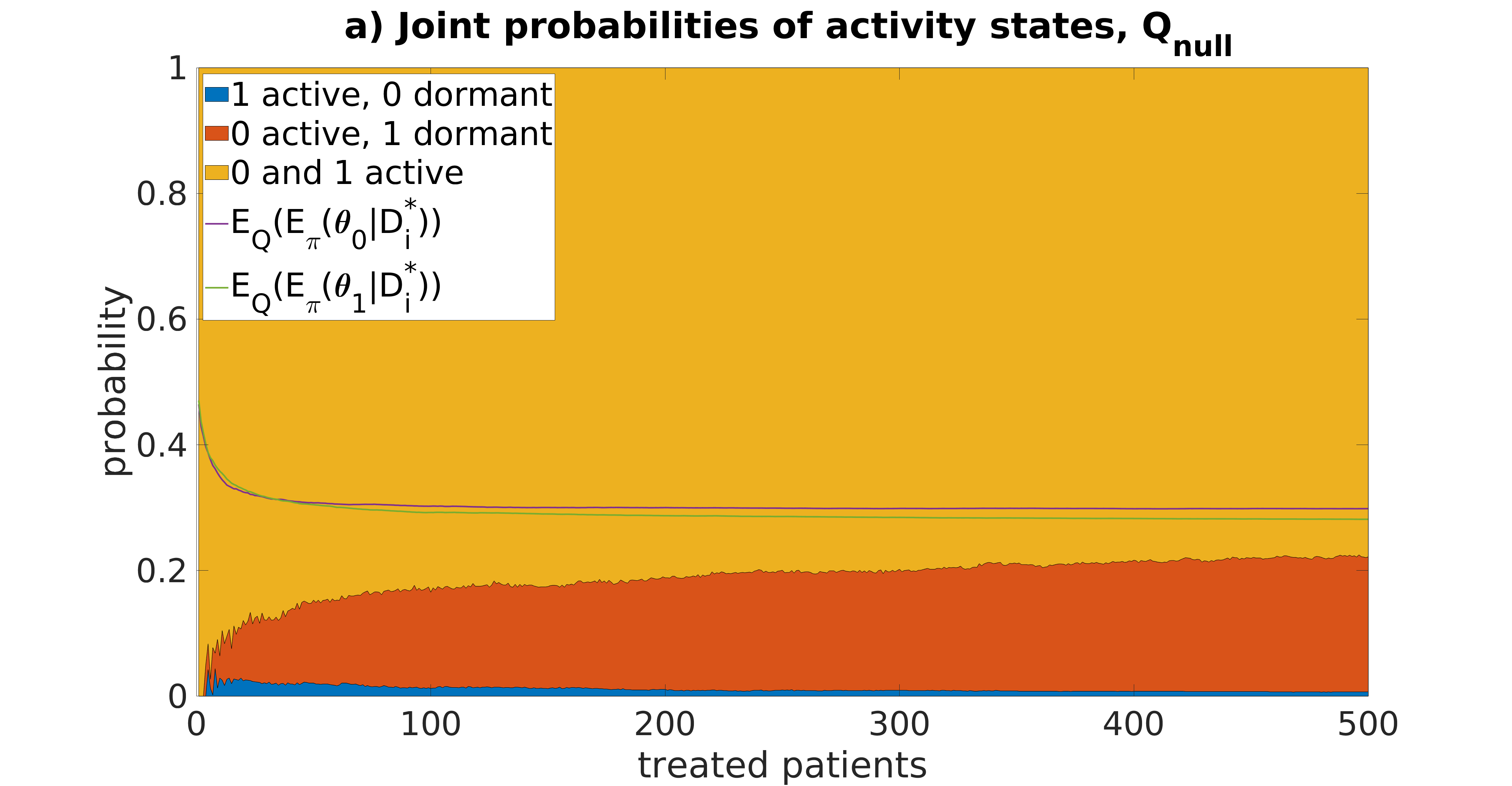}
       \renewcommand\thesubfigure{a} 
         \label{subfig:activity:a:null}
     \end{subfigure}
     \begin{subfigure}[b]{0.5\textwidth}
     \includegraphics[width=\textwidth]{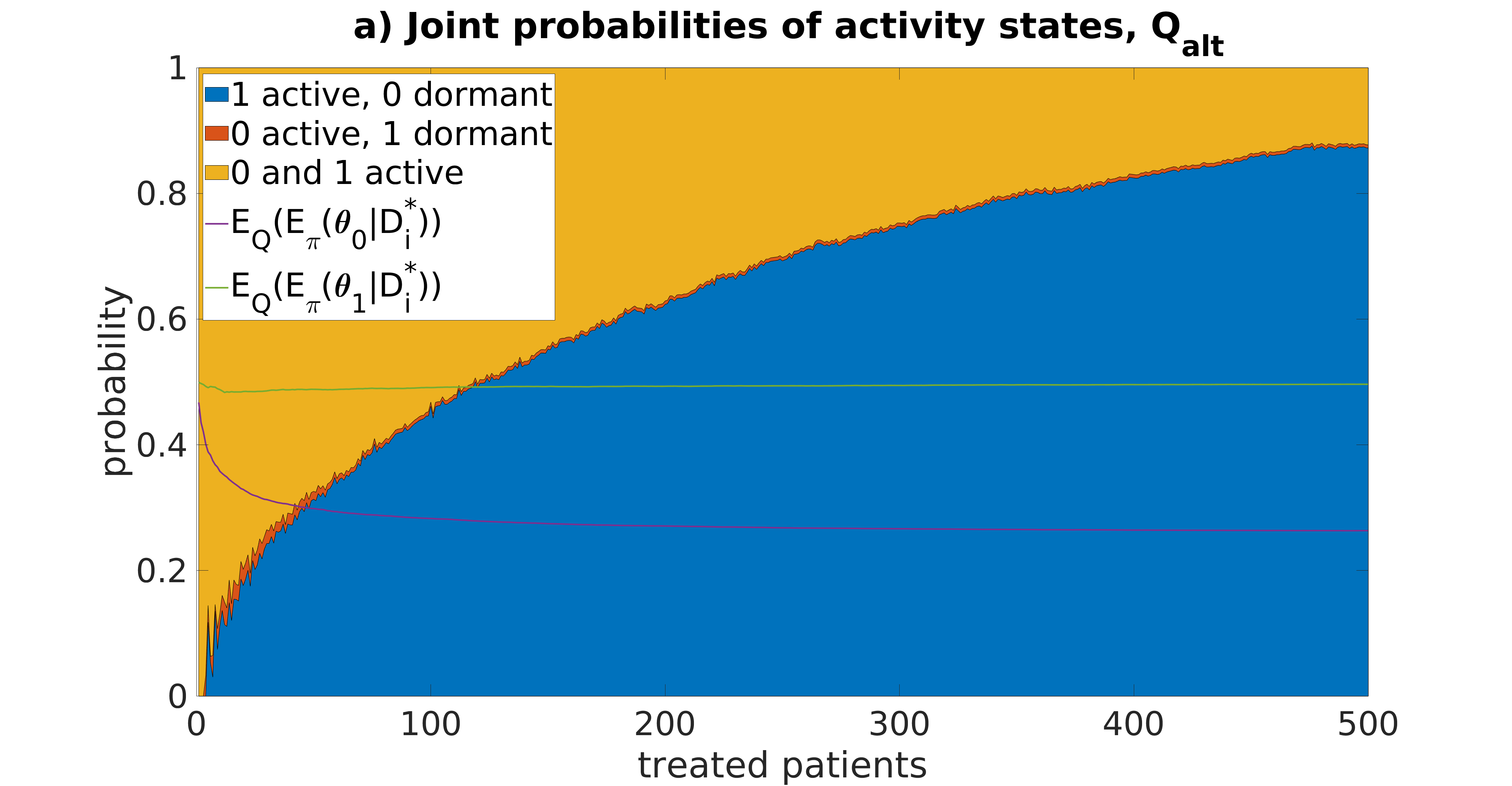}
        \renewcommand\thesubfigure{a} 
         \label{subfig:activity:a:alternative}
     \end{subfigure}
     \hfill
     \begin{subfigure}[b]{0.5\textwidth}
         \includegraphics[width=\textwidth]{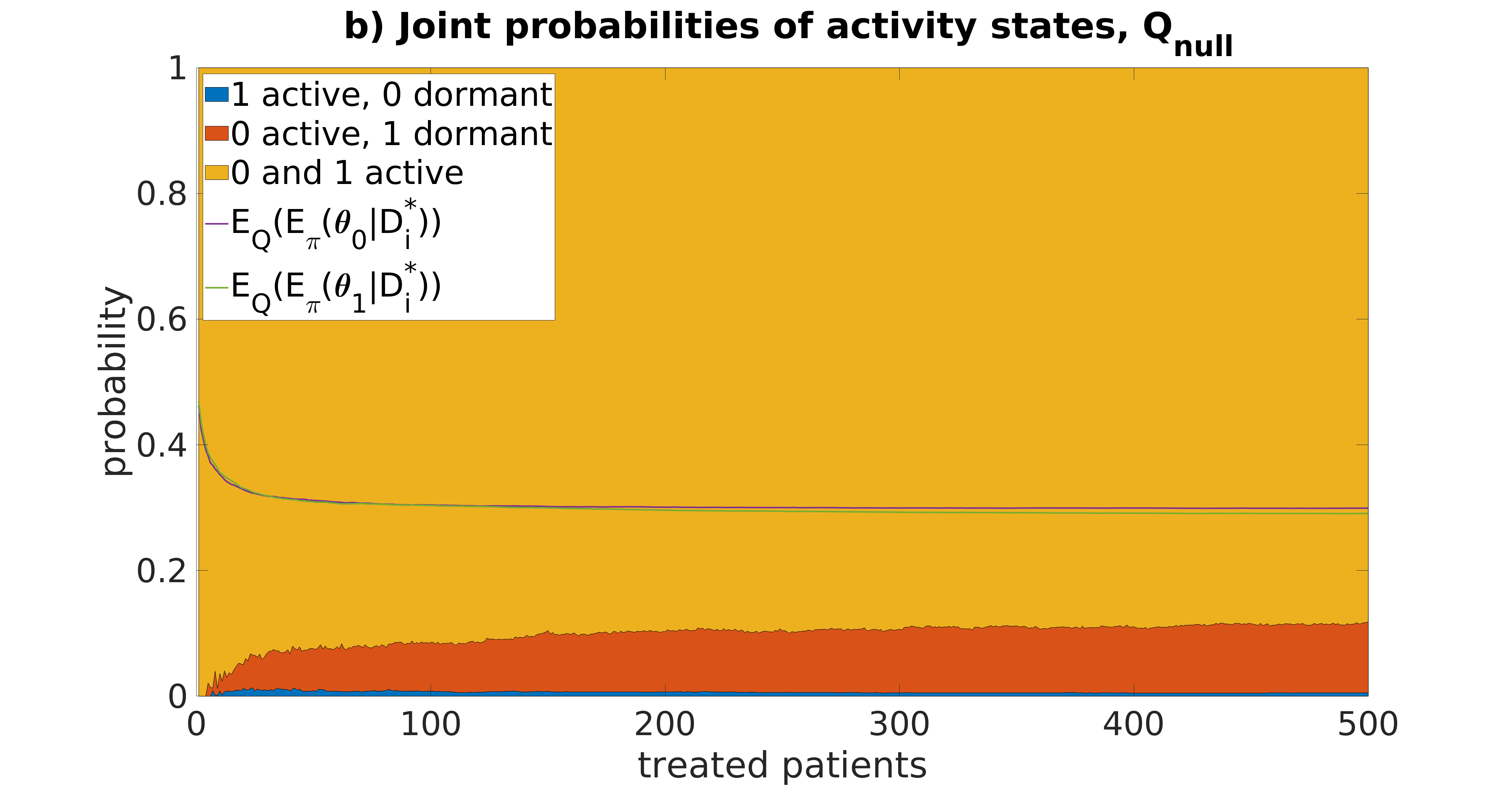}
         \renewcommand\thesubfigure{b}  
         \label{subfig:activity:b:null}
     \end{subfigure}
     \begin{subfigure}[b]{0.5\textwidth}
         \includegraphics[width=\textwidth]{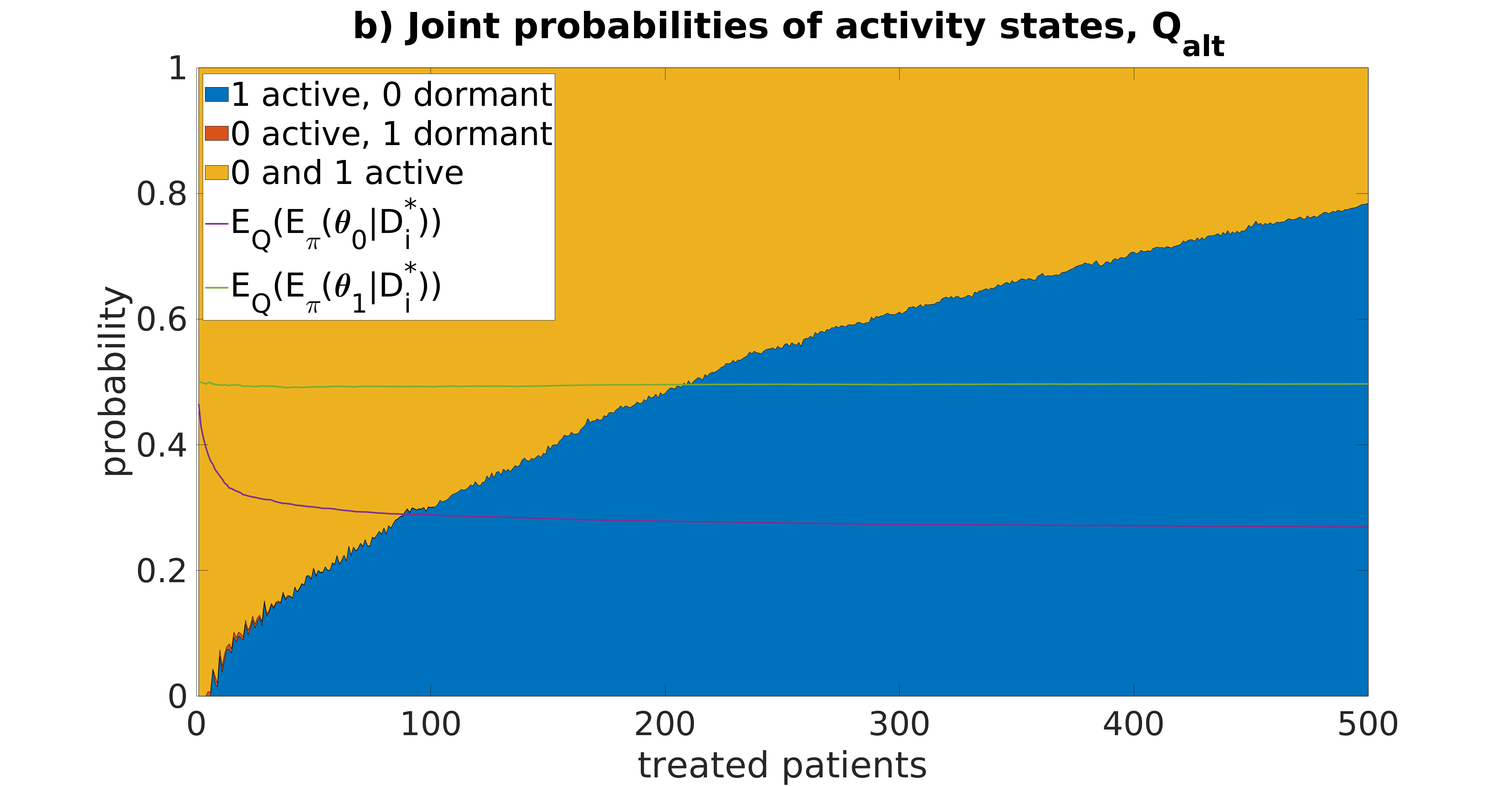}
     \renewcommand\thesubfigure{b}  
         \label{subfig:b:activity:alternative}
     \end{subfigure}
     \hfill
     \begin{subfigure}[b]{0.5\textwidth}

     \includegraphics[width=\textwidth]{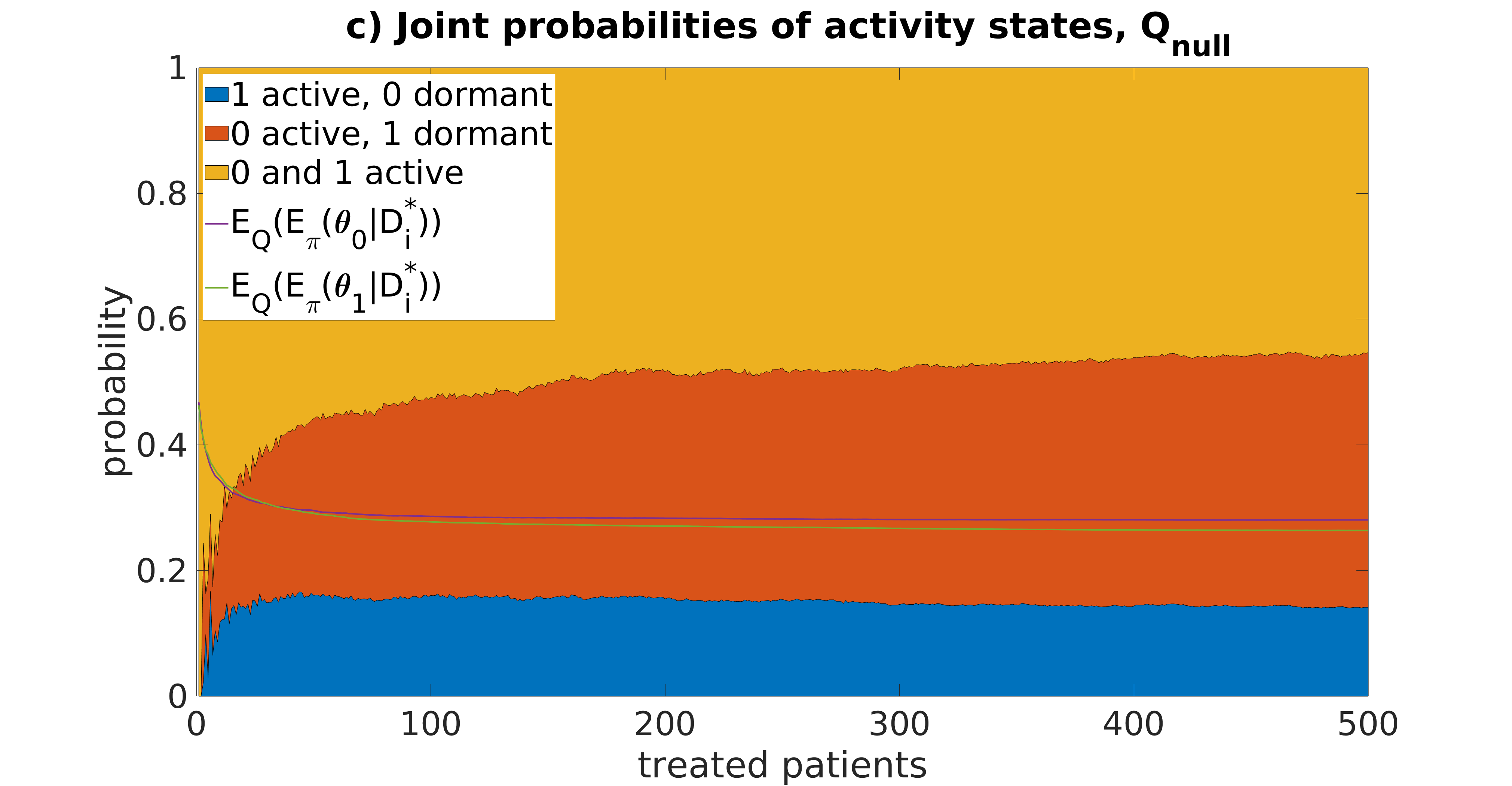}
        \renewcommand\thesubfigure{c}   
         \label{subfig:activity:c:null}
     \end{subfigure}
    \begin{subfigure}[b]{0.5\textwidth}
     \includegraphics[width=\textwidth]{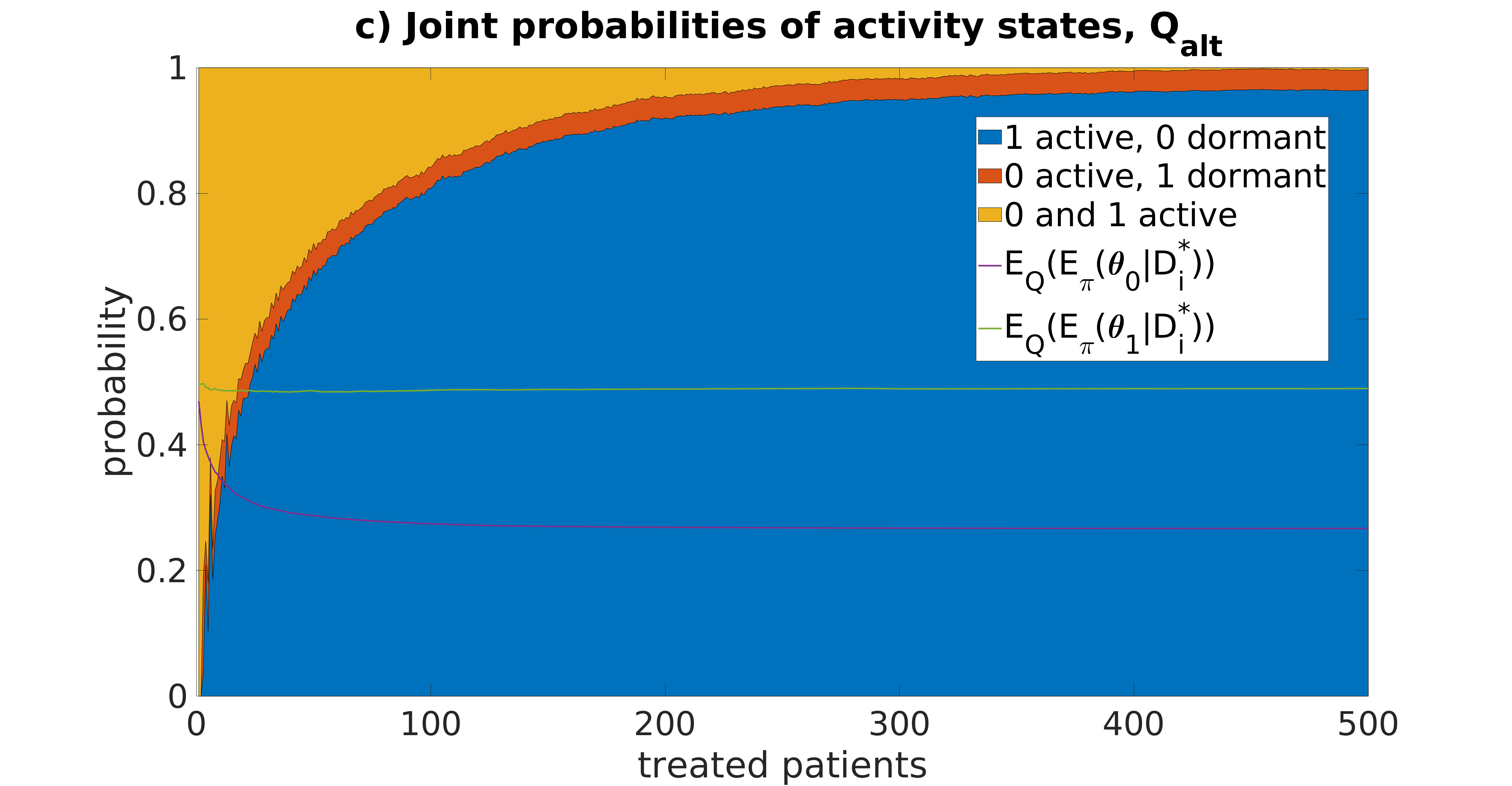}
      \renewcommand\thesubfigure{c}  
         \label{subfig:activity:c:alternative}
     \end{subfigure}
\caption{\small Effect of the choice of the values of the design parameters $\varepsilon$ and $\delta$ on the joint activity states of the two treatment arms when applying Rule 1 for treatment allocation. Joint probabilities of the different combinations  of active and dormant states are shown, as functions of the number $i$ of treated  trial participants. The results are based on   $5000$  simulated data sets of size $N_{\max}= 500$,  under $\Q_{null}$ with true parameter values
$\theta_0=\theta_1=0.3$ (left) and $\Q_{alt}$ with values
$\theta_0=0.3,\theta_1=0.5$ (right). Three combinations of the  design  parameters were used:  {(a)} $\varepsilon=0.1$, $\delta=0.1$ (top), {(b)}  $\varepsilon=0.05$,
$\delta=0.1$ (middle),   {(c)}  $\varepsilon=0.2$,   $\delta=0.05$  (bottom).  Also shown are the expectations  $\E_{\Q_{null}}\bigl(\E_{\pi}\bigl( \bftheta_k\vert D_i^{*} \bigr)\bigr)$ and $\E_{\Q_{alt}}\bigl(\E_{\pi}\bigl( \bftheta_k\vert D_i^{*} \bigr)\bigr), (1\leq i \leq 500, k=0,1)$, computed from these simulations.} 
      \label{fig:combined_Activity_RULE1}
\end{figure}

Next, we study the effect of the choice of the design parameters $\varepsilon$ and $\delta$ in Rule 1  on some frequentist type key characteristics of a trial. Figure \ref{fig:combined_Activity_RULE1} illustrates this effect for the joint distribution of the activity indicators $I_{0}$ and $I_{1}$, considered as a function of the number of treated   patients. Empirical probabilities are shown, based on  $5000$ simulated trials of size $N_{\max}= 500$, under $\Q_{null}$  with true parameter values $\theta_0=\theta_1=0.3$ (left), and under $\Q_{alt}$   with  $\theta_0=0.3,\theta_1=0.5$ (right). 

For $\Q_{alt},$ $\theta_0 + \delta = 0.3 + 0.1 < 0.5 = \theta_1,$ and therefore the posterior probabilities $\P_{\pi} ( \bftheta_0+ \delta \geq \bftheta_{1}\big\vert D_i^{*} ) $ tend to be small, at least for larger values of $i$. When they are below the threshold $\varepsilon$, compliance with Rule 1 forces the control arm to be dormant. We can see this happening in Figure \ref{fig:combined_Activity_RULE1} on the right, where the $\Q_{alt}$ probability of $\{I_{0}=0,I_{1}=1\}$ clearly dominates that of $\{I_{0}=1,I_{1}=0\}$. The effect   is strongest in the liberal parameter setting (c), and weakest but still quite strong in the conservative alternative (b). In contrast, under $\Q_{null}$, with $\theta_0  = \theta_1,$ the configuration $\{I_{0}=1,I_{1}=1\}$  remains the most likely alternative during the entire follow-up, with the strongest tendency to do so in the conservative design (b) and the weakest in the liberal (c). A third aspect to be noted on the left of Figure \ref{fig:combined_Activity_RULE1} is that  the configuration $\{I_{0}=1,I_{1}=0\}$ was always much more likely than $\{I_{0}=0,I_{1}=1\},$ due to the control arm being protected by the positive safety margin $\delta$.

Finally, Figure \ref{fig:combined_Activity_RULE1} shows the expectations of  $\E_{\pi}\bigl( \bftheta_k\vert D_i^{*} \bigr)$ and $\E_{\pi}\bigl( \bftheta_k\vert D_i^{*} \bigr), (1\leq i \leq 500, k=0,1)$, computed from these simulations under $\Q_{null}$ and $\Q_{alt}$. For small $i$ all these values are close to $0.5$, originating from the  \textit{Uniform}$(0,1)$-priors assumed in all simulations. With more data, the curves stabilize close to the true parameter values, but exhibit then a small negative bias. This is an aspect shared by all adaptive methods favoring in treatment allocation arms with relatively more successes in the past, see e.g. \textcite{Villar2015}. Given that the main goal of each on-going trial is the mutual comparison of the different treatments involved, and that this assessment is here made with respect to the joint posterior based on the current trial data, the frequentist property
of a small bias in the estimation of the individual treatment success parameters, in the same direction, does not seem very crucial.

\begin{figure}[!htbp]
     \begin{subfigure}[b]{\textwidth}
     \includegraphics[width=\textwidth]{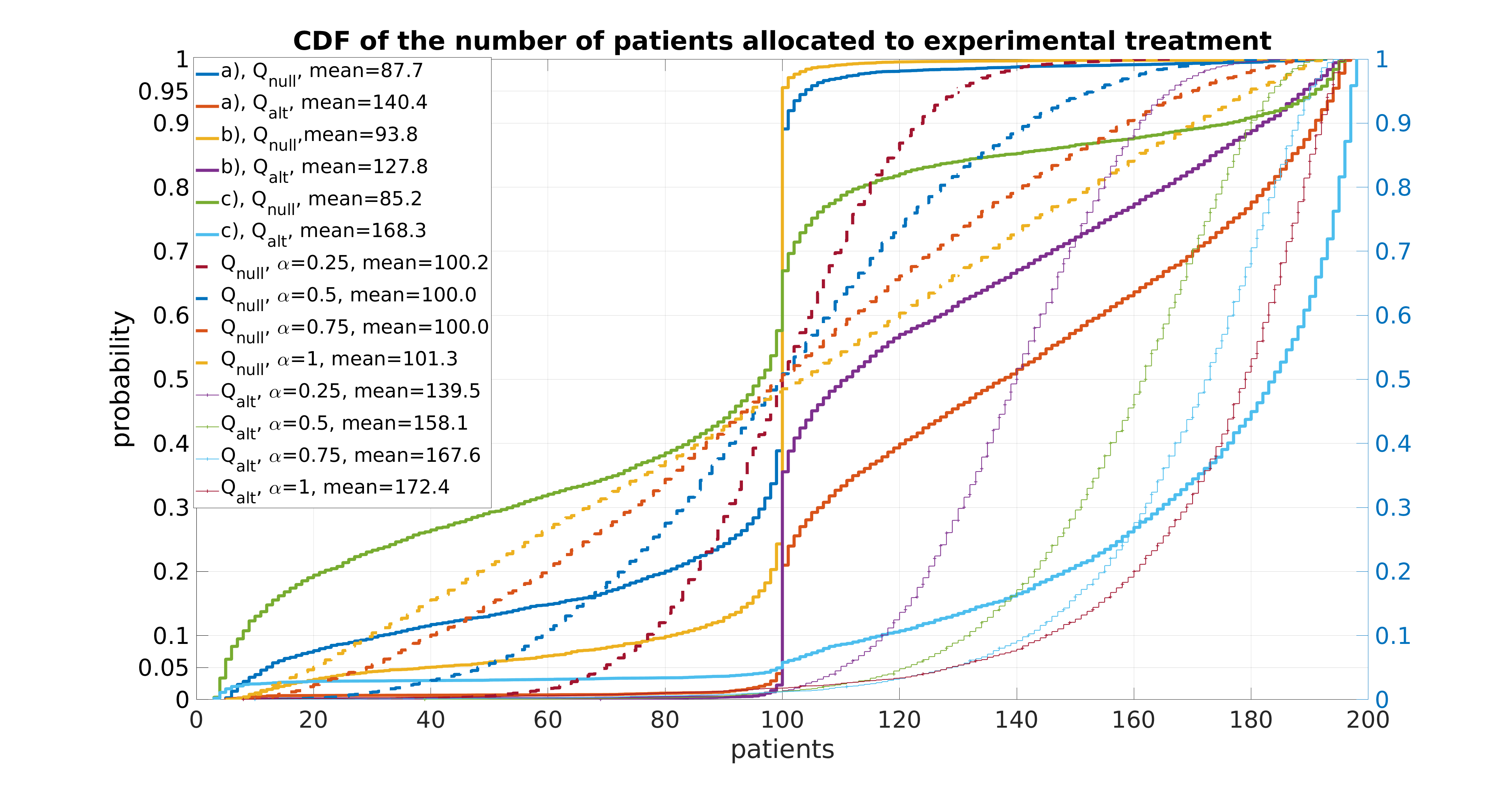}
       \renewcommand\thesubfigure{a}  
     \label{subfig:Ystat:a:null:rule1_200}
     \end{subfigure}
     \vfill 
     \begin{subfigure}[b]{\textwidth}
     \includegraphics[width=\textwidth]{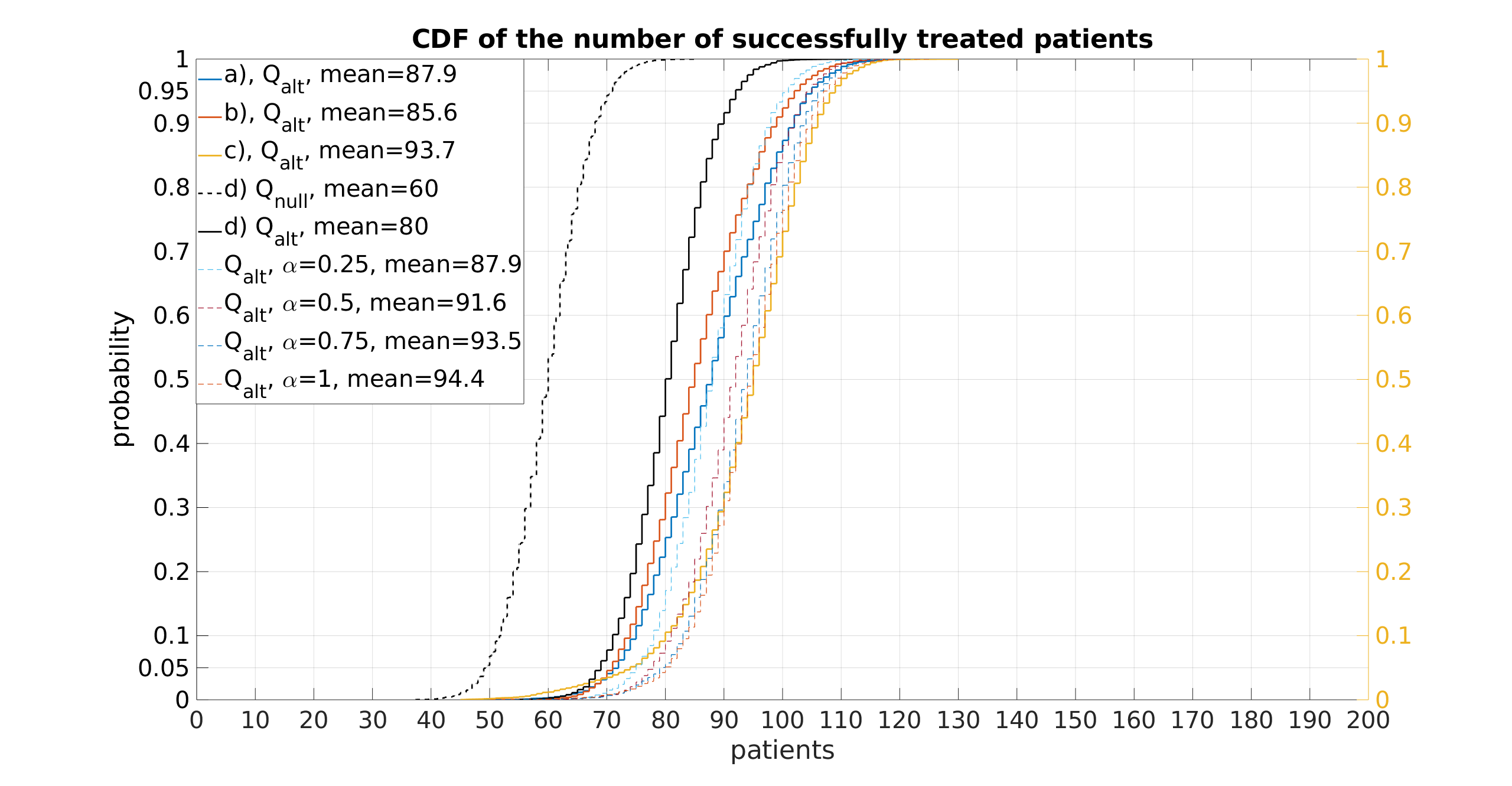}
        \renewcommand\thesubfigure{a} 
 \label{subfig:Ystat:a:alternative:rule1_200}
     \end{subfigure}
\caption{\small Effect of the choice of the  design  parameters $\varepsilon$ and $\delta$ in Rule 1 on the number of   patients allocated to the experimental treatment and on the total number of treatment successes. Cumulative distribution functions of $N_1(200)$ (top) and $S(200)$ (bottom) are shown, based on $5000$ simulated data sets,  under $\Q_{null}$ with true parameter values
$\theta_0=\theta_1=0.3$ and $\Q_{alt}$ with values
$\theta_0=0.3,\theta_1=0.5$. Three combinations of the 
design   parameters were used:  {(a)} $\varepsilon=0.1$,  $\delta=0.1$, {(b)}   $\varepsilon=0.05$, $\delta=0.1$, {(c)}
$\varepsilon=0.2$, $\delta=0.05$. In addition, {(d)} represents a completely symmetric treatment allocation. For comparison we also plot the corresponding CDF under the alternative hypothesis obtained by using fractional Thompson's rule 
with respective parameters
$\kappa=0.25,0.5,0.75$ and $1$.
}
\label{fig:combined_Ystat_RULE1_200}
\end{figure}

A complementary point of view is presented in Figure   \ref{fig:combined_Ystat_RULE1_200} showing the cumulative distribution functions (CDFs) of $N_1(200)$, the number of patients out of the first  $200$ assigned by Rule 1 to the experimental treatment,   and  of   $S(200)$, the total number of successes from both treatments combined. Corresponding results from considering the first $100$ and $500$ patients are shown in Figures \ref{fig:combined_Ystat_RULE1_100} and \ref{fig:combined_Ystat_RULE1} included in the Supplement. 

The CDF's in these figures are based on simulated data sets from $\Q_{null}$ and $\Q_{alt}$ by using the same parameter settings (a), (b) and (c) of Rule 1 as in Figure 2 and, in addition, (d) where adaptive treatment allocation was inactivated by applying threshold value $\varepsilon = 0,$ then leading to a  completely symmetric block randomization. Finally, for a comparison, also shown are the CDFs of these variables when adaptive treatment allocation of patients was applied by using Thompson's rule with fractional exponents $\kappa = 0.25, 0.50, 0.75$ and $1.00.$ Note that $\kappa = 0 $ would correspond to  treatment assignment by tossing a fair coin, and therefore the corresponding CDF of $S(200)$ would be very similar to that obtained under Rule 1 (d).
 
The top part of Figure \ref{fig:combined_Ystat_RULE1_200} shows how the application of Rule 1, under $\Q_{null},$ leads to often allocating exactly half of the patients to both treatment arms, which happens in trial runs during which the  dormant state had not been entered even once. Overall, due to the protective safety margin $\delta >0$, Rule 1 has a tendency of  allocating more patients to the   control arm.  Thompson's rule, in contrast, behaves symmetrically for data coming from $\Q_{null}$. Under $\Q_{alt},$ in which case the true success rate of the experimental treatment is higher than that of the control, after a training period, more than half of the patients will usually be given this better treatment. 

In Rule 1, the mode of control changes abruptly at times at which one of the two treatment arms enters the dormant state. The different versions of Rule 1 can therefore be said to represent a \textit{bang-bang} type of system control. Thompson's rule applies a randomization scheme based on continuously updated posterior probabilities, and in this sense represents a softer control type. Under $\Q_{alt}$, the better performance of the experimental treatment is usually detected rather early in the trial, and then, with more evidence from the data, all adaptive rules use progressively stronger control in directing patients to this better treatment. However, there is a small  probability that, accidentally,   more patients are given the inferior control treatment. It is clear, as is also illustrated by Figure\ref{FIGURE:yksi}, that the risk for this to happen is highest early in the trial when there are only few observed outcomes. In the present simulations, these $\Q_{alt}$-probabilities were, respectively,  $0.041$, $0.023$  and  $0.049$,
under Rule 1 with parameters (a), (b) and (c). 

In   the bottom part of Figure   \ref{fig:combined_Ystat_RULE1_200}, the CDFs for   $S(200)$ under $\Q_{null}$ are identical in all designs, due to both treatment arms having the same true response rate $0.3$. For $\Q_{alt},$
employing the symmetric block randomization scheme Rule 1 (d) gives $\E_{\Q_{alt}}(S(200)) = 80$, and if all patients could be given the better experimental treatment, the resulting optimal expected value would be $100.$ In Figure   \ref{fig:combined_Ystat_RULE1_200}, the expectations $\E_{\Q_{alt}}(S(200))$ for different adaptive schemes range from $85.6$ for Rule 1 (b), to $94.4$ for Thompson's rule with $\kappa = 1.$

\textbf{Employing an initial burn-in period.} The potential problem of accidentally allocating more patients to an inferior treatment arm can be mitigated by delaying the workings of the adaptive mechanism of Rule 1, or Thompson's rule, by employing the symmetric block randomization scheme (d) until a fixed number of patients have been assigned to all treatments. To have an idea of the size of the effect of this modification in the present example, we carried out a simulation study identical to that leading to Figure \ref{fig:combined_Ystat_RULE1_200} except that, of the considered $200$ patients, the first $30$ were divided evenly to the two treatments, $15$ to both.
The result is shown in the Supplement Figure
\ref{fig:combined_Ystat_RULE1_200_16-Mar-2021}.
The probabilities of imbalance in the unwanted direction are now lower, respectively $0.013$, $0.005$ and $0.019$ for Rule 1 (a), (b) and (c). On the other hand, delaying the adaptive mechanism from taking effect  until outcome data from the first $30$ patients are available obviously lowers, in case of $\Q_{alt},$ the expected number of  treatment successes by small amounts. For additional comments on the effects of burn-in, see the Supplement. Alternative versions of burn-in in adaptive designs have been considered, e.g., in \textcite{Thall2007} and \textcite{Thall2015StatisticalCI}.  
 
\subsubsection{Effect of adaptive treatment allocation on  frequentist performance measures} \label{sub:3.1.3}

There are no free lunches, and these potential gains in terms of either more efficacious treatments given to more patients in the trial, or smaller numbers of treated patients needed for being able to select the better treatment, are to be weighed against corresponding potentially stronger statistical inferences that might be obtained from more  balanced designs. For a numerical comparison, we applied a design where adaptive patient allocation was applied following either Rule 1 or Thompson's rule, and an assessment of the results, including the possibility of dropping a treatment arm, was only allowed at the time at which a pre-specified number $i = N_{max}$ of patients had been treated. Here we consider the choice $N_{max}= 200$, reporting the results from experiments with $N_{max}= 100$ and $500$ in the Supplement. 

In a trial with only two treatments, dropping either one is taken to mean selection of the other. The final analysis made at $N_{max}$ need not necessarily use the same threshold values as Rule 1, and therefore we use new notations $\varepsilon_0$ and $\delta_0$ for them. Accordingly, when performing such an analysis at $i = N_{max}$, the control arm is dropped if   $ \P_{\pi} ( \bftheta_0+\delta_0 \geq \bftheta_{1} \vert D_{N_{max}}^{*} ) \leq \varepsilon_0$ and the experimental arm if $\P_{\pi} ( \bftheta_1  \geq \bftheta_{0} \vert D_{N_{max}}^{*} )\leq \varepsilon_0$. Obviously, at most one of these criteria can be satisfied for given data $D_{N_{max}}^{*}$ when $\varepsilon_0 < 0.5$. But it is also possible that neither of them is satisfied, in which case no firm decision concerning treatment selection is made at $N_{max}$.  

Even then, however, there is the possibility of studying the joint posterior $ \P_{\pi} ( (\bftheta_0, \bftheta_1)  \in \cdot   \vert D_{N_{max}}^{*} )$ for the purpose of drawing further inferences from the results of the trial. For example, one can print the posterior CDF $ \P_{\pi} ( \bftheta_1 - \bftheta_{0} \leq x  \vert D_{N_{max}}^{*} ), -1 \leq x \leq 1,$ and then decide, the study protocol permitting this, whether to continue the trial by recruiting more participants. This may then lead to either one of the two selection criteria being satisfied at a later point in time.

\begin{figure}[!htbp] 
     \begin{subfigure}[b]{1.0\textwidth}
    \includegraphics[height=\macroheight,width=\textwidth]{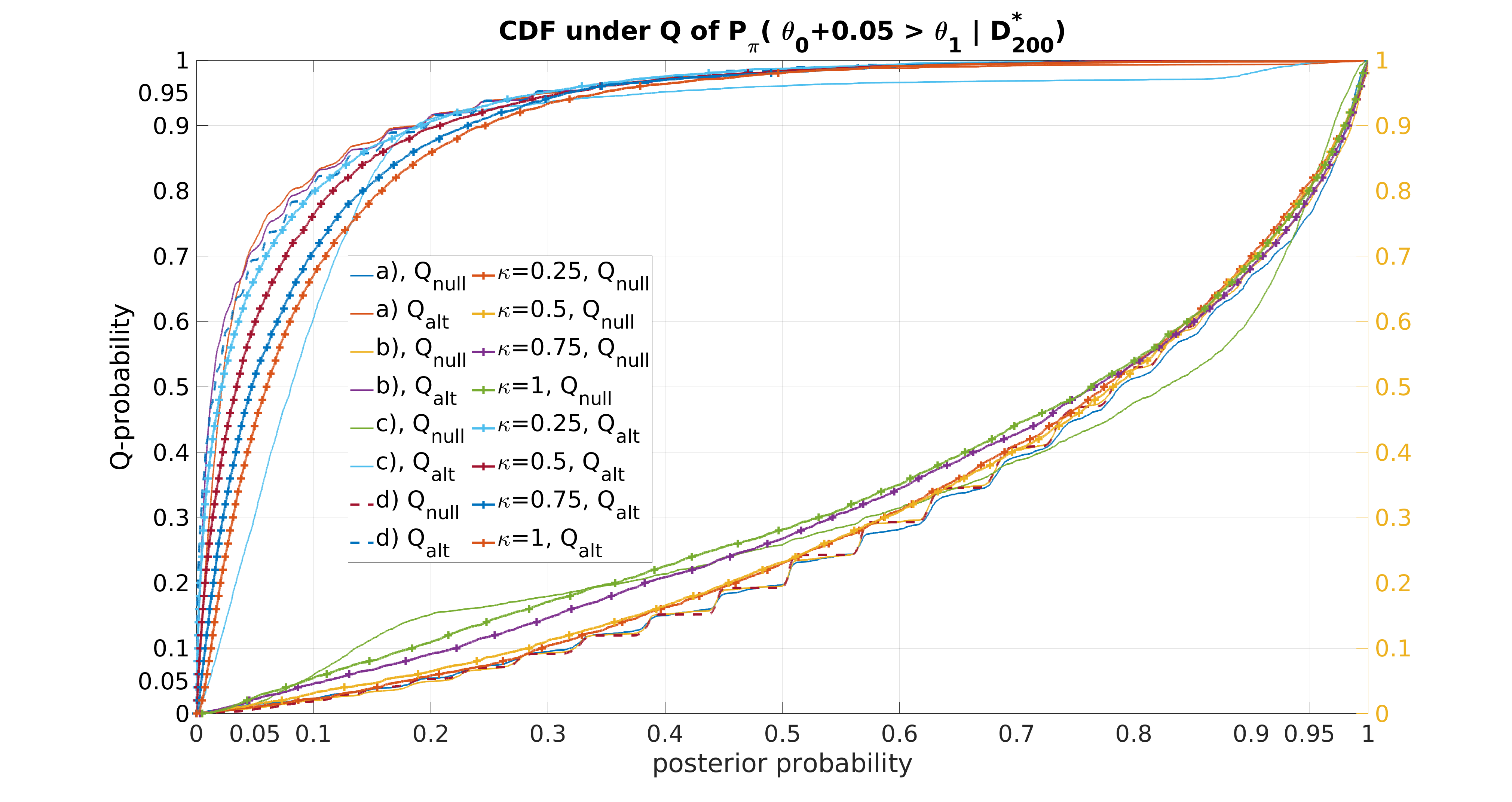}
 \renewcommand\thesubfigure{a}  
         \label{subfig:CDF_of_Pmax_rule1_200:0}
     \end{subfigure} \vfill
    \begin{subfigure}[b]{1.0\textwidth}
     \includegraphics[height=\macroheight,width=\textwidth]{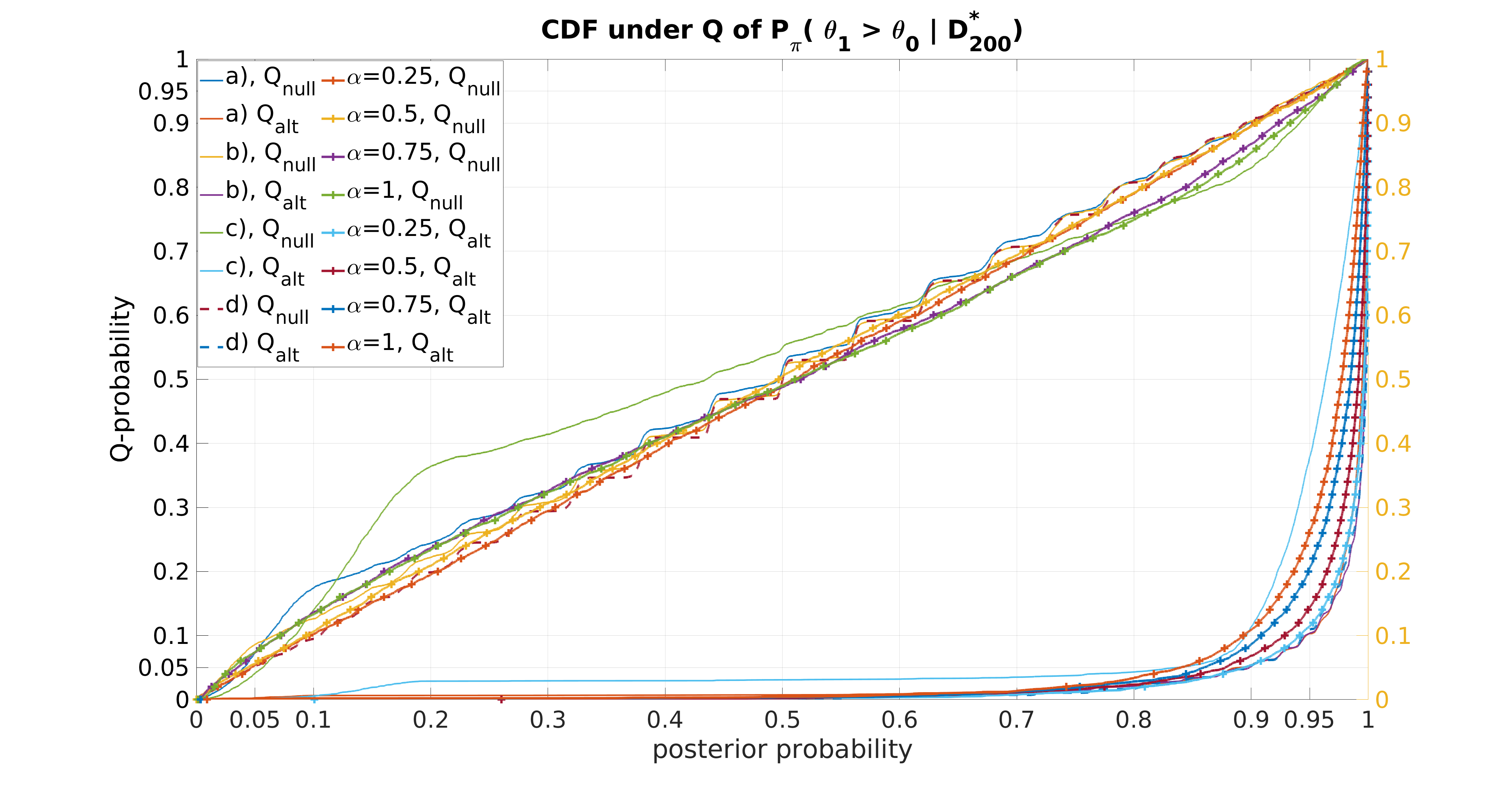}
      \renewcommand\thesubfigure{c}  
\label{subfig:CDF_of_pmax_rule1:1}
     \end{subfigure}\captionsetup{singlelinecheck=off}
\caption[short caption]{
\small   Effect of the  design  parameters $\varepsilon$ and $\delta$ of Rule 1, and $\kappa$ of Thompson's rule, on the CDFs of the posterior probabilities
$\P\bigl( \theta_0 +0.05 \geq \theta_1\big\vert D^*_{200}\bigr)$ (top)
and $\P\bigl( \theta_1 \geq \theta_0\big\vert D^*_{200}\bigr)$ (bottom) in the 2-arm trial of Experiment 1 when applying Rule 1 for treatment allocation and making a final assessment at $i = N_{\max}=200$.  
The results are based on
$5000$ data sets generated under $\Q_{null}$ and $\Q_{alt}$ when using the following combinations of  design  parameters:
 {(a)}  
  $\varepsilon=0.1, 
  \delta=0.1$,
 {(b)}
  $\varepsilon=0.05, 
  \delta=0.1$,
  {(c)}
  $\varepsilon=0.2, 
  \delta=0.05$. 
 }
 \label{fig:CDF_of_pmax_rule1_200}
\end{figure}

We now study how the application of different versions of adaptive treatment allocation influences the strength of statistical inferences, viewed from a frequentist perspective, that can be drawn from trial data in Experiment 1. For this, we consider the probabilities $\Q(\P_{\pi}( \bftheta_0+\delta_0 \geq \bftheta_{1}\vert {\mathbf D}_{N_{max}}^{*})\leq \varepsilon_0) $ and   $\Q(\P_{\pi}(\bftheta_{1} \geq \bftheta_0  \vert {\mathbf D}_{N_{max}}^{*}) \leq  \varepsilon_0)$ for both $\Q = \Q_{null}$ and $\Q =\Q_{alt}$, choosing   $\delta_0 =0.05$ and   $\varepsilon_0 = 0.05$. We applied Rule 1 with  design  parameters (a), (b), (c) and (d), and Thompson's rule with fractional exponents $\kappa = 0.25, 0.50, 0.75$ and $1.0$. Our simulation experiment consisted of $5000$ repetitions of a trial up to $500$ patients. From each simulated data set we then computed numerical values for the posterior probabilities $\P_{\pi}\bigl( \bftheta_1 \geq \bftheta_{0}\big\vert  D_{N_{max}}^{*}\bigr)$ and $\P_{\pi}\bigl( \bftheta_0+\delta_0 \geq \bftheta_{1}\big\vert D_{N_{max}}^{*}\bigr)$, and drew, at $N_{max} = 200$, the resulting CDFs under $\Q_{null}$ and $\Q_{alt}$, shown in Figure   \ref{fig:CDF_of_pmax_rule1_200}. The corresponding figures at $N_{max} = 100$ and $N_{max} = 500$  are included in the Supplement as Figures \ref{fig:CDF_of_pmax_rule1_100}   and \ref{fig:CDF_of_pmax_rule1}.

Under $\Q_{null}$, the CDFs of $\P_{\pi}\bigl( \bftheta_1 \geq \bftheta_0\big\vert {\mathbf D}_{200}^{*}\bigr)$ for different designs, shown in the bottom part of Figure \ref{fig:CDF_of_pmax_rule1_200}, are almost linear, which would correspond to the \textit{Uniform}$(0,1)$ sampling distribution. This is the case particularly in the designs following Thompson's rule, where the two treatment arms are considered symmetrically. For Rule 1 the deviations from linearity are clearer, and most evident in the case of Rule 1 (c). The overall shape of the  CDFs of $\P_{\pi}\bigl( \bftheta_0+0.05 \geq \bftheta_{1}\big\vert {\mathbf D}_{200}^{*}\bigr)$ in the top part of Figure \ref{fig:CDF_of_pmax_rule1_200} is convex, signalling that the $\Q_{null}$-density of these posterior probabilities tends to increase as their values increase. The reason is the threshold $\delta_0 = 0.05$ providing extra protection for the control arm arm against being dropped. 

The CDFs generated under $\Q_{alt}$ behave very differently. Those of $\P_{\pi}\bigl( \bftheta_1 \geq \bftheta_0\big\vert {\mathbf D}_{200}^{*}\bigr)$ in the bottom part of Figure \ref{fig:CDF_of_pmax_rule1_200} show a high concentration of values close to $1$, and those of $\P_{\pi}\bigl( \bftheta_0+0.05 \geq \bftheta_{1}\big\vert {\mathbf D}_{200}^{*}\bigr)$ in the top part of Figure \ref{fig:CDF_of_pmax_rule1_200} a somewhat lower but still high concentration close to $0.$ The main difference between these CDFs stems from the opposite directions of the inequalities between $\bftheta_0$ and $\bftheta_1$, and the difference in concentration is again due to the threshold $\delta_0 = 0.05$.

Based on  these results, we then computed  numerical values for the true and false positive and negative rates, shown in Table \ref{table:1}. More exactly, we use the terms

\textit{false positive rate}  = $\Q_{null}(\P_{\pi}( \bftheta_0+\delta_0 \geq \bftheta_{1}\vert {\mathbf D}_{N_{max}}^{*})\leq \varepsilon_0)$, 

\textit{true negative rate} = $ \Q_{null}(\P_{\pi}(\bftheta_{1} \geq \bftheta_0  \vert {\mathbf D}_{N_{max}}^{*}) \leq  \varepsilon_0)$, 

\textit{true positive rate} = $\Q_{alt}(\P_{\pi}( \bftheta_0+\delta_0 \geq \bftheta_{1}\vert {\mathbf D}_{N_{max}}^{*})\leq \varepsilon_0)$ and 

\textit{false negative rate} = $\Q_{alt}(\P_{\pi}(\bftheta_{1} \geq \bftheta_0  \vert {\mathbf D}_{N_{max}}^{*}) \leq  \varepsilon_0).$ 

In addition, the probabilities
$\Q(\P_{\pi}( \bftheta_0+\delta_0 \geq \bftheta_{1}\vert {\mathbf D}_{N_{max}}^{*}) > \varepsilon_0,  \P_{\pi}(\bftheta_{1} \geq \bftheta_0  \vert {\mathbf D}_{N_{max}}^{*}) > \varepsilon_0)$  are called \textit{inconclusive rates}, respectively, under $\Q =\Q_{null}$ and $\Q=\Q_{alt}$.  

\begin{table} {\small
\begin{center}
\begin{tabular}{ | c | c c c c c c c c | }\hline \hline 
 $\varepsilon_0=0.05,\delta_0=0.05$
 & (a) & (b) & (c) &(d) & $\kappa=0.25$ & $\kappa=0.5$ & $\kappa=0.75$ & $\kappa = 1$
 \\  \hline
$\Q_{null}:$ false positive & 
  0.014 &    0.009    & 0.014 &    0.007  &  0.011 &    0.014   & 0.023  &   0.025
\\ \hline 
 $\Q_{null}:$ true negative & 
  0.074   &  0.086    & 0.040 &    0.052&    0.054  &  0.056 &    0.073 &   0.074
 \\
    \hline $\Q_{null}:$
 inconclusive &
 0.912  &  0.906 &   0.946 &   0.941     &0.935  &  0.929    &0.904   & 0.901
 \\ \hline 
 $\Q_{alt}:$ true positive  & 0.723&    0.711 &   0.303&    0.694 &   0.665   & 0.598  &  0.516 &   0.443 \\
  \hline
$\Q_{alt}:$ false negative & 
0.002  &  0.001 &   0.000 &        $\sim 0$ &        $\sim 0$ &    
$\sim 0$ &    0.001&    0.001 
\\
\hline
$\Q_{alt}:$ inconclusive &
 0.275  &  0.288   & 0.696  &  0.306  &  0.335 &   0.402&    0.483  &  0.555
 \\  
 \hline
\end{tabular}
\end{center}
\caption{\label{table:1}True and false positive and negative rates when applying adaptive treatment allocation with design parameter values $\varepsilon_0 = 0.05$ and $\delta_0 = 0.05$ in a trial of size $N_{max} = 200$.}
}
\end{table}

The following conclusions are now immediate from Table \ref{table:1}. For $N_{max} = 200$, the false positive rates are small, below 2.5 percent, for all considered versions of adaptive treatment allocation. This is true even for the "liberal" design parameters (c) in Rule 1 for which there was a non-negligible probability, about five percent, of serious imbalance in treatment allocation in the unwanted direction. The false negative rates are very small for all considered designs. Under $\Q_{null}$, the trial remains inconclusive with probability at least ninety percent, which is consistent with the fact that then there is no difference between the true response rates $\theta_{0}$ and $\theta_{1}$.  Finally, the true positive rate (\textit{power}) is on the moderate level of approximately seventy percent when applying Rule 1 with design parameter values (a), (b) and (d), and almost as high for Thompson's rule with $\kappa = 0.25$. Recall here that (d) means symmetric block randomization, which can thought to provide a suitable yardstick for such comparisons of power. For larger values of  $\kappa$, for which the adaptive mechanism is stronger, these rates are smaller. Of all considered alternatives, the smallest true positive rate is obtained for the design parameters (c). 

Corresponding tables for $N_{max} = 100$ and $N_{max}=500$ are provided, and commented on, in the Supplement as Table \ref{table:S1} and Table \ref{table:S2}.

\textbf{Employing an initial burn-in period.} We also studied the effect of the burn-in period, described above in subsection 3.1.2, on the  frequentist performance measures   in Table \ref{table:1}. For this, we drew CDFs (not shown) similar to those in Figure \ref{fig:CDF_of_pmax_rule1_200},  and then worked out numerical values for the true and false positive and negative rates. The results, with some comments, can be found in the Supplement Table \ref{table:1:burnin}.  

\textbf{Remarks on other test variants.} Somewhat different numerical values are obtained if the positive safety margin $\delta_0$ protecting the control arm from being dropped is given the value $\delta_0 =0$. With this extra protection removed, the rates of positive findings, both true and false, will naturally increase, while the rates of negative results remain unchanged. Another modification is to change the presently used decision criterion $\P_{\pi}(\bftheta_{1} \geq \bftheta_0  \vert {\mathbf D}_{N_{max}}^{*}) \leq  \varepsilon_0$ for dropping the experimental arm into  $\P_{\pi}(\bftheta_{1} \geq \bftheta_0 + \delta_0 \vert {\mathbf D}_{N_{max}}^{*}) \leq  \varepsilon_0$, in which case it would be symmetric to the condition $\P_{\pi}(\bftheta_0 + \delta_0   \geq \bftheta_1 \vert {\mathbf D}_{N_{max}}^{*}) \leq  \varepsilon_0$ for dropping the control arm. If made, such a conclusion (in effect, declaring \textit{futility}) is made more easily. The true and false negative rates become then larger, while the rates of positive results remain unaltered. For both variants, the inconclusive rates are larger than when applying the original criteria. Numerical values for   these two variants, with $N_{max} = 200$, are provided in Supplement Tables \ref{table:S3} and \ref{table:S4}.

It depends on the concrete context whether either one of these alternative criteria would be considered more appropriate than the version used in the construction of Table \ref{table:1}. All three represent different forms of \textit{superiority} trials. After a suitable modification, the same basic structures would also apply for testing \textit{non-inferiority} and \textit{equivalence} hypotheses (e.g., \textcite{lesaffre2008superiority}).

\subsubsection{Effect of the design parameters on treatment selection}

\begin{figure}[!htbp]
\begin{subfigure}[b]{0.5\textwidth}
     \includegraphics[width=\textwidth]{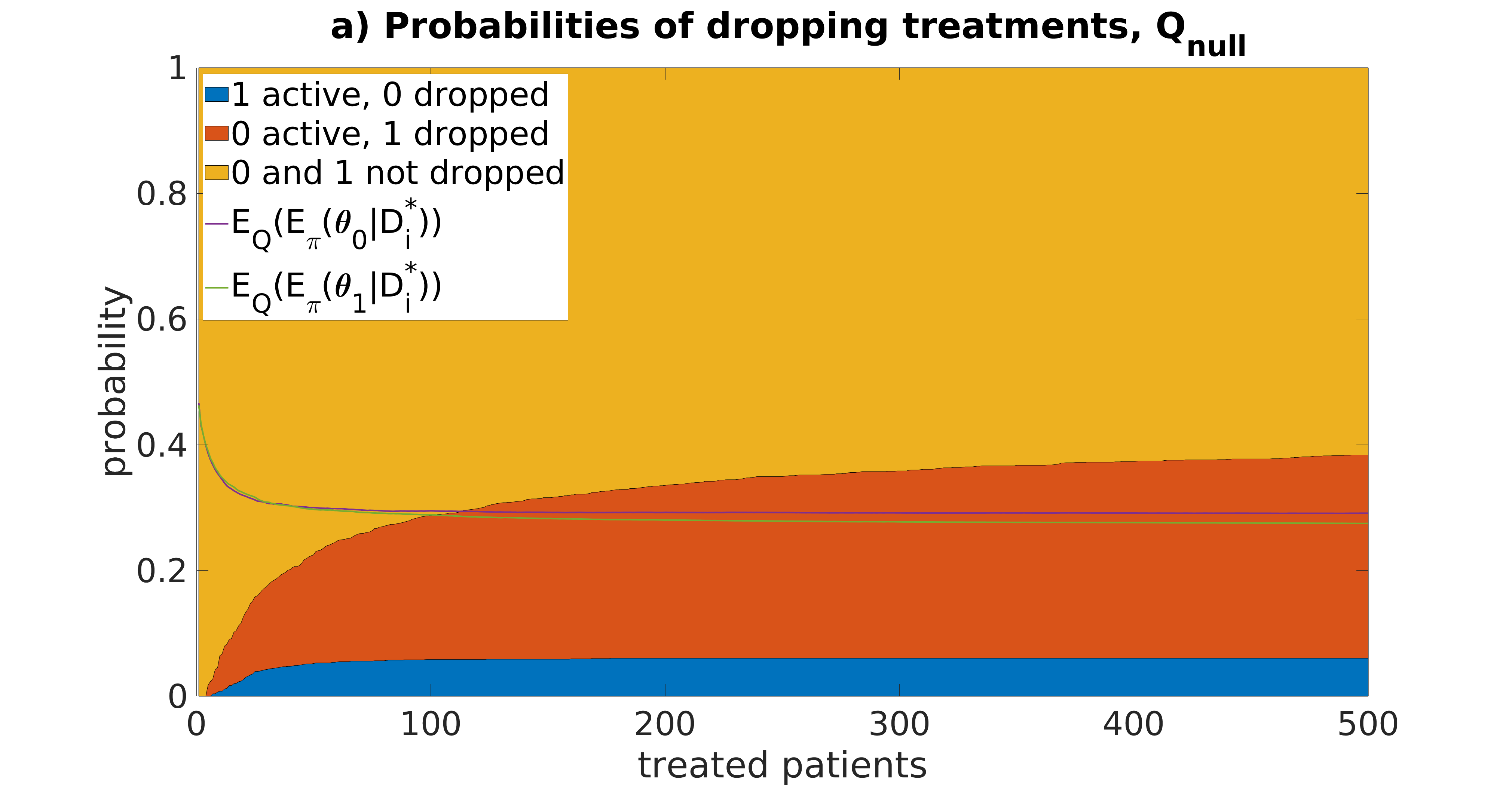}
       \renewcommand\thesubfigure{a}  
         \label{subfig:3activity:a:null}
     \end{subfigure}
     \begin{subfigure}[b]{0.5\textwidth}
     \includegraphics[width=\textwidth]{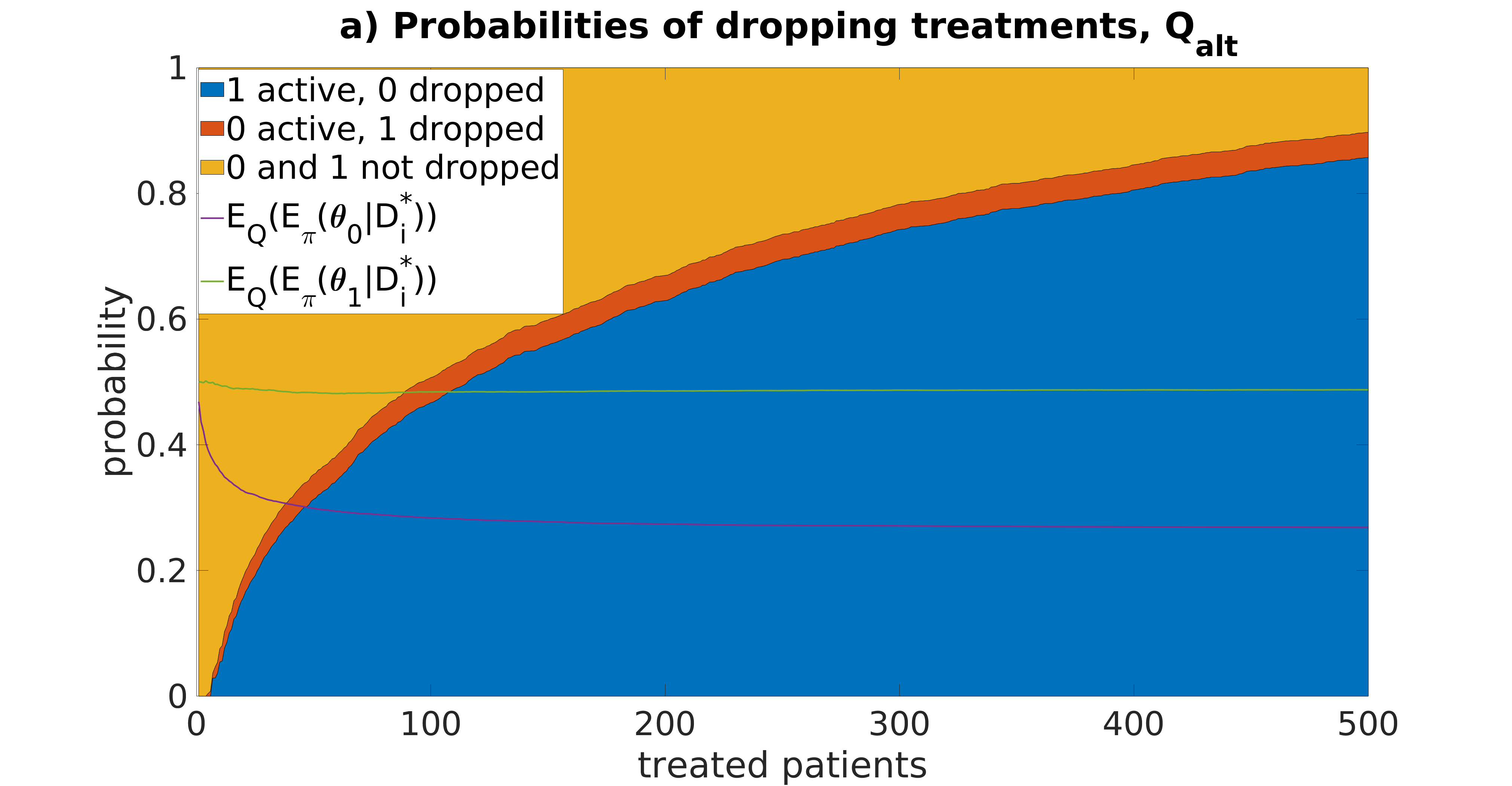}
        \renewcommand\thesubfigure{a} 
         \label{subfig:3activity:a:alternative}
     \end{subfigure}
     \hfill
     \begin{subfigure}[b]{0.5\textwidth}
         \includegraphics[width=\textwidth]{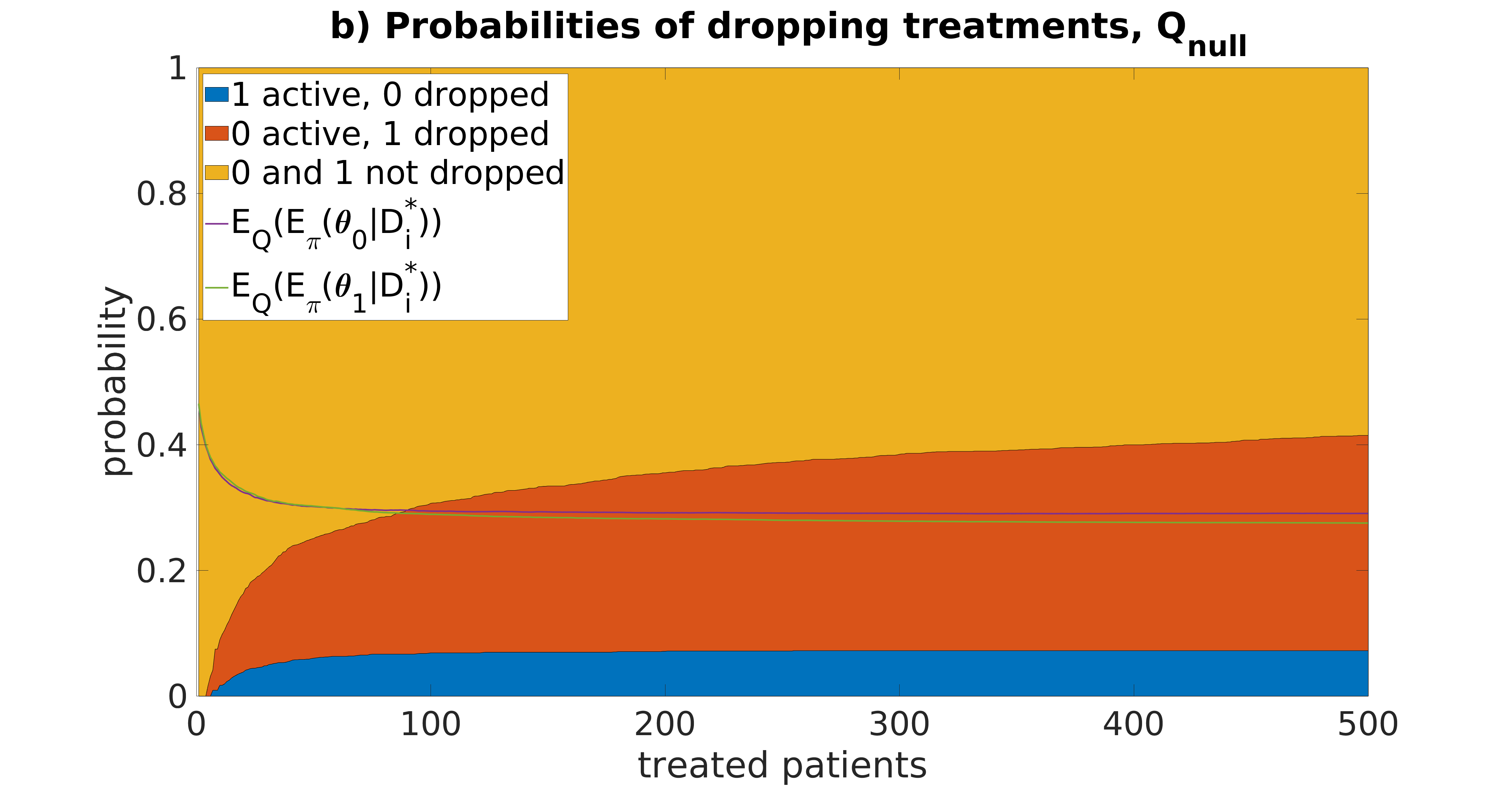}
         \renewcommand\thesubfigure{b}  
         \label{subfig:3activity:b:null}
     \end{subfigure}
     \begin{subfigure}[b]{0.5\textwidth}
         \includegraphics[width=\textwidth]{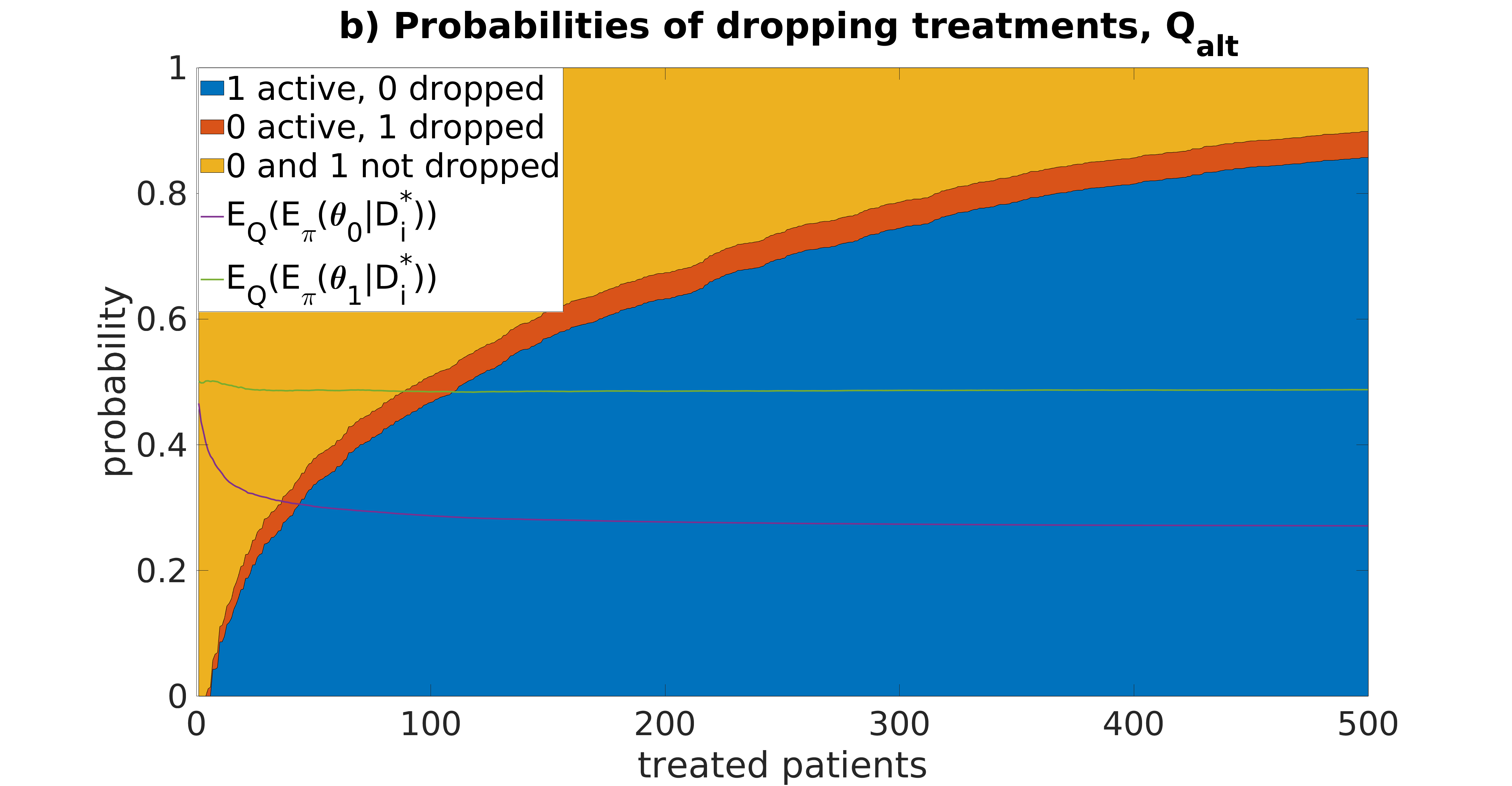}
     \renewcommand\thesubfigure{b}  
         \label{subfig:b:3activity:alternative}
     \end{subfigure}
     \hfill
     \begin{subfigure}[b]{0.5\textwidth}
     \includegraphics[width=\textwidth]{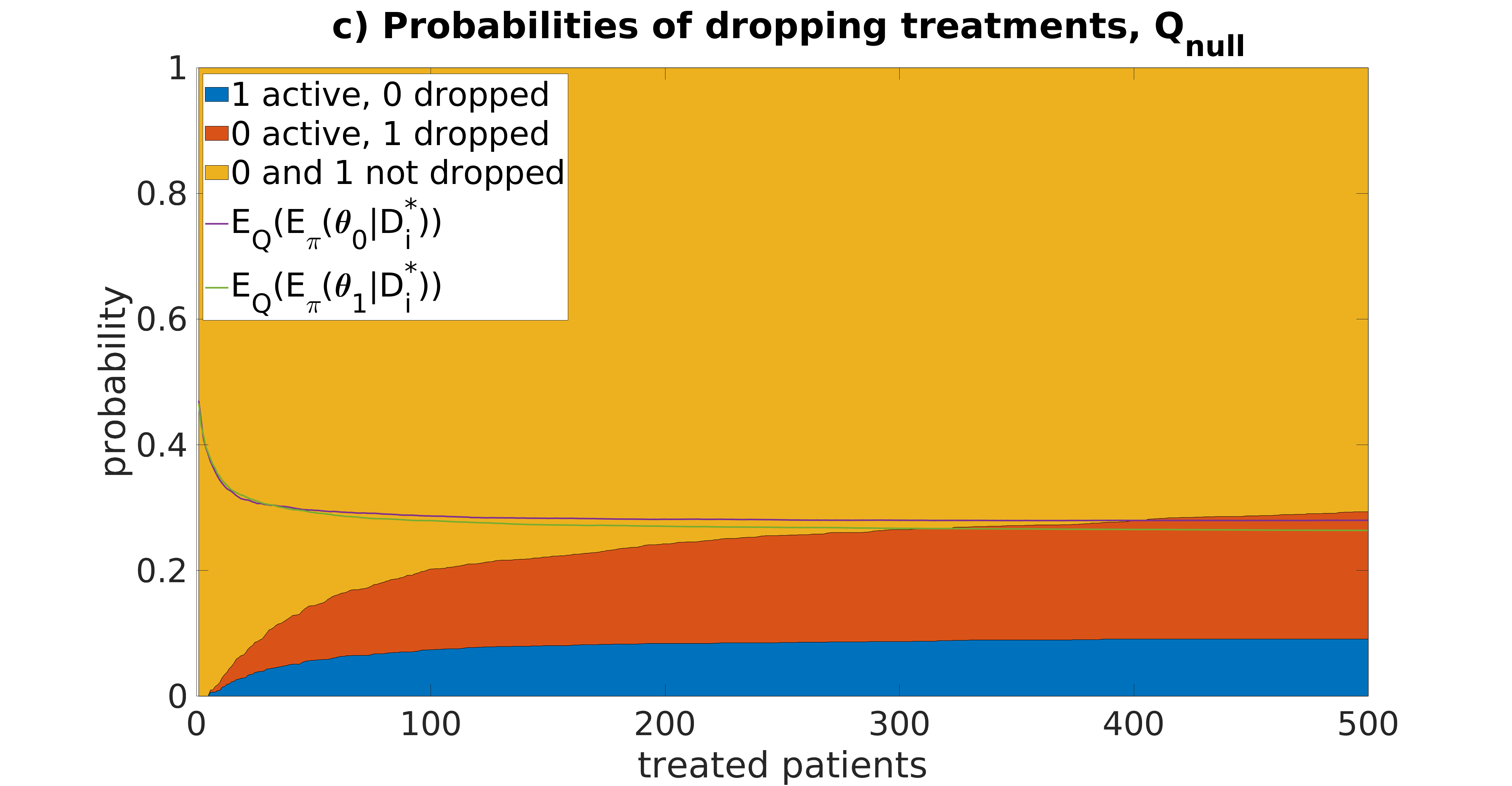}
        \renewcommand\thesubfigure{c}   
         \label{subfig:3activity:c:null}
     \end{subfigure}
    \begin{subfigure}[b]{0.5\textwidth}
     \includegraphics[width=\textwidth]{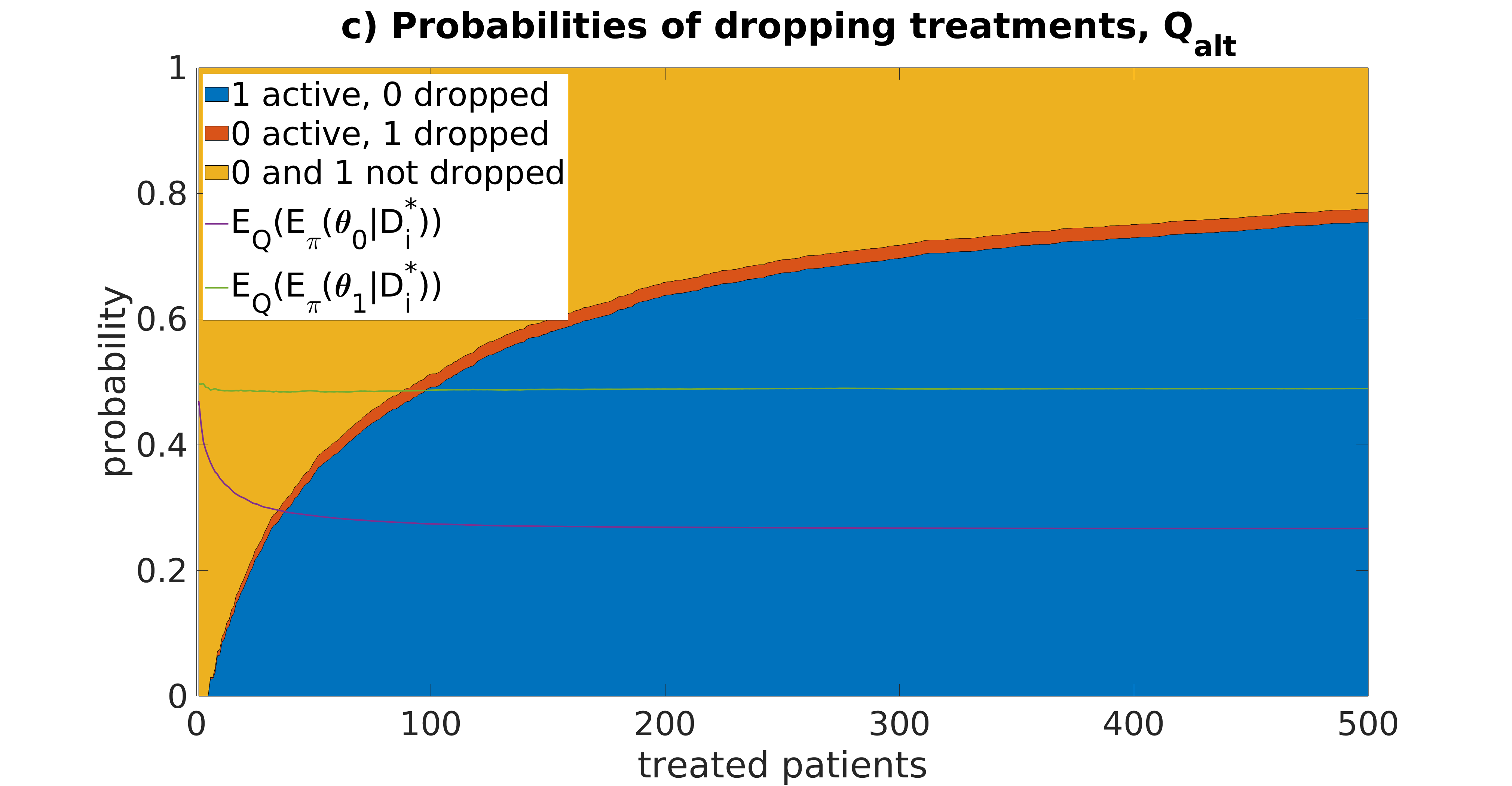}
      \renewcommand\thesubfigure{c}  
         \label{subfig:3activity:c:alternative}
   \end{subfigure}
   
\caption{\small Probabilities of dropping treatments when applying Rule 2, together with expectations of the success parameters, both shown as functions of the cumulative number of treated patients. The results are based on   5000 simulated data sets of size $N_{\max}=500$,  under $\Q_{null}$ with true parameter values $\theta_0=\theta_1=0.3$ (left) and $\Q_{alt}$  with values
$\theta_0=0.3,\theta_1=0.5$ (right). Three combinations of design parameters were considered:
 {(a)}  
  $\varepsilon=0.1, \varepsilon_1=0, \varepsilon_2=0.05$, $\delta=0.1$ (top),
 {(b)}
  $\varepsilon=0.05, \varepsilon_1=0, \varepsilon_2=0.05$, $\delta=0.1$ (middle),
  {(c)}
  $\varepsilon=0.2, \varepsilon_1=0,  \varepsilon_2=0.05$, $\delta=0.05$
  (bottom).  Also shown are the expectations  $\E_{\Q_{null}}\bigl(\E_{\pi}\bigl( \bftheta_k\vert D_i^{*} \bigr)\bigr)$ and $\E_{\Q_{alt}}\bigl(\E_{\pi}\bigl( \bftheta_k\vert D_i^{*} \bigr)\bigr), (1\leq i \leq 500, k=0,1$), computed from these simulations. For more details, see text.  }  
      \label{fig:combined_3_activity_RULE3}
\end{figure}

We then modified the design by employing the adaptive Rule 2 for treatment selection. Figure \ref{fig:combined_3_activity_RULE3}   shows the probabilities $\Q(N_{0,last} \leq i, N_{1,last} > i)$ of having dropped the control arm,  $\Q(N_{0,last} > i, N_{1,last} \leq i)$ of having dropped the experimental arm, and $\Q(N_{0,last} > i, N_{1,last} > i)$ of not having done either of these, all considered at the time $i$  patients had been treated. Note that, since the possibility of dropping both treatment arms in the same trial has been ruled out, the first two probabilities can be written in the shorter form $\Q(N_{0,last} \leq i)$  and $\Q(N_{1,last} \leq i)$. Empirical estimates of these probabilities are shown, based on  5000 simulated samples from $\Q_{null}$   (left) and $ \Q_{alt}$  (right). The earlier threshold values (a), (b) and (c) for $\varepsilon$ and $\delta$ were again applied, but combining them with $\varepsilon _{1} =  0$ and $\varepsilon _{2} =  0.05$ for Rule 2.

In Figure \ref{fig:combined_3_activity_RULE3}, on the left, the curve $\Q_{null}(N_{0,last} \leq i), 1 \leq i \leq 500$, forming the upper boundary of the blue band "1 active, 0 dropped", depicts the false positive rate evaluated at $i$. On the right,   $\Q_{alt}(N_{0,last} \leq i)$ is the true positive rate, or power at $i$. The widths of the brown bands in this figure can be interpreted similarly, with  $\Q_{null}(N_{1,last} \leq i)$ on the left being the true negative rate evaluated at $i$, and  $\Q_{alt}(N_{1,last} \leq i)$ on the right the false negative rate. The latter probabilities are small, at most $0.05$ for the considered designs (a), (b) and (c) for Rule 1. The widths of the yellow bands represent the inconclusive rates at $i.$ 

From this follows that also the 
areas of these colored bands Figure \ref{fig:combined_3_activity_RULE3} have meaningful interpretations in terms of expected values. The area of the blue region from $1$ to $i$ is the expected value, with respect to $\Q_{null}$ (left) and to $\Q_{alt}$ (right), of the number of patients among the first $i$, who were directed to the experimental treatment in the situation in which the control arm had already been dropped. The areas of the brown bands can be interpreted in a similar fashion, with the roles of the two treatments interchanged. The area of the yellow band from $1$ to $i$ is the expected value, again with respect to $\Q_{null}$ (left) and to $\Q_{alt}$ (right), of the random variable $\min\{N_{0,last},N_{1,last},i\}.$  

Finally, Figure \ref{fig:combined_3_activity_RULE3} shows the expectations of  $\E_{\pi}\bigl( \bftheta_k\vert D_i^{*} \bigr)$ and $\E_{\pi}\bigl( \bftheta_k\vert D_i^{*} \bigr), (1\leq i \leq 500, k=0,1)$, computed from these simulations under $\Q_{null}$ and $\Q_{alt}$. Overall, the behaviour of these curves is similar to those in Figure \ref{fig:combined_Activity_RULE1}, although the negative bias seems here slightly larger. Apparently, this difference is due to Rule 2 imposing a stronger control on treatment allocation. 

\subsection{Simulation studies with a 4-arm trial:  Experiment 2}
\label{subsection:no:3.2}

Our second simulation experiment is modeled following the set-up of  Table 7 in \textcite{Villar2015}, describing a trial with  $ K=3 $ experimental arms and a control arm. The  hypotheses were $ H_{0}: \theta _{k} \leq   \theta _{0}    $  for all $k, 1\leq k \leq 3$, and its logical complement $H_{1}:\theta _{k} > \theta _{0}$ for at least one $k, 1\leq k \leq 3$. Considered as a multiple hypothesis testing problem, applying significance level $\alpha = 0.05$ and the Bonferroni correction, $ H_{0}$ was tested separately against each alternative $H_{1k}: \theta _{k} > \theta _{0}$ at level $\alpha/3$. The numerical results shown in Table 7 of \textcite{Villar2015} were based on   using the fixed trial size of $ N_{\max}=80$, together with parameter values $(\theta _{0},\theta _{1}, \theta _{2}, \theta _{3}) = (0.3, 0.3, 0.3, 0.3)$ for computing the family-wise error rate (FWER), and $(\theta _{0},\theta _{1}, \theta _{2}, \theta _{3}) = (0.3, 0.4, 0.5, 0.6)$ for computing the power of concluding $H_{1}$. The small trial size was justified by thinking of a rare disease setting, where the number of patients in the trial could be a high proportion of all patients with the considered condition. Below, we continue using the shorthand notations $\Q_{null}$ and $\Q_{alt}$ for these two parameter settings.  

\subsubsection{Monitoring the operation of Rule 1  in Experiment 2}

\begin{landscape} 
\begin{figure}[!htbp]
\includegraphics[max size={1.6\textwidth}{1.6\textheight}]{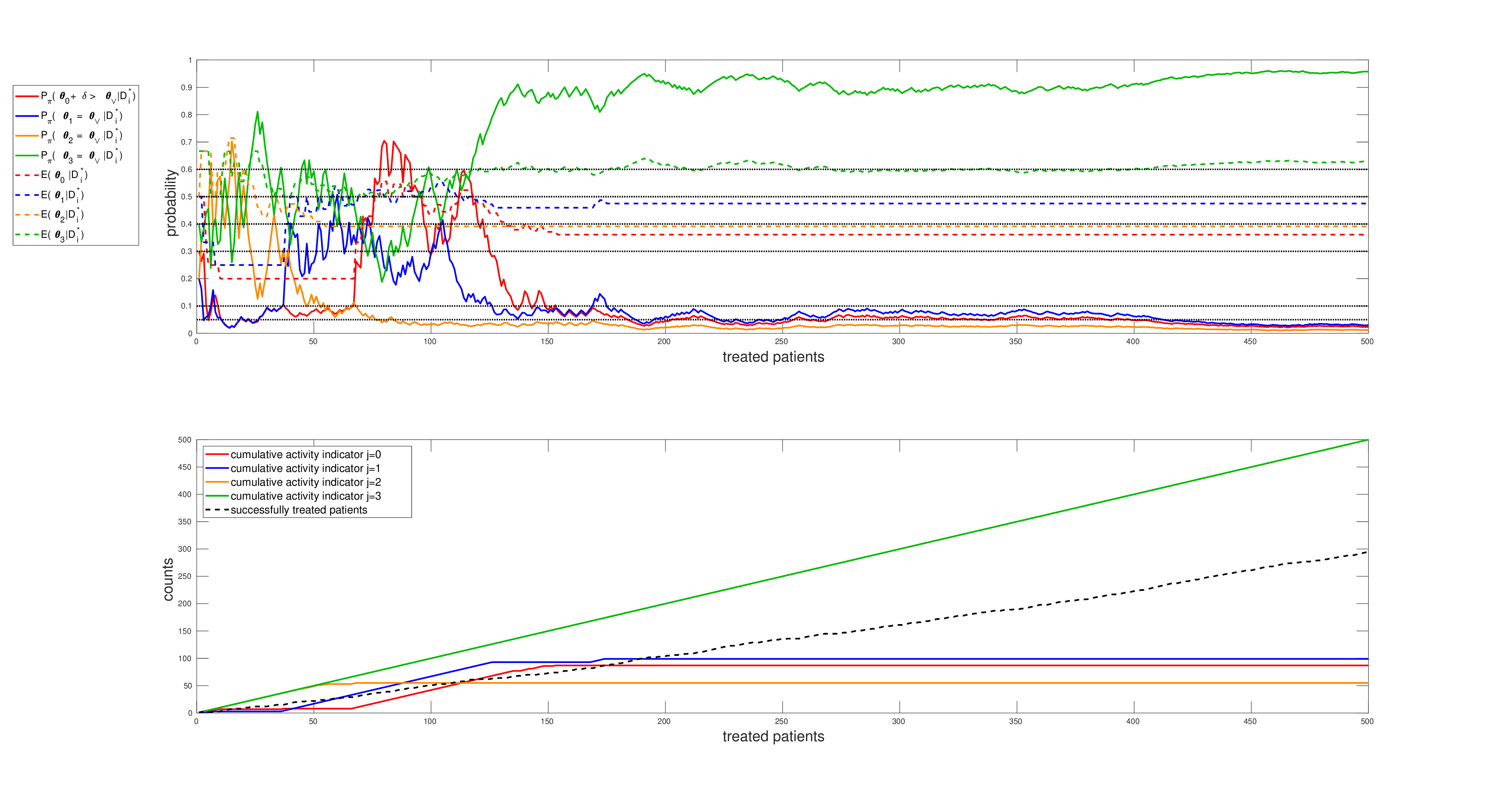}
\caption{\small An example of monitoring the execution of a 4-arm trial in a  simulated clinical trial of size $N_{\max}=500$ when applying Rule 1 for treatment allocation. Top:
Time-evolution of the posterior probabilities
$\P_{\pi}\bigl( \theta_0+\delta \geq \theta_{\vee}\big\vert D_{i}^{*}\bigr)$  and $\P_{\pi}\bigl( \theta_k = \theta_{\vee} 
\big\vert D_{i}^{*}\bigr),1 \leq k \leq 3,$ and of the posterior expectations
$\E_{\pi}\bigl( \theta_k\vert D_{i}^{*}\bigr)$, $0 \leq k \leq 3; 1 \leq i \leq 500. $ Bottom: Cumulative sums of activity indicators $I_k$, $0\le k \le 3$, and of the total number of successful treatments, both as functions of the number of treated patients. The simulation was performed with true parameter values
$\theta_0=0.3,\theta_1=0.4, \theta_2=0.5$ and $\theta_3=0.6$, 
by using  Rule 1 with design parameters  $\varepsilon=0.1$ and
 $\delta=0.1$. For more details, see text. }
      \label{fig:simulation_multi_figure_alternative_RULE_1_01-May-2020}  
\end{figure}
\end{landscape}

As in Experiment 1, we first monitored the execution of this trial, based on a single simulation from $\Q_{alt}$, and thereby applying Rule 1 for treatment allocation with threshold values $\varepsilon=0.1$ and
$\delta=0.1$. Figure 
\ref{fig:simulation_multi_figure_alternative_RULE_1_01-May-2020}  
presents an example based on such simulated data, showing the time-evolution of the posterior probabilities
$\P_{\pi}\bigl( \bftheta_0+\delta \geq \bftheta_{\vee}\big\vert D_i^{*}\bigr)$  and $\P_{\pi}\bigl( \bftheta_k = \bftheta_{\vee}\big\vert D_i^{*}\bigr), 1 \leq k \leq 3$, of the the posterior expectations
$\E_{\pi}\bigl( \bftheta_k\vert D_i^{*}\bigr)$ and of the cumulative sums of the activity 
indicators $I_{k}$, $0\le k \le 3$, all considered at times at which $i$ patients had been treated and up to maximal trial size $ N_{\max} = 500$. 

From the top display we can see how, with some luck in the simulation that was carried out, the posterior probabilities $\P_{\pi}\bigl( \bftheta_0+\delta \geq \bftheta_{\vee}\big\vert D_i^{*}\bigr), \P_{\pi}\bigl( \bftheta_1 = \bftheta_{\vee}\big\vert D_i^{*}\bigr) $ and $\P_{\pi}\bigl( \bftheta_2 = \bftheta_{\vee}\big\vert D_i^{*}\bigr) $ started progressively to take on values below the given threshold $\varepsilon = 0.10$ and finally stayed there during the remaining simulation run. In contrast, after considerable early variation, the posterior probabilities $\P_{\pi}\bigl( \bftheta_3 = \bftheta_{\vee}\big\vert D_i^{*}\bigr)$ corresponding to the highest true response rate $\theta_3=0.6$ stayed consistently above that threshold level, and actually started to dominate the others from approximately $i=120$ onward. The cumulative activity indicators for all treatment arms in the bottom display of Figure \ref{fig:simulation_multi_figure_alternative_RULE_1_01-May-2020} show clearly when each of these arms was active or dormant. In this simulation, there was some back-and-forth movement between these two states, but finally treatment arms $0, 1$ and $2$, respectively after $153,174$, and $68$ treated patients, remained dormant. The dotted line shows the cumulative numbers of successes in the simulation, ending up with the total $S(500) = 294$, not much short of the optimal expected value $300$ that would have been obtained if all $500$ patients had been assigned to the best treatment with success rate $\theta_3 = 0.6$. 

\subsubsection{Effect of the design parameters on treatment allocation }

Next, as in Experiment 1, we studied the effect of the choice of the design parameters $\varepsilon$ and $\delta$ in Rule 1 on some selected frequentist type key characteristics of the trial. For this, we simulated
$2000$ data sets of size
$N_{\max}=500$, under both $\Q_{null}$
and $\Q_{alt}$. The same three combinations of the design parameters were used as before:
{\bf a)} $\varepsilon=0.1$, $\delta=0.1$,
{\bf b)} $\varepsilon=0.05$, $\delta=0.1$,
{\bf c)} $\varepsilon_1=0.2$, $\delta=0.05$. 
For the analysis, $\bftheta_0, \dots, \bftheta_3$ were assumed to be a priori  independent and uniformly distributed on $(0,1)$.

\begin{figure}[!htbp]
     \begin{subfigure}[b]{0.5\textwidth}
     \includegraphics[height=\macroheight,width=\textwidth]{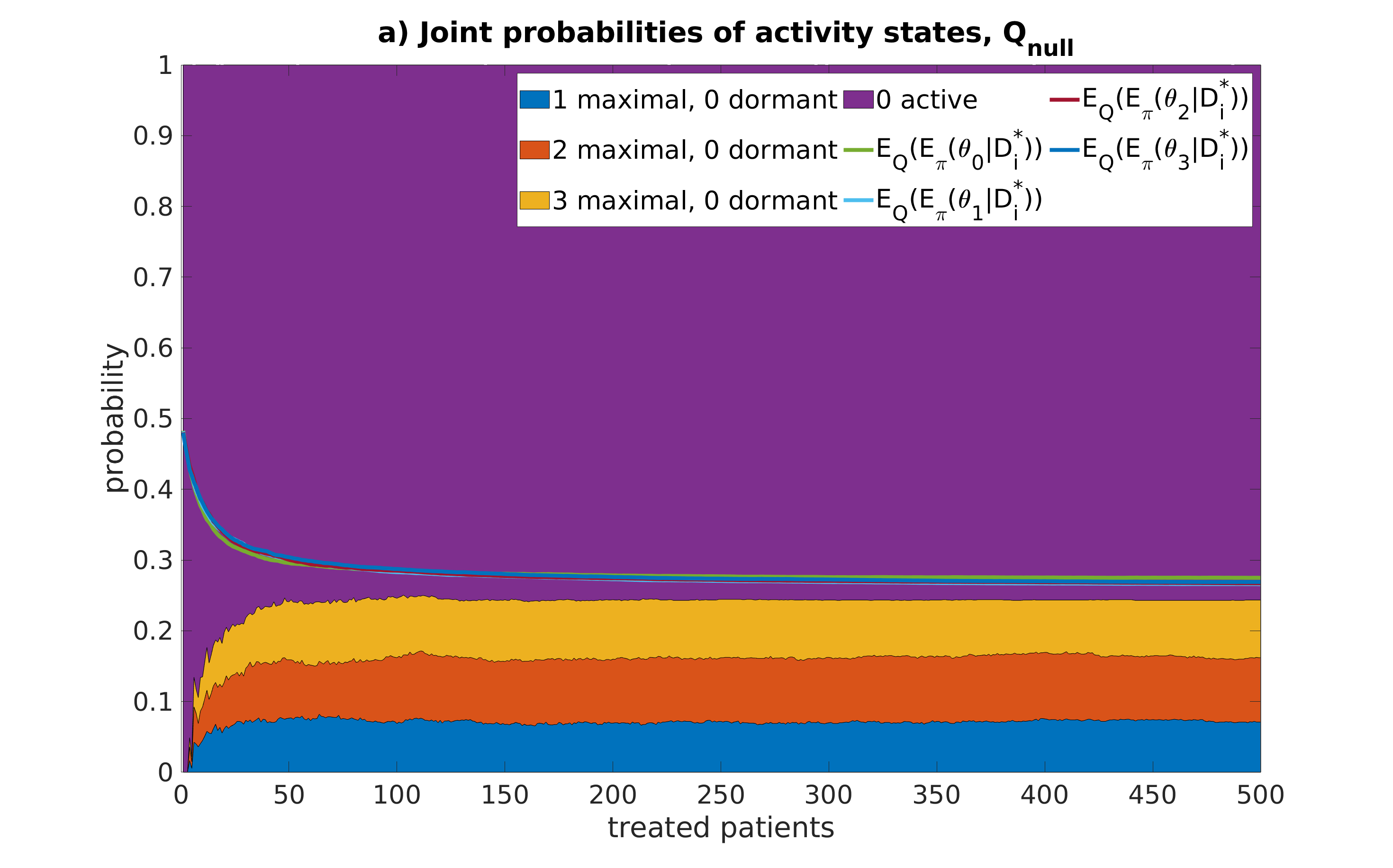}
       \renewcommand\thesubfigure{a}
         \label{subfig:ex2activity:a:null}
     \end{subfigure}
     \begin{subfigure}[b]{0.5\textwidth}
     \includegraphics[height=\macroheight,width=\textwidth]{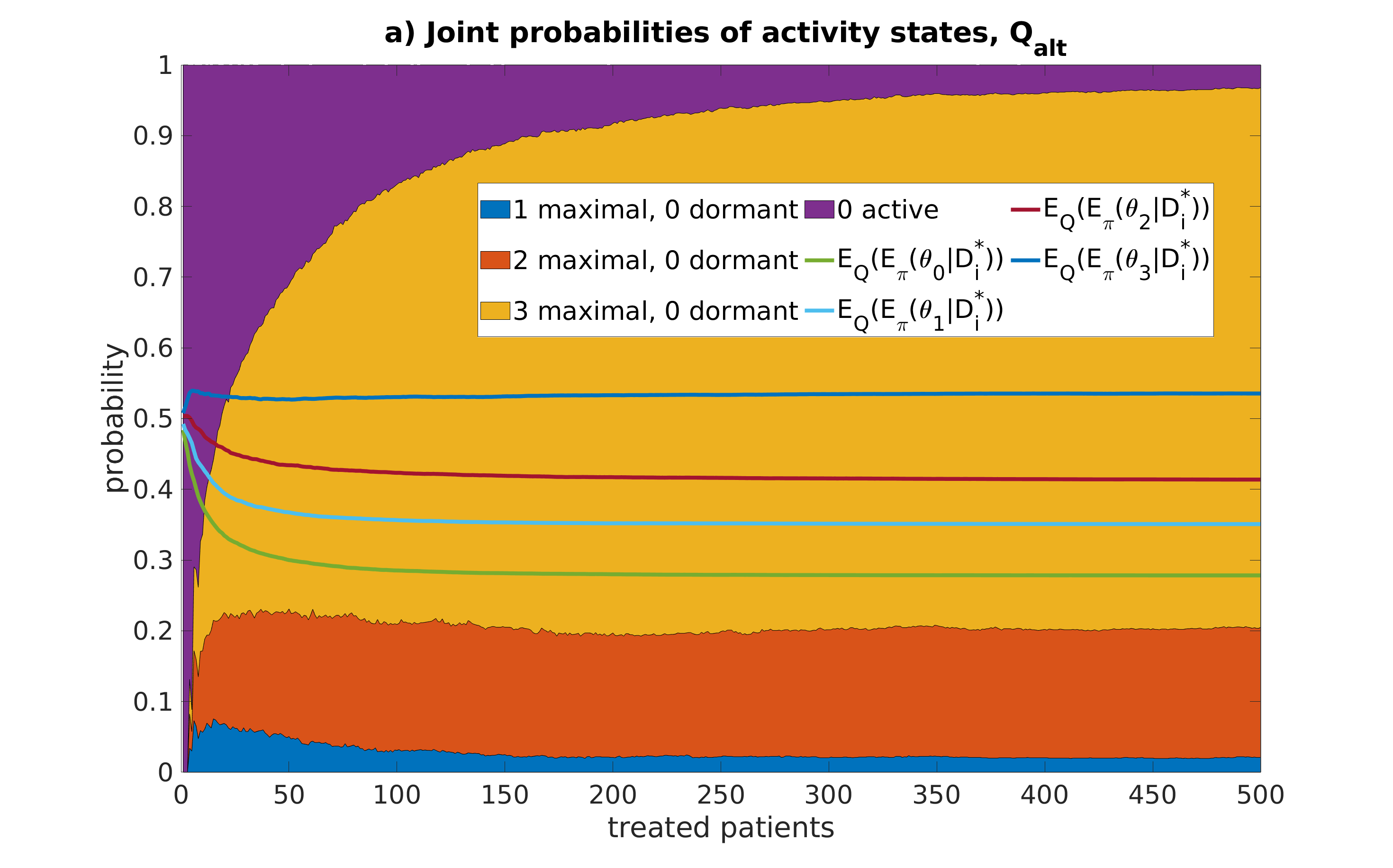}
        \renewcommand\thesubfigure{a} 
         \label{subfig:ex2activity:a:alternative}
     \end{subfigure}
     \hfill
     \begin{subfigure}[b]{0.5\textwidth}
         \includegraphics[height=\macroheight,width=\textwidth]{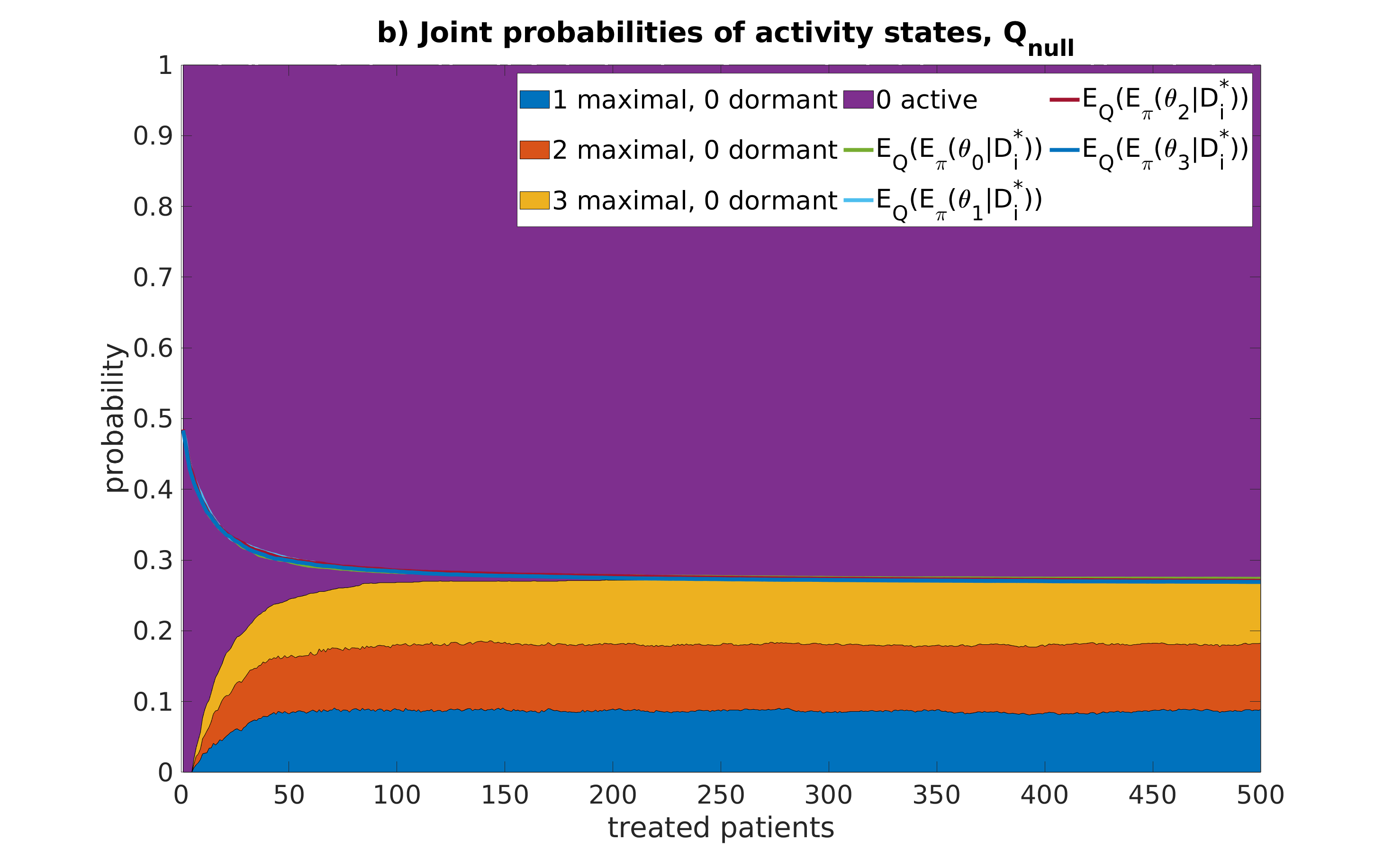}
         \renewcommand\thesubfigure{b}  
         \label{subfig:ex2_activity:b:null}
     \end{subfigure}
     \begin{subfigure}[b]{0.5\textwidth}
         \includegraphics[height=\macroheight,width=\textwidth]{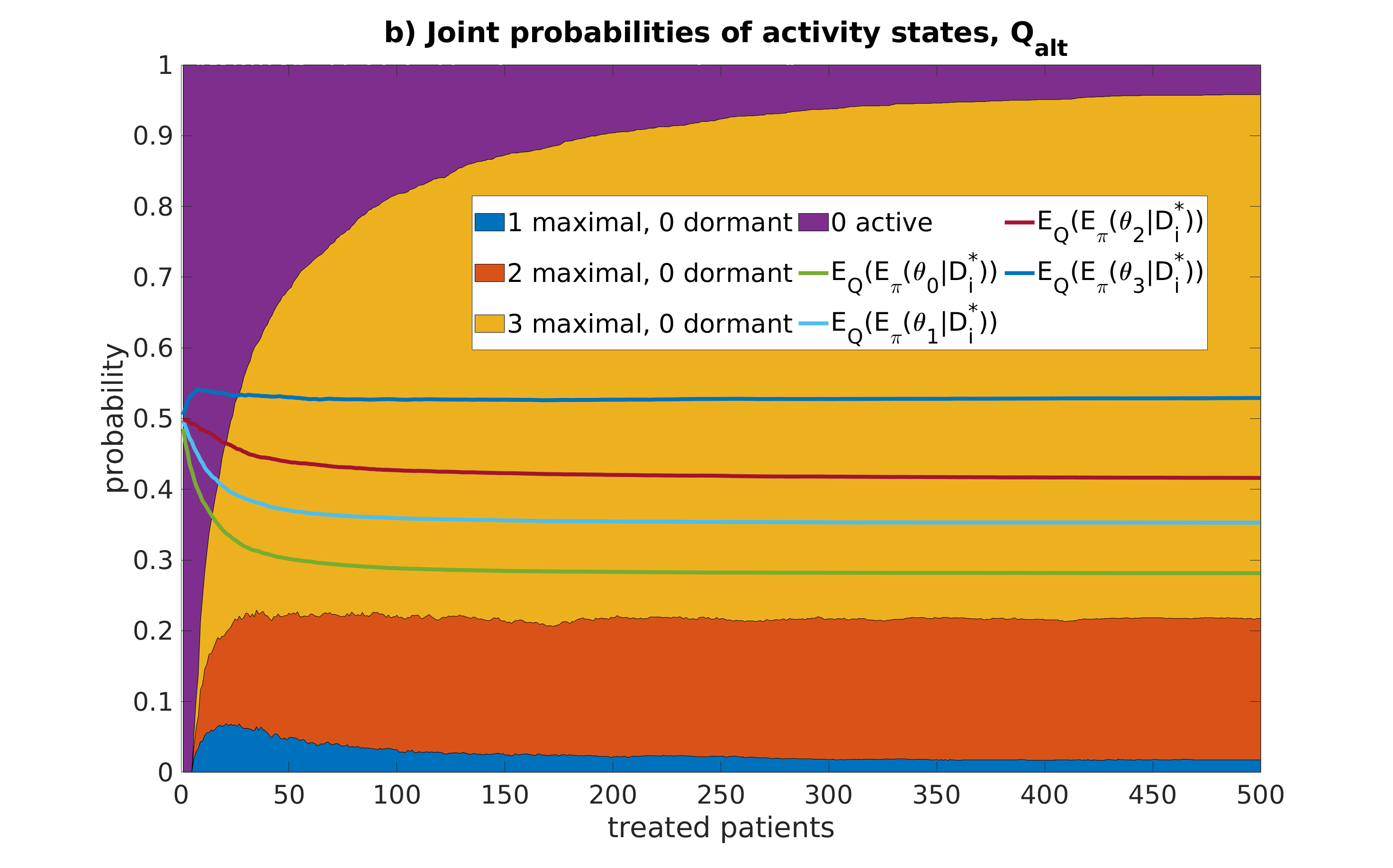}
     \renewcommand\thesubfigure{b}  
         \label{subfig:b:ex2_activity:alternative}
     \end{subfigure}
     \hfill
     \begin{subfigure}[b]{0.5\textwidth}
     \includegraphics[height=\macroheight,width=\textwidth]{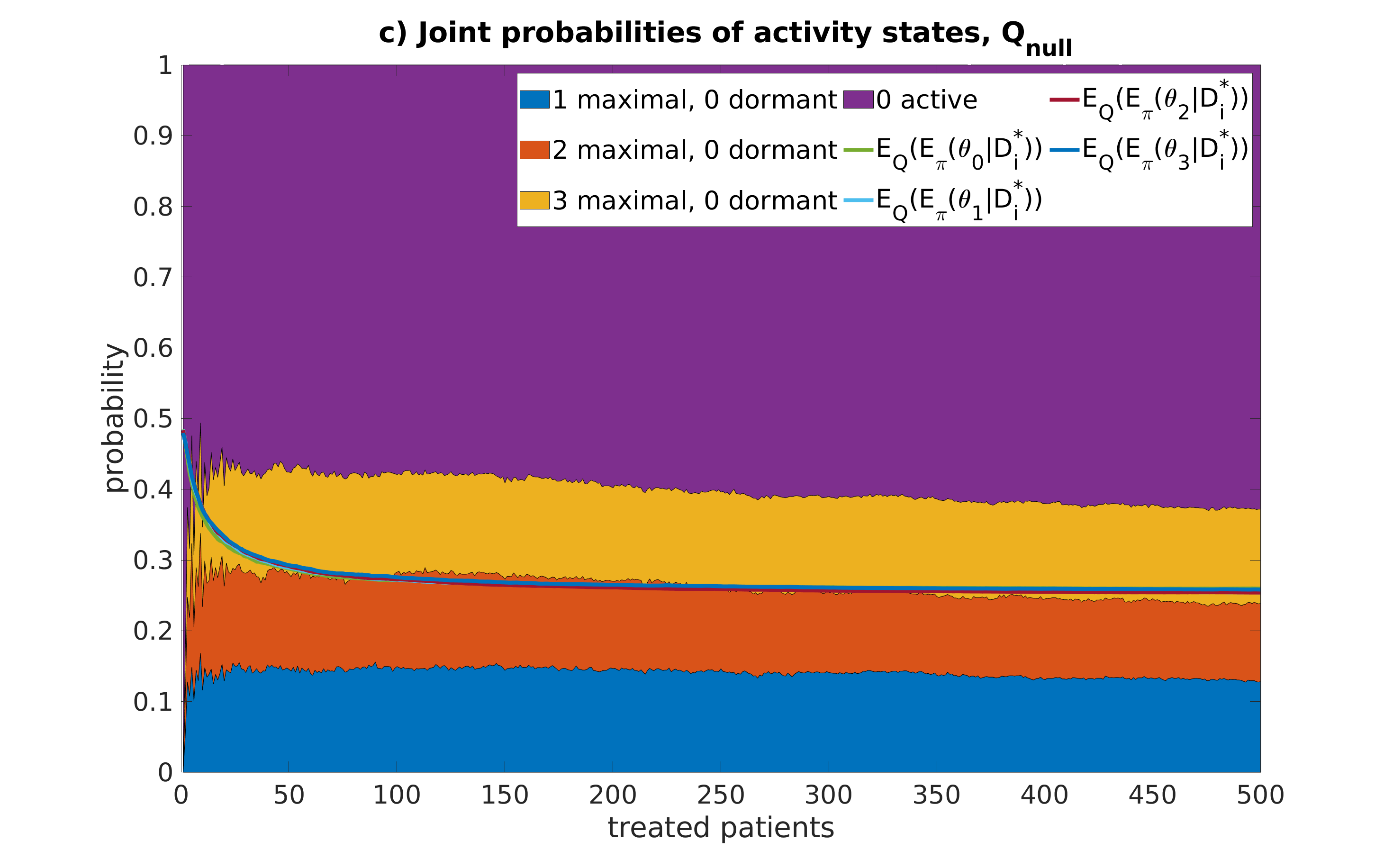}
        \renewcommand\thesubfigure{c}   
         \label{subfig:ex2_activity:c:null}
     \end{subfigure}
    \begin{subfigure}[b]{0.5\textwidth}
     \includegraphics[height=\macroheight,width=\textwidth]{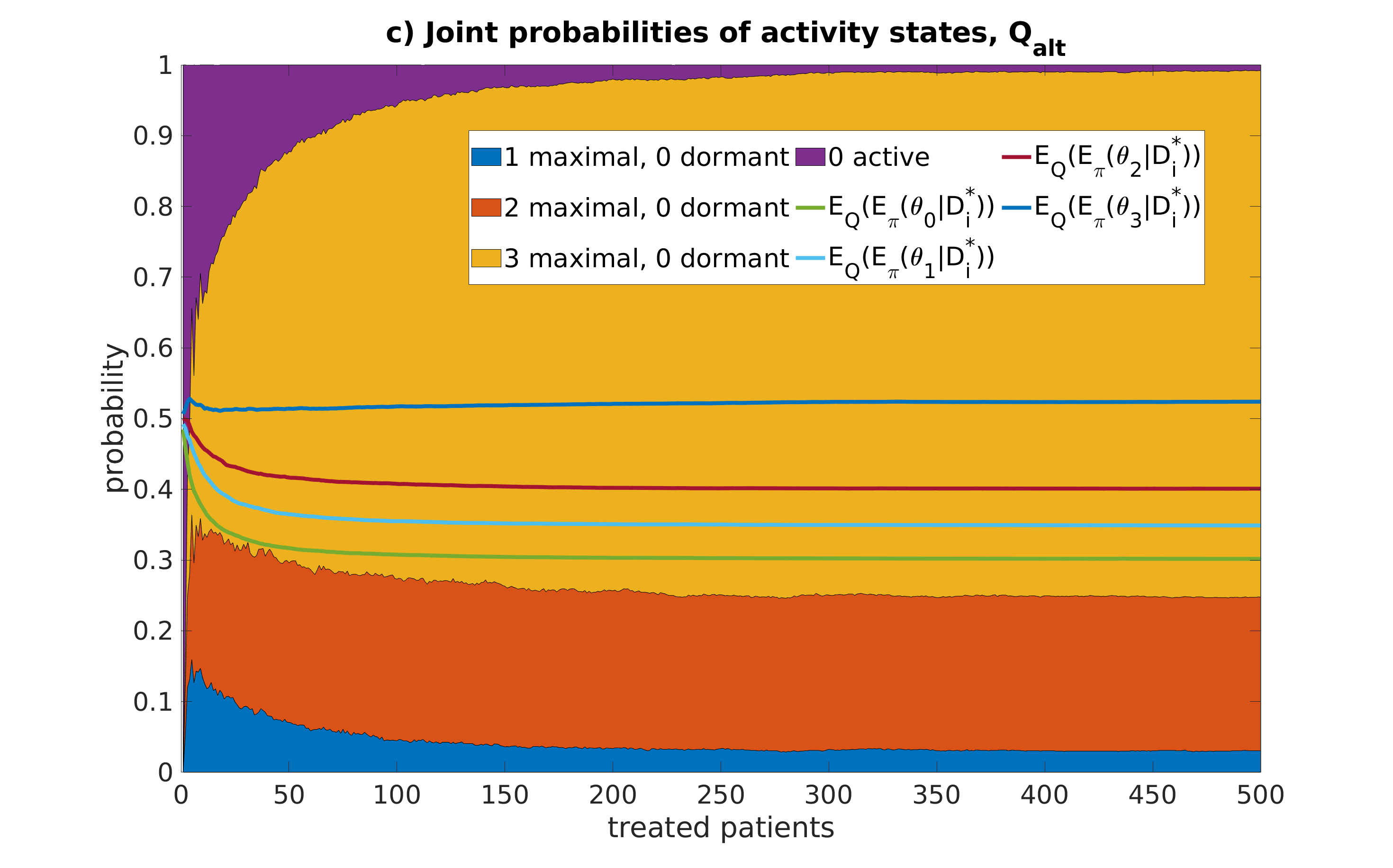}
      \renewcommand\thesubfigure{c}  
\label{subfig:ex2_activity:c:alternative}
     \end{subfigure}\captionsetup{singlelinecheck=off}
\caption[short caption]{
\small
Effect of the choice of the values of the design  parameters $\varepsilon$ and $\delta$ in the 4-arm trial of Experiment 2 when applying Rule 1 for treatment allocation. Joint probabilities of some combinations  of active and dormant states are shown, as functions of the number $i$ of treated patients. The results are based on
$2000$ data sets of size
$N_{\max}=500$, under $\Q_{null}$
(left) and $\Q_{alt}$ (right). Three combinations of the design  parameters were used:
 {(a)}  
  $\varepsilon=0.1$, $\delta=0.1$ (top),
 {(b)}
  $\varepsilon=0.05$, $\delta=0.1$ (middle),
  {(c)}
  $\varepsilon_1=0.2$, $\delta=0.05$
  (bottom).
In the subfigures, the width of  each of the 4 bands corresponds to the $\Q$-probability of a respective event in the box. Also shown are the expectations  $\E_{\Q_{null}}\bigl(\E_{\pi}\bigl( \bftheta_k\vert D_i^{*} \bigr)\bigr)$ and $\E_{\Q_{alt}}\bigl(\E_{\pi}\bigl( \bftheta_k\vert D_i^{*} \bigr)\bigr), (1\leq i \leq 500, 1 \leq k \leq 3$), computed from these simulations. For more details, see text. }     
  \label{fig:combined_ex2_activity_RULE3}
\end{figure}

In a 4-arm trial there would in principle be $2^4 - 1 = 15$ possibilities of forming combinations of active and dormant states at a given $i$, and it would be hard to present such results in an easily understandable graphical form. The main aim of the trial of this type is to find out whether one of the experimental treatments would be better than the others, and in particular, better than the control. In view of this, we call treatment $k$ \textit{maximal at} $i$  if  
$\P_{\pi}\bigl( \bftheta_k=\bftheta_{\vee}
    \vert D_i^*\bigr)
    \ge   \P_{\pi}\bigl( \bftheta_{\ell}=\bftheta_{\vee}
    \vert D_i^*\bigr) \; \forall
    \ell\ne k$, and then focus our attention on events of the form $\{$treatment $k$ is  maximal, control treatment is dormant$\}$. The results are shown in Figure \ref{fig:combined_ex2_activity_RULE3}. 
In the subfigures, the width of  each of the 4 bands at $i$ corresponds to the $\Q$-probability of a respective  event. The three lower bands  represent the $\Q$-probabilities of $\{$treatment $k$ is  maximal at $i$, $I_{0,i}=0\},$ $ 1 \leq k \leq 3,$  and the upper band (violet) the $\Q$-probabilities of $\{I_{0,i}=1\}$.

In the present 4-arm experiment, we can think of all three experimental arms combined as competing, and being evaluated against, the control arm, in a way analogous to the single experimental treatment in Experiment 1. Seen from this angle, the sum of the widths of the three lower bands of Figure \ref{fig:combined_ex2_activity_RULE3} corresponds to the width of the lowest band in Figure \ref{fig:combined_Activity_RULE1}, while that of the top one in the former corresponds to the sum of the top two in the latter.

On the left of Figure \ref{fig:combined_ex2_activity_RULE3}, describing $\Q_{null}$, the violet band corresponding to  $\{I_{0,i}=1\}$ is broader than the other three, not only because the assumed initial state $\{I_{0,1}=1\}$, but  because the control arm is protected by $\delta = 0.1$ against being moved to the dormant state. The other three bands are similar to each other due to the assumed symmetry of the experimental treatments 1, 2 and 3 under $\Q_{null}$. All these probabilities stabilize rather quickly with growing $i$, well before $i=100.$ 

On the right, corresponding to $\Q_{alt}$, the violet band becomes narrower with growing $i$, losing ground mainly to the yellow band, which represents the $\Q_{alt}$-probabilities of the events $\{$treatment $3$ is maximal at $i$, $I_{0,i}=0\}.$ The widths of the three lower bands, yellow, brown and blue, are seen to follow the same order as the corresponding true response parameter values. Approximate values of these probabilities can be read from Figure \ref{fig:combined_ex2_activity_RULE3} as well. For example, considering design (a) at $i=500,$ we get $\Q_{alt}($treatment 3 is maximal at $500, I_{0,500} = 0 ) = 0.763$. Overall, designs (a) and (b) led to very similar $\Q_{alt}$-probabilities, while the more liberal design (c), which allowed for more variability during the early stages of the trial, gave rise to somewhat broader brown and blue bands.

\subsubsection{Effect of the design parameters on treatment selection }
 
We then employed also Rule 2, in order to study the ability of this algorithm to drop possibly inferior treatment arms from the trial and thereby to act as a selection mechanism for those performing better. Using data simulated under $\Q_{null}$ and $\Q_{alt}$, the same three combinations of design  parameters as in Experiment 1 were again considered:
{(a)}  
$\varepsilon=0.1, \varepsilon_1=0, \varepsilon_2=0.05$, $\delta=0.1$,
{(b)}
$\varepsilon=0.05, \varepsilon_1=0, \varepsilon_2=0.05$, $\delta=0.1$,
{(c)}
$\varepsilon=0.2, \varepsilon_1=0,  \varepsilon_2=0.05$, $\delta=0.05$. 

\begin{figure}[!htbp]
     \begin{subfigure}[b]{0.5\textwidth}
     \includegraphics[height=\macroheight,width=\textwidth]{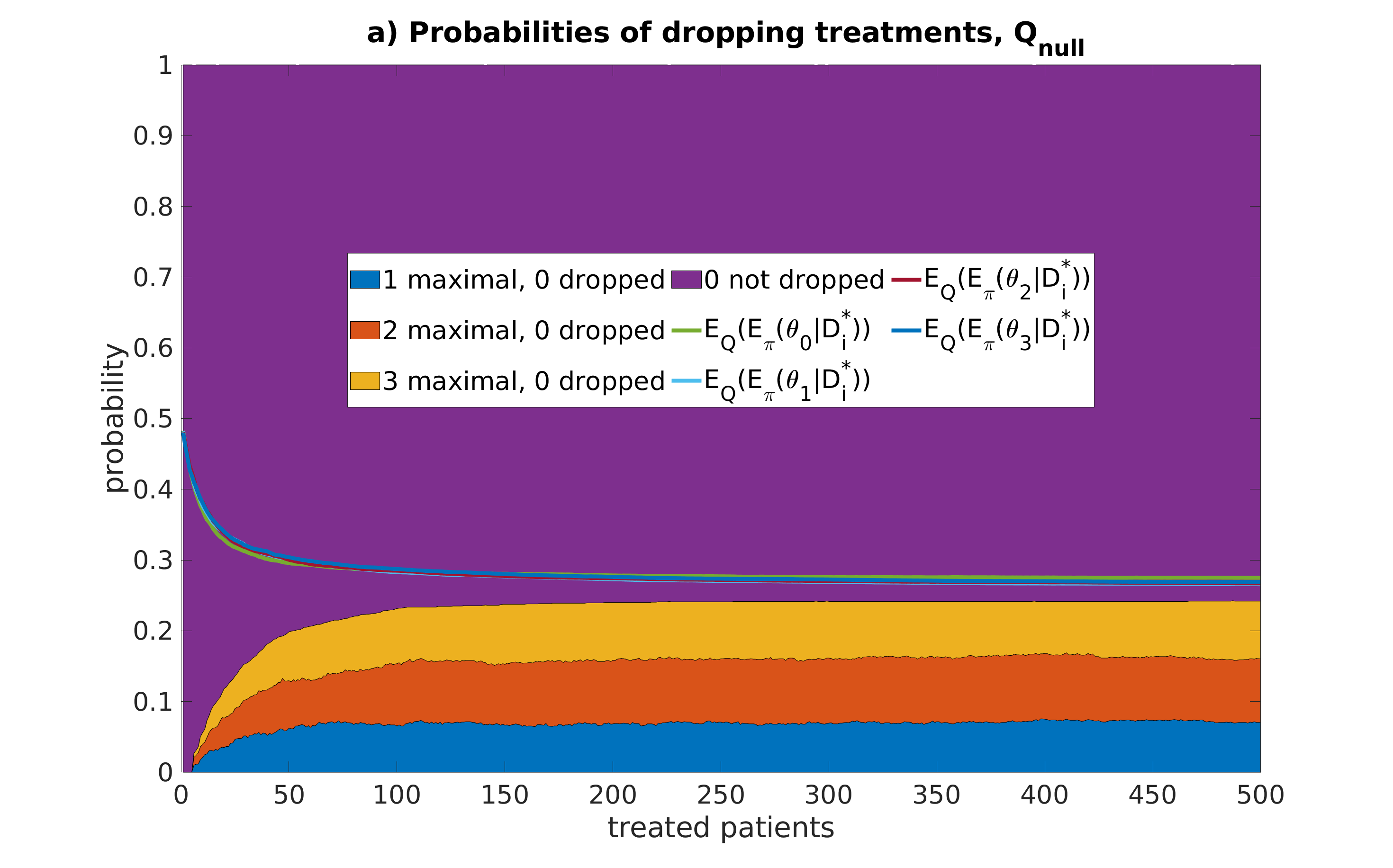}
       \renewcommand\thesubfigure{a}  
         \label{subfig:3ex2activity:a:null}
     \end{subfigure}
     \begin{subfigure}[b]{0.5\textwidth}
     \includegraphics[height=\macroheight,width=\textwidth]{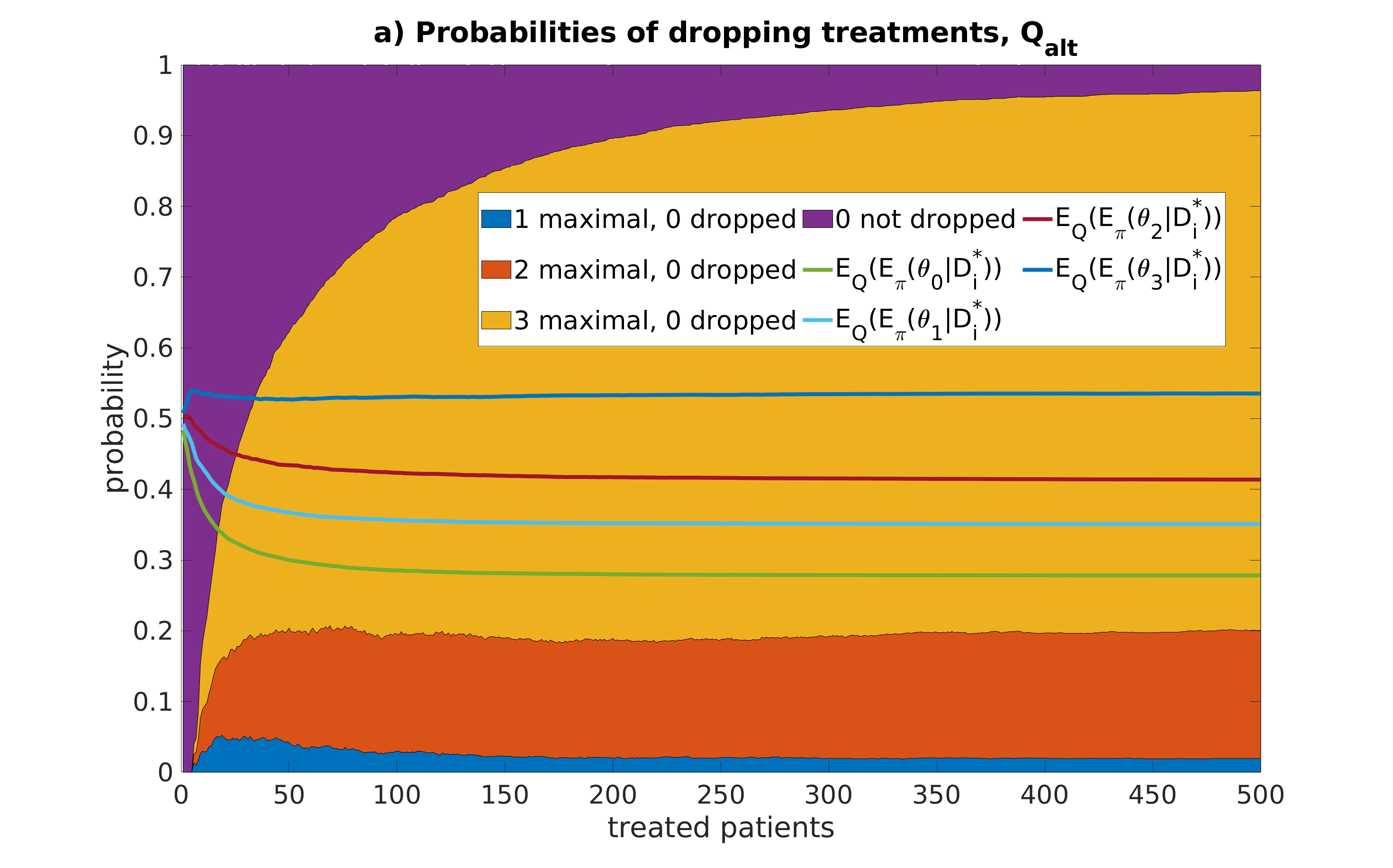}
        \renewcommand\thesubfigure{a} 
         \label{subfig:3ex2activity:a:alternative}
     \end{subfigure}
     \hfill
     \begin{subfigure}[b]{0.5\textwidth}
         \includegraphics[height=\macroheight,width=\textwidth]{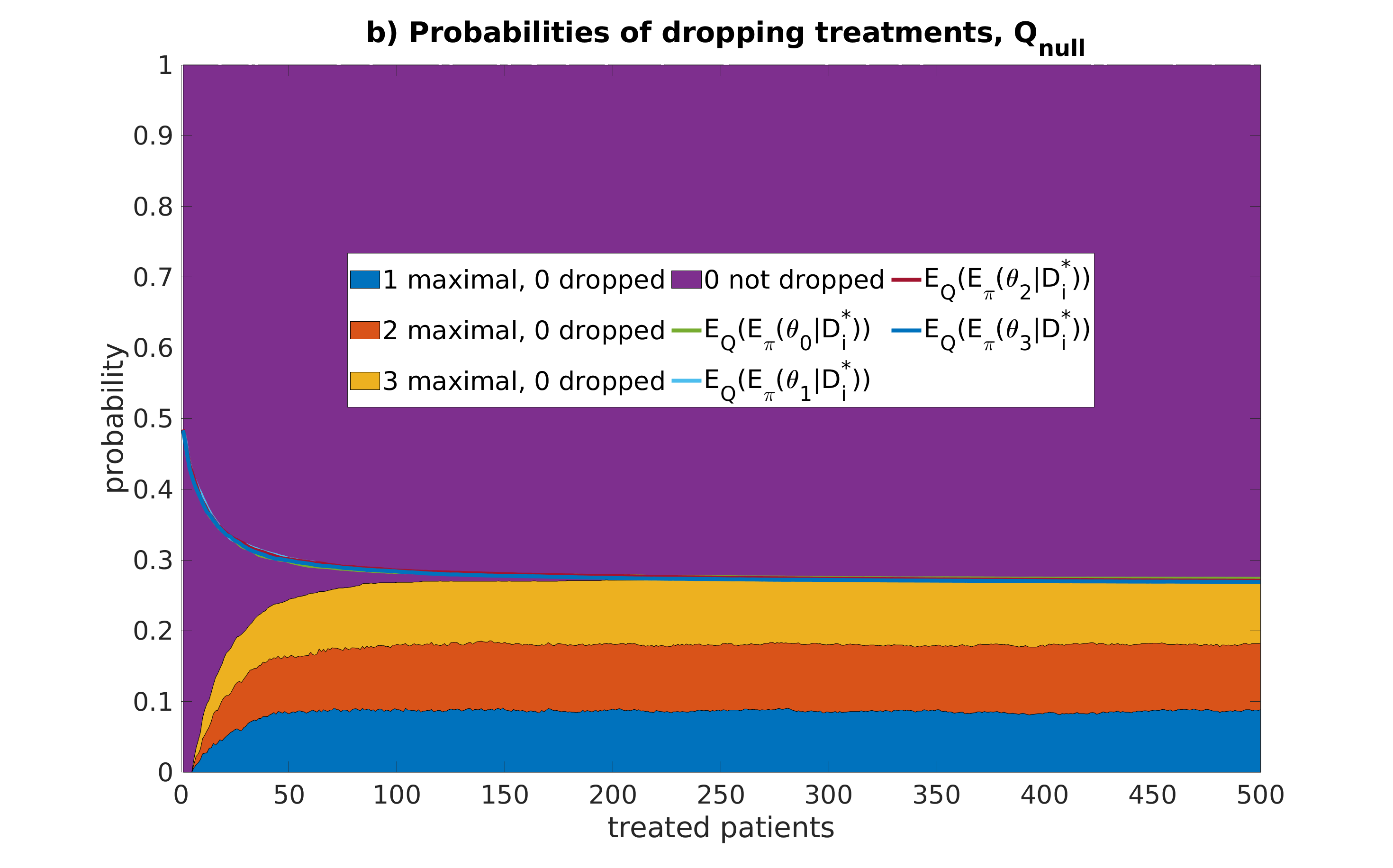}
         \renewcommand\thesubfigure{b}  
         \label{subfig:3ex2_activity:b:null}
     \end{subfigure}
     \begin{subfigure}[b]{0.5\textwidth}
         \includegraphics[height=\macroheight,width=\textwidth]{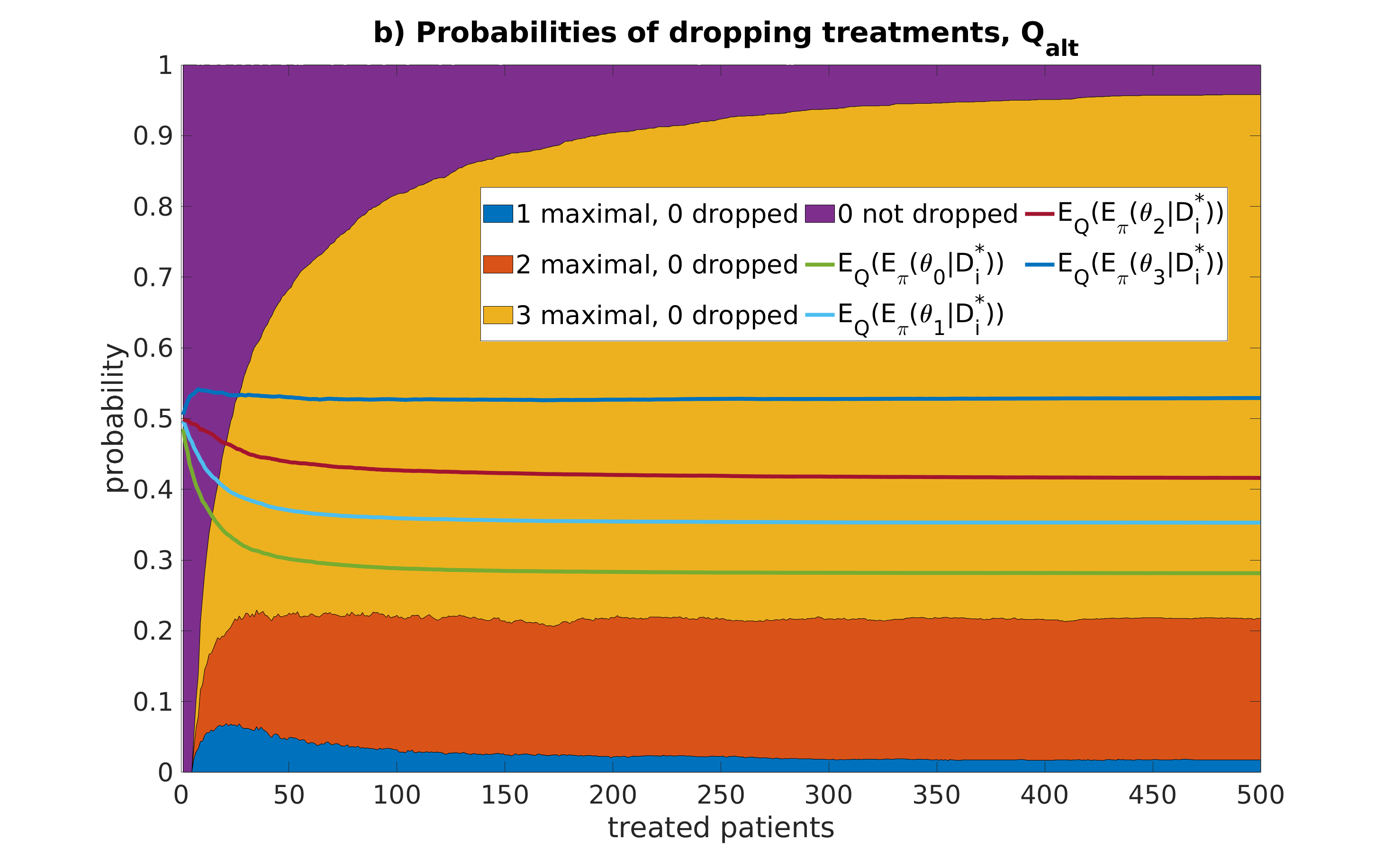}
     \renewcommand\thesubfigure{b}  
         \label{subfig:b:3ex2_activity:alternative}
     \end{subfigure}
     \hfill
     \begin{subfigure}[b]{0.5\textwidth}
     \includegraphics[height=\macroheight,width=\textwidth]{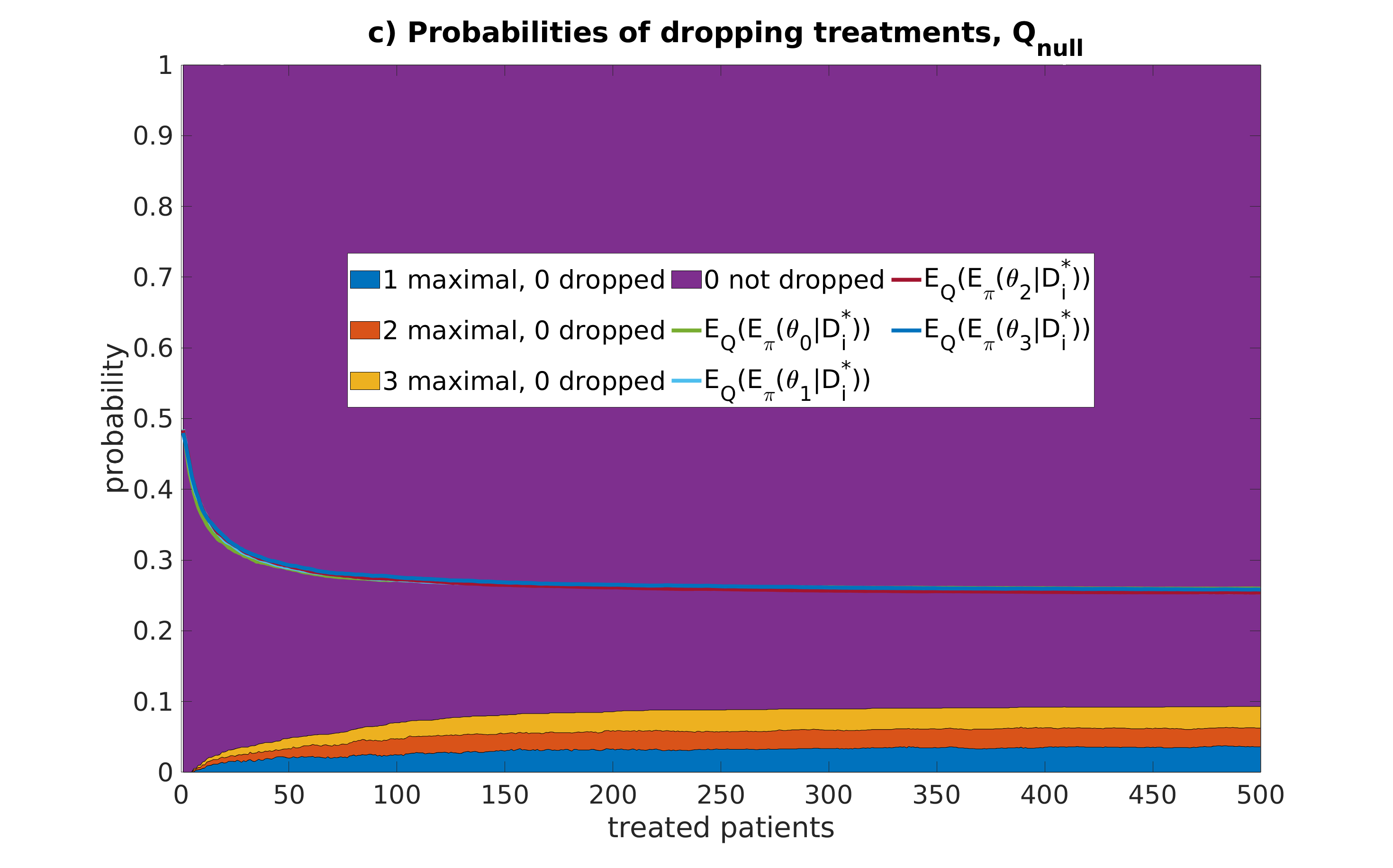}
        \renewcommand\thesubfigure{c}   
         \label{subfig:3ex2_activity:c:null}
     \end{subfigure}
    \begin{subfigure}[b]{0.5\textwidth}
     \includegraphics[height=\macroheight,width=\textwidth]{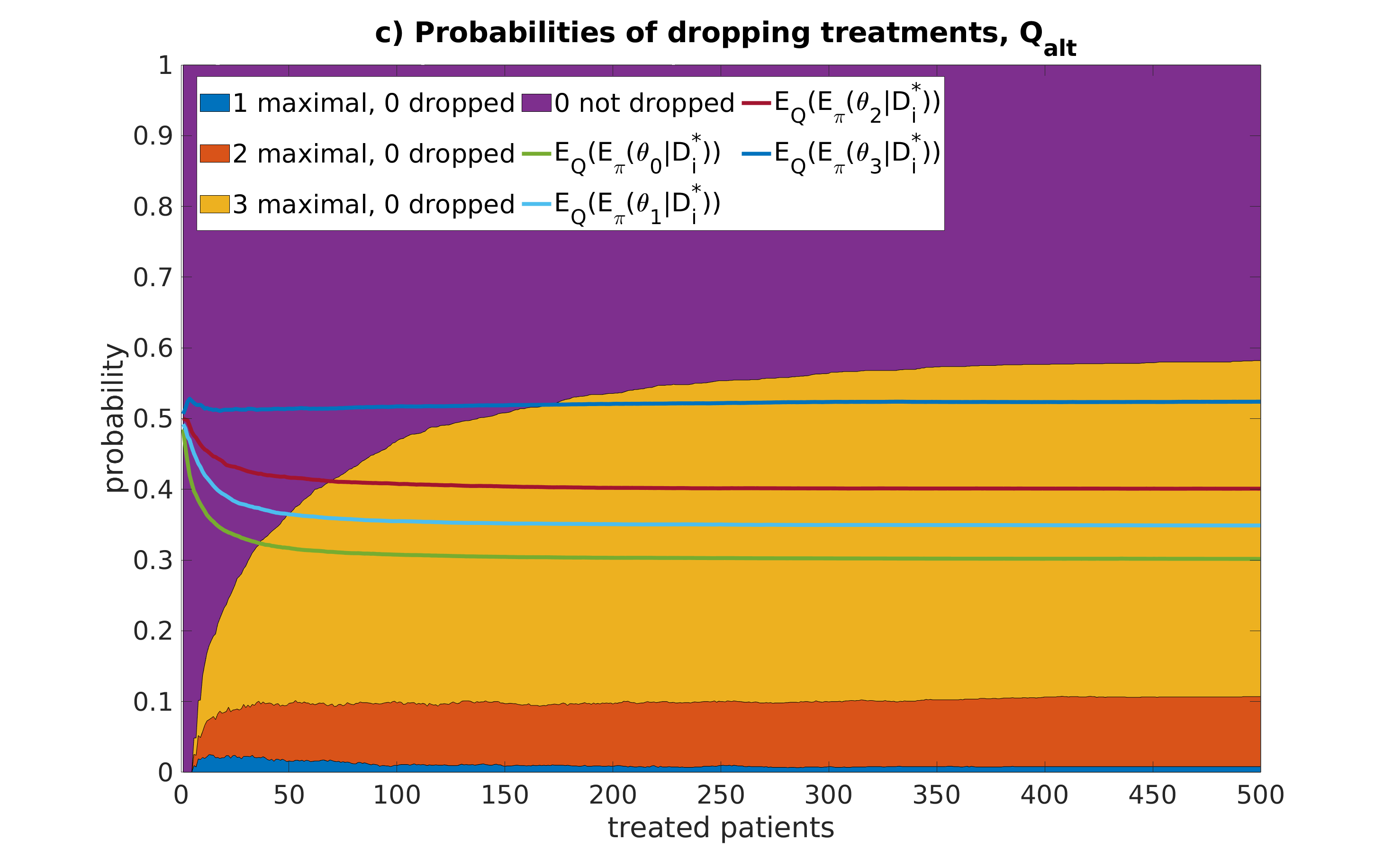}
      \renewcommand\thesubfigure{c}  
\label{subfig:3ex2_activity:c:alternative}
     \end{subfigure}\captionsetup{singlelinecheck=off}
\caption[short caption]{
\small Effect of the design  parameters $\varepsilon$ and $\delta$ in the 4-arm trial of Experiment 2 when applying Rule 2 for treatment selection. Joint probabilities of some combinations  of selected treatments are shown, as functions of the number $i$ of treated patients. The results are based on
$2000$ data sets of size
$N_{\max}=500$, under $\Q_{null}$
(left) and $\Q_{alt}$ (right). Three combinations of  design  parameters were considered:
 {(a)}  
  $\varepsilon=0.1, \varepsilon_1=0, \varepsilon_2=0.05$, $\delta=0.1$ (top),
 {(b)}
  $\varepsilon=0.05, \varepsilon_1=0, \varepsilon_2=0.05$, $\delta=0.1$ (middle),
  {(c)}
  $\varepsilon=0.2, \varepsilon_1=0,  \varepsilon_2=0.05$, $\delta=0.05$
  (bottom). 
In the subfigures, the width of  each of the 4 bands corresponds to the $\Q$-probability of a respective event in the box. Also shown are the expectations  $\E_{\Q_{null}}\bigl(\E_{\pi}\bigl( \bftheta_k\vert D_i^{*} \bigr)\bigr)$ and $\E_{\Q_{alt}}\bigl(\E_{\pi}\bigl( \bftheta_k\vert D_i^{*} \bigr)\bigr), (1\leq i \leq 500, 1 \leq k \leq 3$), computed from these simulations. For more details, see text. 
 }     
  \label{fig:combined_3ex2_activity_RULE3}
\end{figure}

The results are shown in Figure \ref{fig:combined_3ex2_activity_RULE3}. The main distinction to Figure \ref{fig:combined_ex2_activity_RULE3} is that, in the definition of the four colored bands, the events $\{I_{0,i} = 0\}$ have here been replaced by $\{N_{0,last} \leq i\}$. The widths of the three lowest bands therefore  represent the $\Q$-probabilities of the events $\{$treatment $k$ is maximal at $i, N_{0,last} \leq i\}$, $1 \leq k \leq 3$, while that of the violet band  is the $\Q$-probability of $\{N_{0,last} > i\}$.  As in the case of Experiment 1, these events have operational meanings comparable to corresponding key concepts used in hypothesis testing. Thus, on the left of Figure \ref{fig:combined_3ex2_activity_RULE3}, the sum of the three lower bandwidths at $i$ is the false positive rate when observing outcome data from $i$ patients. If its size is of major concern to a person considering the design from a frequentist perspective, it can be reduced smaller in a similar fashion as suggested in Section 3.1.2 in the context of Experiment 1, by employing a form of a burn-in period and activating adaptive treatment allocation only after some fixed number of patients have been treated in all four arms.

\section{Extensions for handling delayed outcome data} 
\label{section:no:3}

Data of the kind considered in Sections 2 and 3, where binary outcomes are determined and observed soon after the treatment is delivered, may be rare in practical applications such as drug development. 
More likely, it takes some time until a response to a treatment can measured in a useful manner. For example, the status of a cancer patient could be determined one month after the treatment was given. Incorporation of such a delay into the model is not technically very difficult, but it necessitates explicit introduction of the recruitment or arrival process, in continuous time, of the patients to the trial. 
A somewhat different problem arises if the outcome itself is a measurement of time, such as time from treatment to relapse or to death in a cancer trial, or to infection in vaccine development. 
When such information would be needed for adaptive treatment allocation, part of the data are typically right censored. Both types of extensions of the basic  Bernoulli model in Section 2 are considered briefly below.

\subsection{Fixed delay from treatment to binary outcome}

We now consider a model, where a binary outcome is systematically measured after a fixed time period has elapsed from the time at which the patient in question received the treatment. Modelling such a situation, rather obviously, requires that the model is based on a continuous time parameter. 

Let, therefore,  $ t>0 $  be a continuous time parameter, and denote by  $ U_{1}<U_{2}< \ldots <U_{i}< \ldots  $  the arrival times of the patients to the trial, again using  $ i=1, 2, \ldots  $  to index the  participants. 
We then assume that the treatment is always given immediately upon arrival, and that the outcome  $ Y_{i} $  is measured at time  $ V_{i}=U_{i}+d $, where  $ d>0  $ is fixed as part of the design. Let $  N(t) = \sum _{i \geq 1}^{ }1_{ \{ U_{i} \leq t \} },t>0,  $ be the counting process of arrivals.
At time $t$, outcome measurements are available from only those  patients who arrived and were treated before time $t-d$. Therefore, the adaptive rule for assigning a treatment to a  participant arriving at time $t$ can utilize only the data 
\begin{align*} 
 D_{t}= \{U_{i}, A_{i},C_{i} \left( t \right),C_{i} \left( t \right) Y_{i} :i \leq N(t)  \},  \end{align*}
where the indicator $C_{i} \left( t \right) = 1_{\{U_{i}< t-d\}}$ signals that $Y_{i}$ has been measured by time $t$. 
 
With a minor change from \eqref{eq:no:2.4}, 
let 
\begin{align}\label{eq:no:4} N_{k,1} \left( t \right) = \sum _{i=1}^{N \left( t \right) }C_{i} \left( t \right)1_{ \{ A_{i}=k,Y_{i}=1 \} }, \; N_{k0} \left( t \right) = \sum _{i=1}^{N \left( t \right) }C_{i} \left( t \right)1_{ \{ A_{i}=k,Y_{i}=0 \} },\; 0 \leq k \leq K, 0 < t \leq T_{\max}. \end{align}
As before, we assume that the arrival process is not informative about the model parameters, that the  participants are conditionally exchangeable given their respective treatment assignments, and that the assignment rule is the same as in Section \ref{section:no:2}. 
The main distinction between the model with instantaneous response times  and the present one with delayed measured outcomes is that, in the former case, once the outcome on an arriving  patient becomes known, there is no additional information in the data until the next  patient arrives and is treated. In the present situation, however, during such a time period some other  patients, who had arrived earlier, may complete the required duration $d$ from treatment to measured outcome and thereby provide new information to the data that are available. That information can then be utilized when deciding on the treatment for the next arriving  patient. 

By inspection we find that the basic product form of the likelihood expression \eqref{eq:no:2.3} can be retained in this case. More concretely, the only change needed in the algorithms of Rule 1 and Rule 2 is that, instead of $\Lik_{n}(\theta) \leftarrow \Lik_{n-1}(\theta)\times\theta_{r(n)}^{Y_{N(n)}} \left( 1- \theta_{r(n)} \right)^{1 - Y_{N(n)} } ,$ the inductive step for updating the likelihood becomes  
\begin{align} \Lik_{n}(\theta) \leftarrow \Lik_{n-1}(\theta) \prod_{k=0}^{K} 
\theta_{k}^{N_{k,1} \left(U_{N(n)}\right) - N_{k,1} \left(U_{N(n)-1} \right)}
\left( 1- \theta_{k} \right)^{N_{k,0} \left(U_{N(n)}\right) - N_{k,0} \left(U_{N(n)-1} \right)}. \end{align}
  
\subsection{ The case of time-to-event data}
Time-to-event data can arise in several different ways. For example, the times from treatment to  relapse or death are often used as primary endpoints in cancer trials. 
Below we show how Rule 1 and Rule 2 need to be modified to apply for such data.  

Let  $ U_{i}  $ be the time of treatment and  $ V_{i} $ the time of response for patient $i$, and let $ X_{i}=  V_{i}-U_{i} $. Changing the notation slightly, we now denote by  $ N(t) = \sum _{i \geq 1}^{ }
1_{\{ U_{i} \leq t \}}, \; t>0,  $  the process counting the arrivals to the trial.
If the data are collected at time $ t $, and   $U_{i} \leq t$ and $ V_{i}>t $ hold for patient  $ i $, the response time  $ X_{i}  $ will be right censored.
Observed in the data are then the times  
$ Y_{i} \left( t \right) = \left[ (V_{i}\wedge t) - U_{i} \right]^{+} $ and the indicators 
$ C_{i} \left( t \right) =1_{\{ V_{i} \leq t \}} =1_{\{ X_{i}=Y_{i}(t) \}}. $  

Suppose now that the original response times  $ X_{i}$ arising from treatment $ k $, i.e., those for which  $ A_{i}=k, $  are independent and distributed according to some distribution $F(x \vert \theta _{k})$ with respective parameter value $ \theta _{k}>0,\; k=0, 1, \ldots ,K $. Denote the corresponding densities by $f(x \vert \theta _{k})$. As above, we assume that the arrival process is not informative about the model parameters, and that the  participants are conditionally exchangeable given their respective treatment assignments. Then the likelihood expression corresponding to data
\begin{align*} D_{k,t}= \left\{ U_{i},A_{i}, Y_{i} \left( t \right) ,C_{i}\left  ( t \right) :i \leq N(t), A_{i}=k \right\} ,  
\end{align*}
collected from treatment arm  $k$ up to time  $t$, has the familiar form
\begin{align} \label{eq:no:5} 
L \left( \theta _{k} \vert D_{k,t} \right)
=  \prod_{i=1}^{N(t) }
f(X_{i} \vert \theta _{k})^{C_{i} \left( t \right)1_{\{A_{i}=k\}} }(1 - F(Y_{i}\left( t \right) \vert \theta _{k}))^{(1 - C_{i} \left( t \right))1_{\{A_{i}=k\}} }.
 \end{align}
Such data are in the survival analysis literature commonly referred to as data with \textit{staggered entry}. Due to the assumed conditional independence of the response times across the different treatment arms, given the respective parameters  $ \theta _{k},  $ the combined data
 \begin{align*}
  D_{t}= \bigcup_{k=0}^{K}
  D_{k,t}= \{ U_{i},A_{i}, Y_{i} \left( t \right) ,C_{i} \left( t \right):i \leq N(t)\}\end{align*} 
  give rise to the product form likelihood 
 \begin{align} 
 \label{eq:no:6} 
 L \left(  
 \theta  \vert D_{t} \right) =
 \prod_{k=0}^{K}
 L \left( \theta _{k} \vert D_{k,t}
 \right),  
 \end{align}
where $\theta=(\theta_{0},\theta_{1},\ldots,\theta_{K})$.   
Upon specifying a prior for $\theta$, the posterior probabilities corresponding to the data $D_{t}$ can then be computed and utilized in Rule 1 or Rule 2.

\textbf{Remarks.} It is well known that, in Bayesian inference, \textit{Gamma}-distributions are conjugate priors to the likelihood arising from exponentially distributed survival or duration data, with $\theta _{k}$ representing the corresponding intensity parameters.
This holds also when such data are right censored, in which case the likelihood \eqref{eq:no:5} corresponding to $D_{k,t}$ has the Poisson form, with $\sum _{i=1}^{N \left( t \right) } C_{i} \left( t \right) 1_{\{ A_{i}=k \}} $ being the number of measured positive outcomes and  $\sum _{i=1}^{N \left( t \right) }Y_{i} 
 \left( t \right) 1_{\{ A_{i}=k \}}$ the corresponding \textit{Total Time on Test} (TTT) statistic. 
Assuming independent 
\textit{Gamma}$\left(\theta _{k}\; \vert\; \alpha_{k}, \beta_{k} \right)$-priors for the respective treatment arms  $ k=0,1, \ldots ,K, $  the posterior for  $ \theta _{k} $  corresponding to data  $ D_{k,t} $  becomes
 \begin{align}  \label{eq:no:7}
 p \left( \theta _{k} \vert  D_{k,t} \right) =  \mbox{\textit{Gamma}}\biggl(   
\theta _{k}\; \vert \; \alpha _{k}+
 \sum _{i=1}^{N \left( t \right) }
 C_{i} \left( t \right) 1_{\{ A_{i}=k \}}   ,\beta _{k}+
 \sum _{i=1}^{N \left( t \right) }Y_{i} 
 \left( t \right) 1_{\{ A_{i}=k \}}  \biggr) , \end{align}
and the joint posterior  
$ p \left( \theta \; \vert \; D_{t} \right)$ is the product distribution of these independent marginals.

When considering the application of Rule 1 or Rule 2 in this exponential response time model, the natural target would often be to decrease, rather than increase, the value of the intensity parameter corresponding to an experimental treatment in the trial. Moreover, for measuring the degree of such potential improvements, use of hazard ratios, or relative risks, seems often more appropriate than of absolute differences.  Criteria such as $\P_{\pi}\left(\bftheta _{k}  \geq 
\theta_{low} \big\vert D_{n}\right)<\varepsilon_1$
and $\P_{\pi} \left( \bftheta _{0} + \delta \geq \max\limits_{\ell\in\T} \bftheta_{\ell}  \big\vert  D_{n} \right)<\varepsilon_2 $  applied   previously in Rule 2 should then be replaced by corresponding requirements of the form $\P_{\pi}\left(\bftheta _{k}  \leq 
\theta_{high} \big\vert  D_{t} \right)\;<\varepsilon_1$   
and $\P_{\pi} \left(\rho \bftheta _{0} \leq \min\limits_{\ell\in\T} \bftheta_{\ell}  \big\vert  D_{t} \right)<\varepsilon_2,$ where $\rho < 1$ is a given safety margin protecting the control arm from inadvertent dropping. Writing $\rho = \exp{\{-\delta\}}$ and using $\eta_k = -\log{\theta_k} $ as model parameters brings us back to the absolute scale, with the last inequality becoming the requirement $\bfeta_0 + \delta \geq \max\limits_{\ell\in\T}{\bfeta_{\ell}}.$

\subsection{Notes on application to vaccine trials}

An important and timely special case of  time-to-event data are data coming from large scale Phase III vaccine trials. When a newly developed vaccine candidate has reached the stage when it is tested in humans for efficacy, the  trial participants are usually healthy individuals and the control treatment is either placebo or some existing vaccine that has been already approved for wider use. In such trials adaptive treatment allocation is less likely to be an issue, whereas it would be important to arrive at some reasonably  definitive conclusion about efficacy already before reaching the planned study endpoint $N_{max}$. For this reason, in the recent trials for testing COVID-19 candidate vaccines in humans, the design has allowed for from two to five `looks` into the data before trial completion, usually defined as times at which some pre-specified number of infections have been observed. To our knowledge, most of these trials have applied frequentist group sequential methods for testing, adjusting the targeted significance level by suitably defined spending functions. This standard practice is followed in spite of that, arguably, in trials for experimental vaccines such as the COVID-19 candidates, for which Phase II has been already successfully completed, Type 1 errors could be considered less worrisome than Type 2 errors.

Entertaining the idea that such vaccine trials had been designed by using the Bayesian framework as presented above in $4.2$, this task could have been accomplished by applying Rule 2 and thereby selecting suitable values for its design parameters  $\rho, \theta_{high}, \varepsilon_1, \varepsilon_2$ and $N_{max}$, letting finally $\varepsilon = \varepsilon_2$ to inactivate the separately defined adaptive mechanism for treatment allocation. For example, considering the case of a single experimental vaccine, the value $\rho = 0.4$ would signify the target of sixty percent decrease in the value of the intensity parameter $\theta_1$ compared to the placebo control $\theta_0$, and thereby a corresponding reduction in the expected number of infected individuals among those vaccinated.  

The trial could then be run, and it would stop with declared \textit{success} if a posterior probability $\P_{\pi}\bigl(\rho\bftheta_0 < \bftheta_{1} \big\vert D_i^{*}\bigr) < \varepsilon_2$ were obtained for some $i \leq N_{max}$. On the other hand, \textit{futility} would be declared if either $\P_{\pi}\bigl(\bftheta_1 \leq \theta_{high} \big\vert D_i^{*}\bigr) < \varepsilon_1$ or $\P_{\pi}\bigl(\rho \bftheta_0 \geq \bftheta_{1} \big\vert D_i^{*}\bigr) < \varepsilon_2$ were established for such $i$. In either case, the monitoring of these probabilities could in principle be done in an open book form, and not just in a few `looks` made at pre-planned check points.

A somewhat different approach to modeling and analyzing vaccine trial data can be outlined as follows. Suppose that  the design is fixed by allocating, at time $t = 0,$ $n_1$ individuals to the vaccination group and $n_0$ individuals to the placebo group. Denote by $0 < T_{1,1} < T_{1,2} < ... $ the times at which the individuals in the former group become infected and by $0 < T_{0,1} < T_{0,2} < ... $ the corresponding times in the latter group. Expressed in terms of counting processes, $N_{1}(t) = \sum _{m\geq1}{ 1}_{
\{T_{1,m} \leq t \}}$ and $N_{1}(t) = \sum _{m\geq1}{ 1}_{
\{T_{1,m} \leq t \}}$ count the number of infections up to time $t$ in these two groups. We then assume that infections occur at respective rates $(n_1 - N_{1}(t-))\lambda_{1}(t)$ and $(n_0 - N_{0}(t-))\lambda_{0}(t)$, where $\lambda_{1}(t)$ and $\lambda_{0}(t)$ are unknown functions of the follow-up time $t$. In practice, $n_1$ and $n_0$ are large, of the order 10.000 or more, while $N_{1}(t)$ and $N_{0}(t)$ can during the observation interval be at most a few hundred. Therefore, $\{N_{1}(t); t \geq 0\}$ and $\{N_{0}(t); t \geq 0\}$ can be approximated quite well by Poisson processes with respective intensities $n_1\lambda_{1}(t)$ and $n_0\lambda_{0}(t)$.  

Suppose that these processes are (conditionally) independent given their intensities. Then the likelihood corresponding to the data $D_t = \{N_{0}(s), N_{1}(s); s \leq t\},$ combined from both groups and up to time $t$, gets the familiar Poisson-form expression 
\begin{align} \label{eq:no:8} 
L(\lambda_{0},\lambda_{1} \vert D_t) = \prod_{k=0}^1\  \exp
\biggl\{-\int_{0}^t{n_k\lambda_{k}(s)ds}\biggr\} \prod_{m \leq N_{k}(t)}n_k\lambda_{k}(T_{k,m}).\end{align} 
Assuming that the processes $\{T_{0,m}; t \geq 1\}$ and  $\{T_{1,m}; t \geq 1\}$ do not have exact ties, we now consider their superposition $\{0 < T_1 < T_2 < ...\}$ and the corresponding counting process $N(t) = N_{0}(t) + N_{1}(t) = \sum _{m\geq1}{ 1}_{
\{T_m \leq t \}}$, which then has intensity  $n_0\lambda_{0}(t) + n_1\lambda_{1}(t).$ In what follows,  for the purposes of statistical inference, this superposition is decomposed back into its components. For this, we define a sequence $\{\delta(T_m); m  \geq 1\}$ of indicators, letting $\{\delta(T_m) = 1\}$ if $\{N_0(T_m) - N_0(T_{m}-) = 1\}.$ Expressed in concrete terms, the event  $\{\delta(T_m) = 1\}$ occurs if the $m^{th}$ individual in the trial who was recorded as being infected  happens to  belong to the placebo group, and $\{\delta(T_m) = 0\}$ if to the vaccination group. It is well known that the conditional probability of these events,  given $\lambda_{0}(.), \lambda_{1}(.)$ and $\{N(T_m) - N(T_{m}-) = 1\},$ are equal respectively to
$  n_0 \lambda_{0}(T_m)(n_0 \lambda_{0}(T_m) + n_1 \lambda_{1}(T_m))^{-1}$ and $ n_1 \lambda_{1}(T_m)(n_0 \lambda_{0}(T_m) + n_1 \lambda_{1}(T_m))^{-1}$. 

Estimation of the function $\lambda_{0}(.)$, describing the infection pressure in the non-vaccinated population, may be possible by utilizing data sources that are external to the trial, but estimation of $\lambda_{1}(.)$ would be hard. This problem can be circumvented if we are ready to impose a proportionality assumption, according to which, although the rates at which infections occur in the vaccination and placebo groups generally vary in time, their ratio is a constant $\rho > 0.$ Expressed in symbols, we assume then that   $\lambda_{1}(t) = \rho \lambda_{0}(t), t \geq 0$. The smaller the value of $\rho$, the better protected, according to this model, the vaccinated individuals are. The value $1 - \rho$ is what is commonly called \textit{vaccine efficacy at reducing infection susceptibility}, abbreviated as $VE_{S}$ (e.g., \textcite{halloran2010design}).

The postulated proportionality property appears to be reasonable if all trial participants are vaccinated approximately at the same time, in which case $t$ refers to time from vaccination, and if both groups, due to randomization, can be assumed to be exposed to approximately the same infection pressure. If the trial participants have been recruited from different geographical regions with highly varying levels of infection pressure, a stratified analysis based on a common vaccine efficacy value might still be possible. However, if  vaccination takes place over a longer time period, it becomes difficult to differentiate from each other the effects of infection pressure, varying in the population with calendar time, and that of individual level susceptiblity, which is likely to depend on the build-up of the immune response and thereby on the time from vaccination.  

A different matter, which has received much attention recently in connection of COVID-19 vaccine trials, is the dependence of $\rho$ on age, due to the immune response in the older age groups generally developing more slowly. Stratification of the analyses by using some age threshold has been applied, but the selected thresholds have varied. This is a problem for statistical analysis as long as the numbers of infected individuals in some age groups remain low. 

Supposing now a common value for $\rho,$ there are two alternative approaches to be selected from: Either (i) considering joint inferences on the pair $(\lambda_{0}(.), \rho)$, using the "full" likelihood \eqref{eq:no:8} for this purpose and introducing a separate model for a description of $\lambda_{0}(.)$, or (ii) following the path well known from the context of the Cox proportional hazards model and employing a corresponding \textit{partial likelihood} expression (e.g., \textcite{Yip2000APL}). In a stratified analysis, the (partial) likelihood expressions would become products across the considered strata. Here we consider briefly the approach based on partial likelihood. A comparative assessment of these approaches is beyond the scope of this presentation.

By inserting the assumed form $\lambda_{1}(.) = \rho \lambda_{0}(.)$ of the intensity  $\lambda_{1}(.)$ into \eqref{eq:no:8}, it can be written, after some re-arrangement and cancellation of terms, in the form
\begin{align*}  
L(\lambda_{0},\rho \vert D_t) = &   \exp\biggl\{-(n_0 + n_1\rho)
\int_{0}^t
\lambda_{0}(s)ds\biggr\} 
(n_0 + n_1\rho)^{
N(t)}\prod_{m \leq N(t)}
\lambda_{0}(T_m)& \\ & \times
\prod_{m \leq N(t)}
\biggl(\frac{n_0 }{n_0 + n_1\rho} \biggr)^{\delta(T_m)} \biggl(\frac{n_1 \rho}{n_0 + n_1\rho}\biggr)^{1 - \delta(T_m)}. &
\end{align*} 
The latter product in this expression simplifies further into
\begin{align} \label{eq:no:9}
L_{part}(\rho \vert D_t) = {\biggl(\frac{n_0}{n_0 + n_1\rho } \biggr)^{\sum\limits_{m \leq N(t)}\delta(T_m)}}
{\biggl(\frac{n_1 \rho}{n_0 + n_1\rho }\biggr)^{\sum\limits_{m \leq N(t)}(1 - \delta(T_m))}}  =  \theta^{N_0(t)}(1 - \theta)^{N(t) - N_0(t)},  
\end{align}
where we have denoted $\theta = n_0(n_0 + n_1\rho )^{-1}.$ This is the sought-after partial likelihood and, parameterized in this way, it has the familiar Binomial form.  The word \textit{partial} signifies the fact that the parts in the "full" likelihood that were omitted in the derivation of \eqref{eq:no:9} also contain the unknown model parameter $\rho$. We now proceed by employing the approximation where the partial likelihood is treated as if it were the "full".  On specifying a $Beta(\;.\; \vert \; \alpha, \beta)$-prior for $\theta$, and using the conjugacy property of the \textit{Beta-Binomial} distribution family, we would get the posterior 
$p(\theta \;\vert \; D_{t}) = \mbox{\textit{Beta}}(   
\theta \; \vert \; \alpha + N_0( t), \beta + N(t) - N_0(t)),$ and further the posterior for $\bfrho$ by noting that $\rho = n_0(1 - \theta)/n_1\theta$. 

However, a \textit{Beta}-prior may not be fully appropriate for this particular application. More naturally we could postulate, for example, the \textit{Uniform}$(0,1)$ prior for $\rho$. It would correspond to the assumption that infectivity in the vaccine group cannot be larger than in the placebo group, but all  values of vaccine efficacy between $0$ and $100$ percent are a priori equally likely. This would entail for $\theta$ a prior density, which is no longer of \textit{Beta}-form. With the conjugacy property lacking in this case, the posterior can nevertheless be computed easily by applying Markov Chain Monte Carlo sampling. 

While adaptive treatment allocation appears to be less of an issue in vaccine trials, there will be more interest in how, and when, results from such trials could be appropriately reported. At times such as the current SARS-CoV-2 pandemic, there is much pressure towards  making the results from vaccine trials available as soon as a pre-specified level of certainty can be assured.   Again, consistent with the likelihood principle, all monitoring of posterior probabilities could be done in an open book form, and not just in a few 'looks' at pre-planned check points. For example, the trial could be run, and it could stop with declared success at time $t$ if the posterior probability   $\P_{\pi}(VE_S \geq ve^* \vert D_t) > 1 - \varepsilon_1$ were obtained, with $ve^*$ a pre-specified minimal target value and $\varepsilon_1$ having a small value such as $0.05$ or $0.01$. (To compare, according to the 
WHO guidelines
for evaluation of COVID-19 vaccines (\textcite{who_2020}), for a candidate vaccine  the primary efficacy endpoint point estimate in a placebo-controlled efficacy trial should be at least 50 percent, and the lower bound of the appropriately alpha-adjusted  confidence  interval  around  the  primary  efficacy  endpoint  point  estimate should be larger than 30 percent. Note that, while such a criterion defines a stopping time with respect to the internal history of the trial, it violates the likelihood principle.) A similar criterion could be set up for declaring futility.  

To give an  example from a recent real study, Moderna, Inc. announced on November 30, 2020 (\textcite{moderna_2020})
a primary efficacy analysis
of their Phase III COVID-19 Vaccine Candidate. The announcement,
based on a randomized, 1:1 placebo-controlled study of $30.000$ participants, reported $185$ infections in the placebo group and $11$ in the vaccine group, leading to the point estimate $11/185 = 0.059$ of $\rho$ and thereby efficacy estimate 0.941. We computed the posterior density $p(\rho \;\vert \; D_t)$ of $\bfrho$, using these data $N_0{(t)}=185$ and $N_1{(t)}=11$ and assuming the uniform prior for $\bfrho$ as described above. The result, together with the $95$ percent HPDI $(0.030,0.105)$, is shown in Figure  9. The corresponding HPDI for $VE_S = 1 - \rho$ is then $(0.895,0.970)$.

 \begin{figure} \includegraphics[height=\macroheight, width=\textwidth]{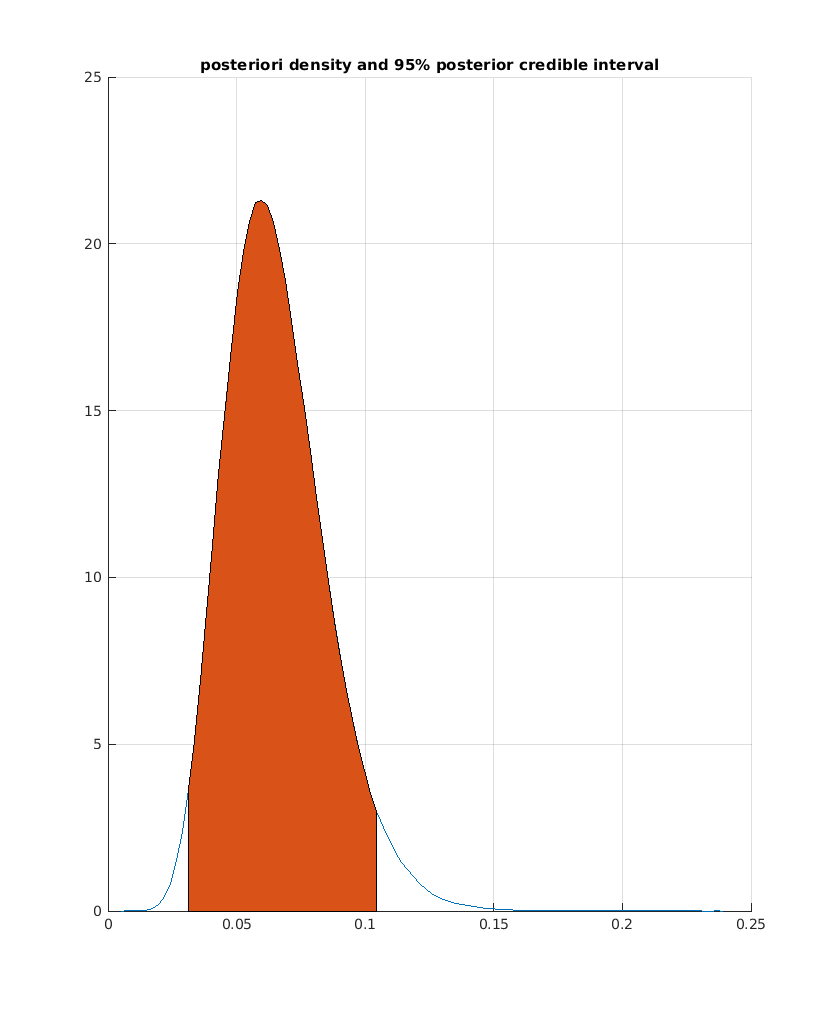} 
 \caption{Posterior density of $\bfrho$ based on Moderna, Inc. COVID-19 primary efficacy data, with posterior mode  at $0.0595$ and $95\%$ HPD interval
 $(0.030,0.105).$} 
 \end{figure} 
 
\textbf{Remarks.} A practical advantage of the Poisson process approximation entertained above is that only the numbers $N_0(t)$ and $N_1(t)$ are needed for computing the posterior of $\bfrho$ at time $t$. If $n_0$ and $n_1$ are not large enough to justify such an approximation, statistical inference based on partial likelihood is still possible, but it then necessitates monitoring of the sizes of the two risk sets. The exact times of infection are not required, but the ordering in which members of either the placebo or of the vaccine groups become infected needs to be known. As in the case of the Cox proportional hazards model, the partial likelihood expression is then somewhat more involved and the computations more slow.

In the above approach and analysis we have assumed that the risk set sizes are reduced only due to the trial participants becoming infected. This may not be so, as there may be various other reasons why they may be lost from follow-up. If the resulting right censoring concerns a large proportion of the participants, this has to be accounted for in the analysis. It does not create a conceptually difficult problem, but it requires that the sizes of the risk sets,  both in the vaccine and the placebo groups, are known at the times at which new infections are registered. The simple power form expression \eqref{eq:no:9} for partial likelihood is then not valid any more, and needs to be replaced by the product
\begin{align} \label{eq:no:10}
L_{part}(\rho \vert D_t) = \prod_{T_m\leq t} {\biggl(\frac{R_{0,T_m}}{R_{0,T_m}+R_{1,T_m}\rho}\biggr)^{\delta(T_m)}}
{\biggl(\frac{R_{1,T_m} \rho}{R_{0,T_m} + R_{1,T_m}\rho }\biggr)^{ 1 - \delta(T_m)}}   ,  
\end{align}
where $R_{0,T_m}$ and $R_{1,T_m}$ are the sizes of the two risk sets at time $T_m$. It is, in fact, a simple form of the familiar expression used for the Cox proportional hazards model, connected to the latter by the transformation $\rho = \exp{\{-\beta\}}.$

Currently, several vaccines against COVID-19 have been successfully tested in placebo controlled Phase III trials and, somewhat depending on the country, have then been approved by the relevant regulatory authorities for wider use in their respective population. In addition to the original efficacy trials, there are now several studies on the  population level effectiveness of COVID-19 vaccines (e.g., \cite{dagan2021bnt162b2}, \cite{vasileiou2021interim}). On the other hand, in the present situation in which several vaccines that are demonstrably efficacious against both infection and the more serious forms of COVID-19 disease are available, it is difficult to find support, for a number of different reasons, to   additional large-scale placebo controlled trials for testing new candidate vaccines, cf. \cite{krause2020covid}. 

A possible alternative to such testing would be to use one or more of these existing vaccines as controls, and then make a comparative study. Such a design presents two major challenges, however. The first difficulty is demonstrated clearly by the Moderna study described briefly above: Of the approximately $15.000$ individuals in the vaccine group only 11 were infected during the trial. If the candidate vaccine has at all comparable efficacy, as would naturally be desirable, the number of infected individuals in the vaccine group of a similar size, and assuming a comparable infection pressure in the study population, could not be expected to be much larger. With such small frequencies from both treatment arms in the trial, it would not be possible to arrive at a sufficiently firm conclusion concerning the desired target of \textit{superiority} or \textit{non-inferiority}, and this would be the case regardless of the statistical paradigm that were applied for such purpose. 

To overcome this problem, it would therefore be almost mandatory to seek regulatory approval to a design in which healthy volunteers, some vaccinated by the candidate and some by an already approved vaccine, say \textit{Vaccine*}, used as a control treatment, are exposed to the virus under a carefully specified protocol. The possibility of a \textit{human challenge} design, albeit with placebo controls, was already discussed at the time when no efficacious vaccine was available 
(\cite{who_2020},
\cite{eyal2020human}, \cite{Richards502}), and it is still considered relevant now (\cite{eyal2021test}).
One could anticipate that in a challenge trial, naturally depending on the level of viral exposure that would be applied, a much smaller number of participants  would be needed for reaching a statistically valid conclusion on comparability. If desired, such a design could be extended to involve more than a single candidate and/or control vaccine. Note that adaptive sequential recruitment and Bayesian  decision making, as exemplified by Rule 2, would find here their natural place: It would not be necessary to fix the group sizes in advance; the trial could be run with newly recruited individuals until the desired level of certainty, as specified in the design, has been reached.

A second issue arising in the context of such a design concerns statistical modeling and inference in a situation in which information comes from different data sources: While the design may lead to an efficacy estimate where the candidate vaccine is compared to another in routine use, this estimate cannot be readily converted to a corresponding $VE_S$-estimate, where the candidate vaccine is compared to placebo. For practical consideration, this latter estimate could be the one of most interest. An approximate solution to this problem could be provided by assuming that the relative $VE_S$-efficacy measures obtained from  different trials, viz. an 'old' trial for testing \textit{Vaccine*} vs. placebo, and the 'new' trial for testing the candidate vaccine vs. \textit{Vaccine*}, act multiplicatively on each other, which would correspond to the structure of the Cox proportional hazards model. This would then yield a synthetic $VE_S$-estimate for comparing the candidate vaccine to placebo, with a corresponding posterior derived by applying Bayesian inferential tools providing an uncertainty quantification. The relevance of this idea of combining estimates from different trials needs to be given careful scrutiny, however, and in particular since the dominant virus variant may have changed in between. This approach will be studied in more detail elsewhere.

\section{Discussion}\label{section:no:4}
	
Clinical trials are an instrument for making informed decisions. In Phase II trials, the usual goal is to make a comparative evaluation on the success rates  of one or more experimental treatments to a standard or control, and in multi-arm trials, also to each other. More successful treatments among the considered alternatives, if found, can then be selected for further study, possibly in Phase III. 

With this as the stated goal for a trial, the conclusions should obviously be drawn as fast as possible, but not jumping ahead of the evidence provided by the acquired data. Both aspects can be accounted for by applying a suitable adaptive design, allowing for a continuous monitoring of the outcome data, and then utilizing in the execution of the trial the information that the data contain.  Still, there is always the antagonism  \textit{Exploration} versus \textit{Exploitation}: From the perspective of an individual patient in the trial, under postulated exchangeability, the optimal choice of treatment would be to receive the one with the largest current posterior mean of the success rate, as this would correspond to the highest predictive probability of treatment success. However, as demonstrated in \textcite{Villar2015}, this \textit{Current Belief} (CB) strategy leads to a very low probability of ultimately detecting the best treatment arm among the considered alternatives and would therefore be a poor choice when considering the overall aims of the trial.  

Finding an appropriate balance between these two competing interests is a core issue in the design and execution of clinical trials, and can realistically be made only in each concrete context. For example, in trials involving medical conditions such as uncomplicated urinary infections, or acute ear infections in children, use of balanced non-adaptive 1:1 randomization to both symptomatic treatment and antibiotics groups appears fully reasonable. A very different example is provided by the famous ECMO trial on the use of the potentially life-saving technique of extracorporeal membrane oxygenation in treating newborn infants with severe respiratory failure (e.g., \textcite{Bartlett479}, \textcite{wolfson2003development}).    While statisticians advising clinical researchers have the responsibility of making available the best methods in their tool kit, there may well be overriding logistic, medical or ethical arguments which determine the final choice of the trial design.  It has been even suggested that randomized clinical trials as such can present a scientific/ethical dilemma for clinical investigators, see \textcite{royall1991ethics}.

Bayesian inferential methods are naturally suited to sequential decision making over time. In the present context, this involves  deciding at each time point    whether to continue accrual of more  participants to the trial or to stop,  either temporarily or permanently, and if such accrual is continued,  selecting  the treatment arm to which the next arriving participant is assigned. The current joint posterior distribution of the success parameters captures then the essential information in the data that is needed for such decisions. 
 
The posterior probabilities used for formulating Rule 2, when considered as functions of the accumulated data $D_{n}$, can  be viewed as test statistics in sequential tests of null hypotheses against corresponding alternatives. This link between the Bayesian and the  frequentist  inferential approaches makes it possible to compute, for the selected design parameters, the values of traditional performance criteria such as false positive rate and power.  In the present approach, specifying a particular value for the trial size has no real theoretical bearing, and would serve mainly as an instrument for resource planning. Instead, the emphasis in the design is on making an appropriate choice of the operating characteristics, the $\varepsilon$'s  and $\delta$, which control the execution of the trial, and on the direct consideration of posterior probabilities of events of the form $ \{ \bftheta _{k}= \bftheta_{\vee}\} $ and $ \{\bftheta _{0}+\delta \geq \bftheta _{\vee} \} $ when monitoring outcome data from the trial. 

An important difference to the methods based on classical hypothesis testing is that posterior probabilities, being conditioned on the observed data, are directly interpretable and meaningful concepts as such, without reference to their quantile value in a sampling distribution conditioned on the null. This is true regardless of whether the trial design applies adaptive treatment allocation and selection while the trial is in progress, or whether only a final posterior analysis is performed when an  initially prescribed  number of trial participants have been treated and their outcomes observed. 

Large differences between the success parameters, if present, will often be detected early  without need to wait until reaching a planned maximal trial size. On the other hand, if the joint posterior stems from an interim analysis, it forms a principled basis for predicting, in the form the consequent posterior predictive distribution, what may happen in the future if the trial is continued  (e.g., \textcite{SPIEGELHALTER19868}, \textcite{yin2012phase}). Note, however, that future outcomes are uncertain even in the fictitious situation in which the true values of the success parameters were known. Therefore, from the perspective of decision making, the predictive distribution involves only "more uncertainty" than the posterior, not less. 

Another advantage of the direct consideration of posterior probabilities is that the joint posterior of the success parameters may contain useful empirical evidence for further study even when no firm final conclusion from the trial has been made.  This is in contrast to classical hypothesis testing, where, unless the observed significance level is below the selected $\alpha$-level so that the stated null hypothesis is rejected, the conclusion from the trial remains hanging in mid-air, without providing  much guidance on whether some parts of the study would perhaps deserve further experimentation and consequent closer assessment.

The standard paradigm of null hypothesis significance testing (NHST), and particularly the version where the observed  $p$-value is compared mechanistically to a selected $\alpha$-level such as $0.05$, have been criticised increasingly sharply in the recent statistical literature (e.g., \textcite{doi:10.1080/00031305.2016.1154108}, \textcite{greenland2016statistical}). In spite of this, the corresponding strong emphasis on  controlling the frequentist Type 1 error rate at a pre-specified fixed level has been largely adopted in the Bayesian clinical trials literature as well (e.g.,  \textcite{shi2019control}, \textcite{stallard2020comparison}). These error rates are conditional probabilities, evaluated from a sampling distribution under an assumed null hypothesis $\Q_{null}$ and in practice computed during the design stage when no actual outcome data from the trial are yet available. In contrast, in the Bayesian clinical trials methodology as outlined here, error control against false positives is performed continuously while the trial is run by applying  bounds of the form  $\P_{\pi}\bigl( \bftheta_0+\delta \geq \bftheta_{\vee}\big\vert D_i^{*}\bigr) < \varepsilon_2$, where the considered posterior probabilities are conditioned on the currently available trial data $D_i^{*}$. For this reason, in our view, calibration of Bayesian trial designs on a selected fixed frequentist Type 1 error rate (e.g., \textcite{Thall2015StatisticalCI}) does not form a natural basis for comparing such designs. More generally, the role of testing a null hypothesis and the consequent emphasis on Type 1 error rate should not enjoy primacy over other relevant criteria in  drawing concrete conclusions from a clinical trial   (\textcite{greenland2020analysis}). Even posterior inferences  alone are not sufficient  for rational decision making in such a context, and should therefore optimally be combined with appropriately selected utility functions (e.g., D.V. Lindley  in \textcite{grieve1994bayesian}). 

If the trial is continued into Phase III, this can be done in a seamless fashion by using the joint posterior of the selected   treatments from Phase II as the prior for Phase III.  In particular, if some treatment arms have been dropped during Phase II, the trial can be continued into Phase III as if the selected remaining treatments had been the only ones present from the very beginning. Recall, however, from the remarks made in Section 2 that such treatment elimination, as encoded into Rule 2, contains a violation of the likelihood principle. 

If Rule 2 is employed in Phase III, and considering that Phase III trials are commonly targeted at providing confirmatory evidence on the safety and efficacy of the new experimental treatment against the current standard treatment used as a control, it may be a reasonable idea to lower the threshold values $\varepsilon_1$  and $\varepsilon_2$ from their levels used in Phase II, and thereby apply stricter criteria for final approval.

No statistical method is uniformly superior to others on all accounts. Important criticisms against the use of adaptive randomization in clinical trials have been presented, e.g., in \textcite{Thall2015StatisticalCI}. There, computer simulations were used to compare adaptive  patient allocation based on  Thompson’s rule (\textcite{thompson1933likelihood}, \textcite{Villar2015}) in its original and fractional forms, in a  two-arm 200-patient clinical trial, to an equally randomized group sequential design.   The main argument against using methods applying adaptive randomization was their potential instability, that is, there was, in the authors' view, unacceptably large (frequentist) $\Q$-probability of allocating  more patients to the inferior treatment arm, the opposite of the intended effect. Although these simulations were restricted to Thompson's rule, the criticism in \textcite{Thall2015StatisticalCI} was directed more generally towards applying adaptive randomization and would therefore in principle apply  to our Rules 1 and 2 as well. The results from our limited simulation experiments, shown in graphical form in Figure \ref{fig:combined_Ystat_RULE1_200} and  Figures \ref{fig:combined_Ystat_RULE1_100} and \ref{fig:combined_Ystat_RULE1}  in the Supplement, do not support such a firm negative conclusion, however. This holds at least  provided that the possibility of actually dropping a treatment arm is deferred to a somewhat later time from the beginning of the trial, and  that in such assessment the deviations from balance in the opposite directions are not weighted   completely differently. A precautionary approach to the design, from a frequentist perspective, could apply a sandwich structure, starting with a symmetric burn-in, followed by an adaptive treatment allocation realized by Rule 1 or Thompson's rule, and finally coupling in Rule 2 for actual treatment selection.

Another criticism presented in \textcite{Thall2015StatisticalCI}  was that, for trial data collected from an adaptive design, the considered tests had lower power than in a corresponding equally randomized design, and particularly so if the tests were calibrated to have the same Type 1 error rate. This question was discussed   in subsection \ref{sub:3.1.3} and in the corresponding part of the Supplement. In these experiments,  adaptive treatment allocation methods based on Rule 1 (a) and (b), and on Thompson's rule with fractional power $\kappa = 0.25$, demonstrated frequentist performance quite comparable to what was observed when applying the fully symmetric block randomization design (d). 

All adaptive methods favoring treatment arms with relatively more successes in the past will inevitably introduce some degree of bias in the estimation of the respective success parameters, see \textcite{Bauer1994EvaluationOE} and \textcite{Villar2015}. A comprehensive review of the topic is provided in \textcite{robertson2021point}. Here, we have only considered this matter briefly in our simulation experiments, and instead emphasized the, in our view, more important aspect of the mutual comparison of the performance of different treatment arms in the trial. All biases in these experiments were relatively small and in the same direction, downward, and therefore unlikely to have had a strong influence on the  conclusions that were drawn.  

Our main focus has been on trials with binary outcome data, where individual outcomes could be measured soon after the treatment was delivered. More complicated data situations  were outlined in Section 4. The important case of normally distributed outcome data was by-passed here; there is a large body of literature relating to it, e.g., \textcite{spiegelhalter1994bayesian} and \textcite{gsponer2014practical}. A complication with the normal distribution is that, unless the variance is known to a good approximation already from before, there are two free parameters to be estimated for each treatment. If a suitable yardstick  at the start is missing, many observations are needed before it becomes possible to separate the statistical variability of the outcome measures from a true difference between treatment effects.  

In principle, the logic of Rules 1 and 2 remains valid and these rules can be applied for different types of outcome data, requiring only the ability to update the posterior distributions of the model parameters of interest when more data become available. The computation of the posteriors is naturally much less involved if the prior and the likelihood are conjugate to each other. Vague priors, or models containing more than a single parameter to be updated, will necessarily require more outcome data before adaptive actions based on Rule 1 or Rule 2 can kick in. 

If such updating is not done systematically after each individual outcome is measured, for example, for logistic reasons, but less frequently in batches, Rule 1 and Rule 2 can still be used at the times at which the batches are completed. The same holds if updating is done at regularly spaced points in time. Such thinning of the data sequence has the effect that some of the actions that would have been otherwise implied by Rule 1 and Rule 2 are then postponed to a later time or even omitted. In designing a concrete trial, one then needs to find an appropriate balance between, on one hand, the costs saved in logistics and computation, and on the other, the resulting loss of information and the effect this may have to the quality of the inferences that can be drawn. 

\textbf{Acknowledgements} 

We are grateful to Jukka Ollgren for comments and encouragement, and to Mikko Marttila for useful suggestions on the text. E.A. thanks Arnoldo Frigessi and David Swanson for support and useful discussions during an early stage of this work.

\printbibliography

\title{Supplement for the
article: 
Adaptive treatment allocation and selection in multi-arm  clinical trials: a Bayesian perspective
 }

\beginsupplement
\maketitle
\appendix


\section{Additional figures to subsection \ref{sub:3.1.2}} \label{supplement:A}

\begin{figure}[h!tbp]
     \begin{subfigure}[b]{\textwidth}
     \includegraphics[width=\textwidth]{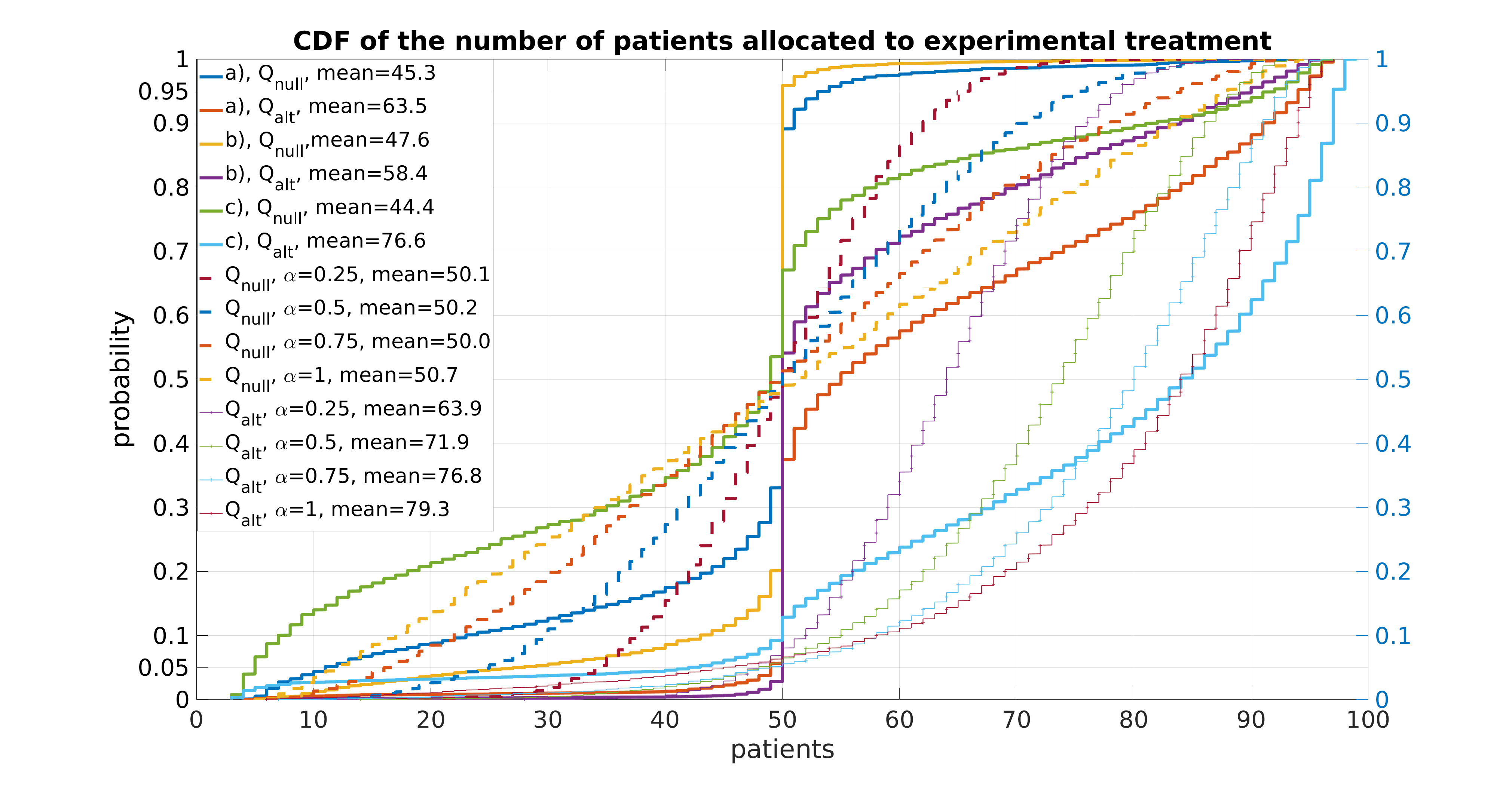}
       \renewcommand\thesubfigure{a}  
     \label{subfig:Ystat:a:null:rule1_100}
     \end{subfigure}
     \vfill 
     \begin{subfigure}[b]{\textwidth}
     \includegraphics[width=\textwidth]{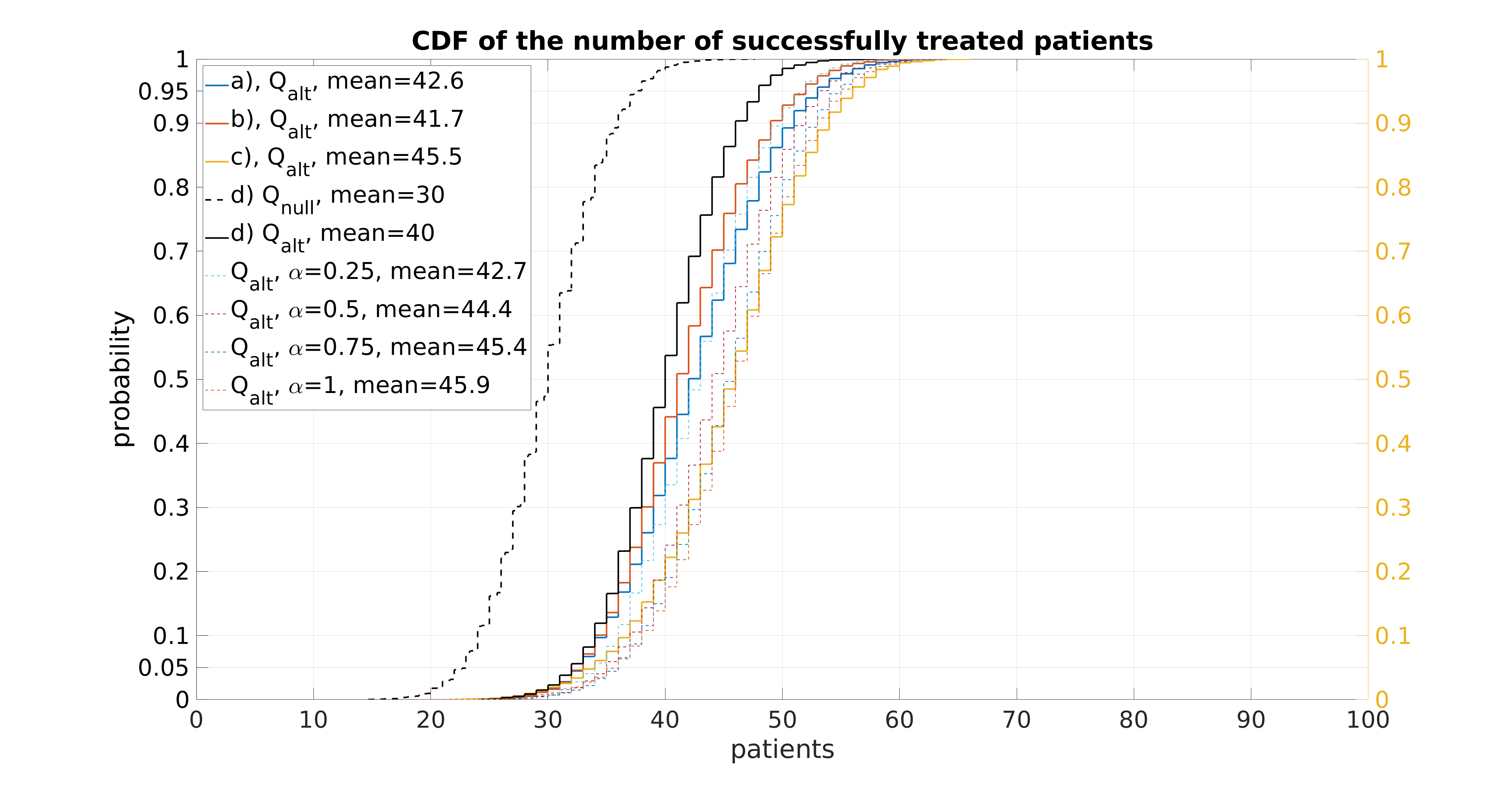}
        \renewcommand\thesubfigure{a} 
 \label{subfig:Ystat:a:alternative:rule1_100}
     \end{subfigure}
\caption{\small Effect of the choice of the threshold parameters $\varepsilon$ and $\delta$ in Rule 1 on the number of   patients allocated to the experimental treatment and on the total number of treatment successes. Cumulative distribution functions of $N_1(100)$ (top) and $S(100)$ (bottom) are shown, based on $5000$ simulated data sets,  under $\Q_{null}$ with true parameter values
$\theta_0=\theta_1=0.3$ and $\Q_{alt}$ with values
$\theta_0=0.3,\theta_1=0.5$. Three combinations of the 
design parameters were used:  {(a)} $\varepsilon=0.1$,  $\delta=0.1$, {(b)}   $\varepsilon=0.05$, $\delta=0.1$, {(c)}
$\varepsilon=0.2$, $\delta=0.05$. In addition, {(d)} represents a completely symmetric treatment allocation. For comparison we also plot the corresponding CDF under the alternative hypothesis obtained by using fractional Thompson's rule 
with respective parameters
$\kappa=0.25,0.5,0.75 $ and $1$.
}
\label{fig:combined_Ystat_RULE1_100}
\end{figure}

\begin{figure}[h!tbp]
     \begin{subfigure}[b]{\textwidth}
     \includegraphics[width=\textwidth]{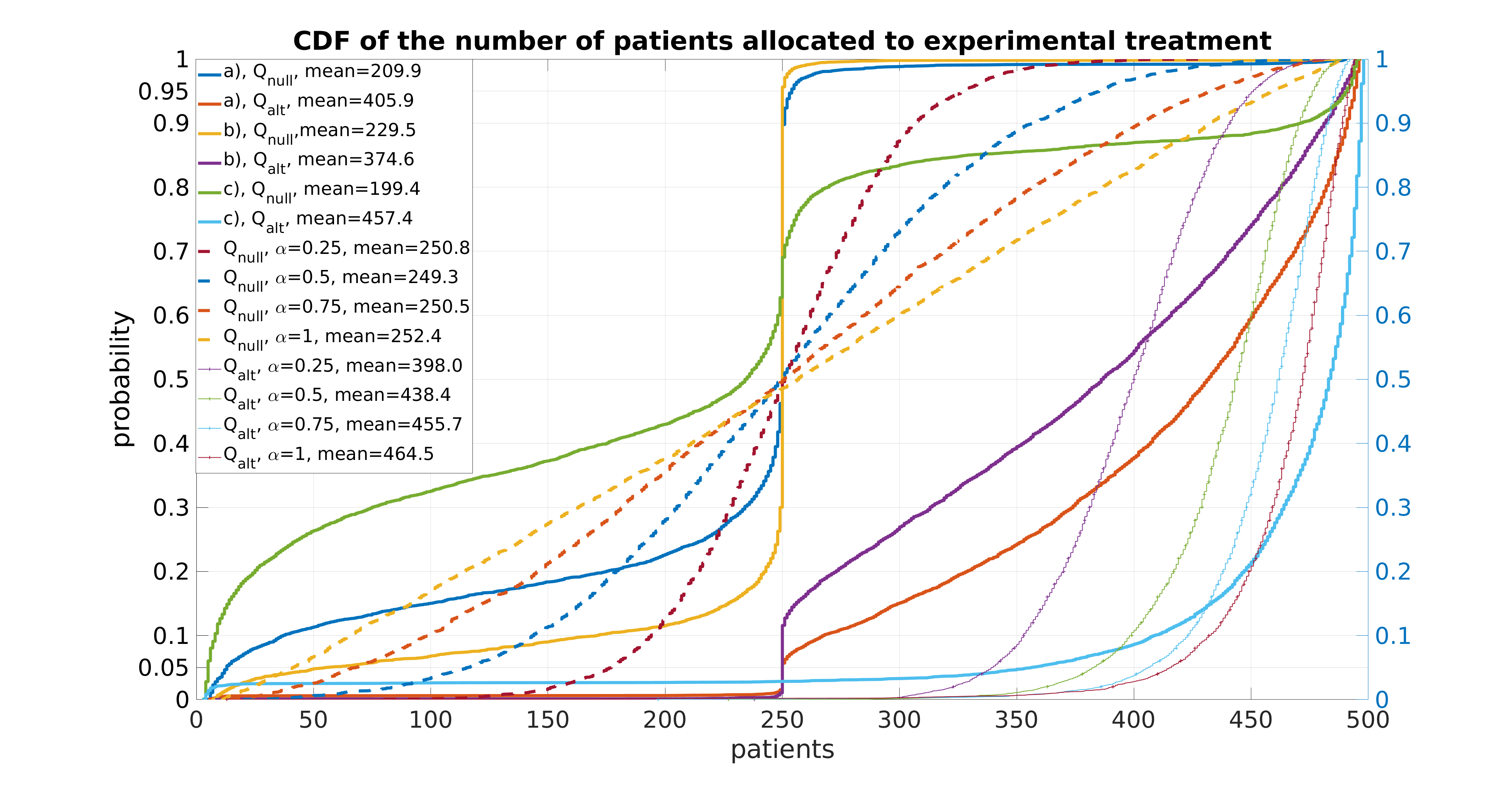}
       \renewcommand\thesubfigure{a}  
     \label{subfig:Ystat:a:null:rule1}
     \end{subfigure}
     \vfill 
     \begin{subfigure}[b]{\textwidth}
     \includegraphics[width=\textwidth]{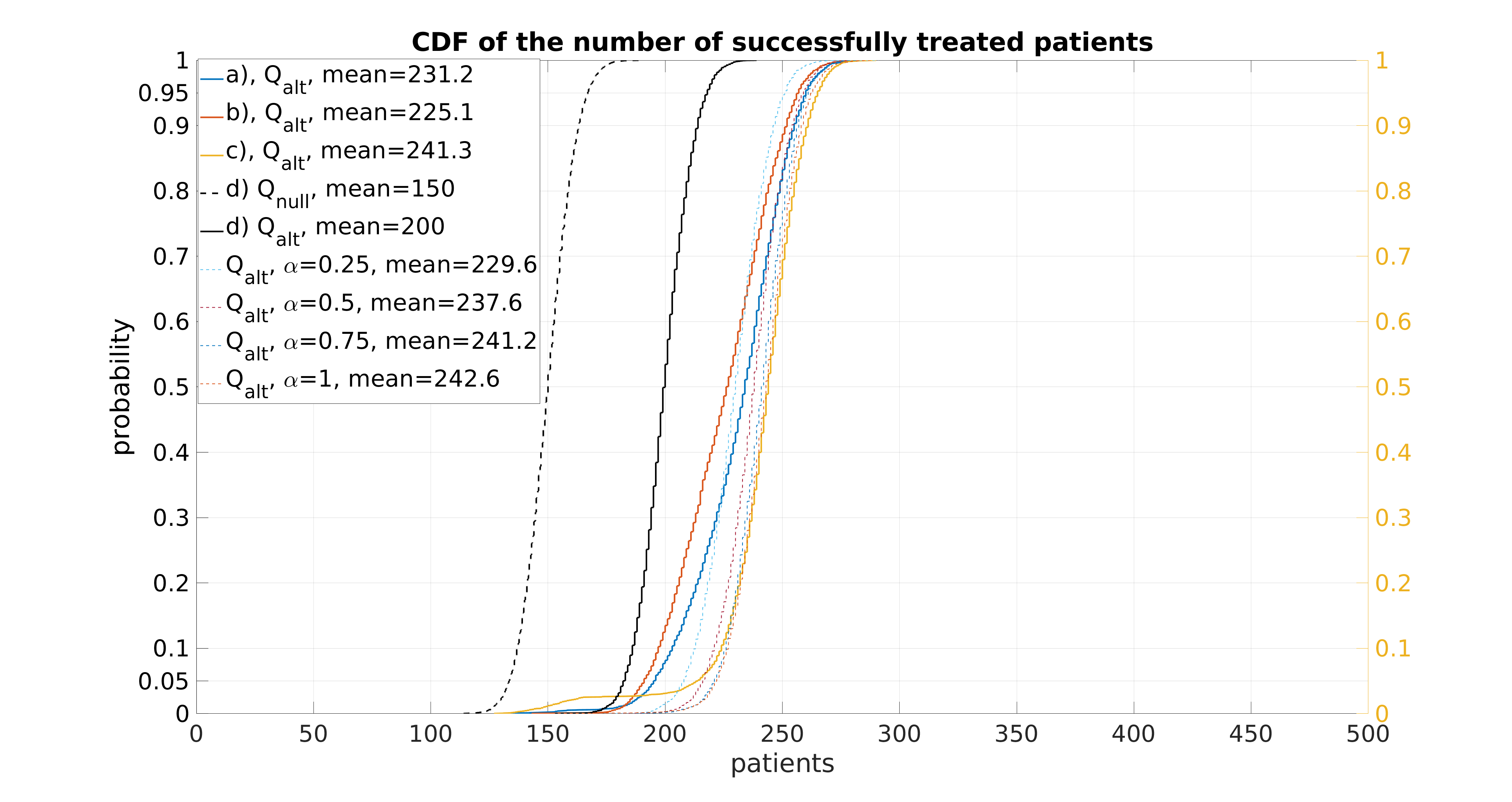}
        \renewcommand\thesubfigure{a} 
 \label{subfig:Ystat:a:alternative:rule1}
     \end{subfigure}
\caption{\small Effect of the choice of the threshold parameters $\varepsilon$ and $\delta$ in Rule 1 on the number of   patients allocated to the experimental treatment and on the total number of treatment successes. Cumulative distribution functions of $N_1(500)$ (top) and $S(500)$ (bottom) are shown, based on $5000$ simulated data sets,  under $\Q_{null}$ with true parameter values
$\theta_0=\theta_1=0.3$ and $\Q_{alt}$ with values
$\theta_0=0.3,\theta_1=0.5$. Three combinations of the 
design parameters were used:  {(a)} $\varepsilon=0.1$,  $\delta=0.1$, {(b)}   $\varepsilon=0.05$, $\delta=0.1$, {(c)}
$\varepsilon=0.2$, $\delta=0.05$. In addition, {(d)} represents a completely symmetric treatment allocation. For comparison we also plot the corresponding CDF under the alternative hypothesis obtained by using fractional Thompson's rule 
with respective parameters
$\kappa=0.25,0.5,0.75$ and $1$.
}
\label{fig:combined_Ystat_RULE1}
\end{figure}

Figures \ref{fig:combined_Ystat_RULE1_100} and \ref{fig:combined_Ystat_RULE1} below complement Figure  \ref{fig:combined_Ystat_RULE1_200} in the main paper, where  we illustrated the effect of the design parameters of Rule 1 and Thompson's rule on treatment allocation in a two-arm trial with $N_{max}=200$, and on the consequent total number of treatment successes. Here we do the same  for $N_{max}=100 $ in Figure \ref{fig:combined_Ystat_RULE1_100} and for $N_{max}=500$ in  Figure \ref{fig:combined_Ystat_RULE1}.

For data generated under $\Q_{null}$, the overall shape of the CDFs in Figures \ref{fig:combined_Ystat_RULE1_100} and   \ref{fig:combined_Ystat_RULE1} remains remarkably close to that in Figure \ref{fig:combined_Ystat_RULE1_200}, where the trial size was $N_{max}=200$. The differences become more evident when considering    $\Q_{alt}$, in which case the adaptive rules can use their potential to assign more patients to the treatment with higher true success rate. But learning from data takes time, and therefore the gains from using such adaptive rules become progressively more evident as the trial size increases. Thus,  the expected number of successes can be increased by approximately ten percent by employing a strong adaptive treatment allocation rule when $N_{max}=100$, by fifteen percent when $N_{max}=200$, and twenty percent when  $N_{max}=500$. 

Another point of interest in the case of $\Q_{alt}$ is the probability of unwanted imbalance,   allocating more patients to the inferior control arm than to the better experimental one. The highest risk for this to happen is in the case of  Rule 1 (c), for which it was found to be approximately five percent when $N_{max}=200$.  The corresponding   percentage for $N_{max}=100$ is ten and for  $N_{max}=500$ three. Rule 1 (c) appears to be the only design, among those considered, for which there is a non-negligible probability that the imbalance turns out to be serious. For the other designs, including different versions of Thompson's rule, the probabilities are much smaller, and very small for $N_{max}=500$.

\begin{figure}[!htbp]
     \begin{subfigure}[b]{\textwidth}
     \includegraphics[width=\textwidth]{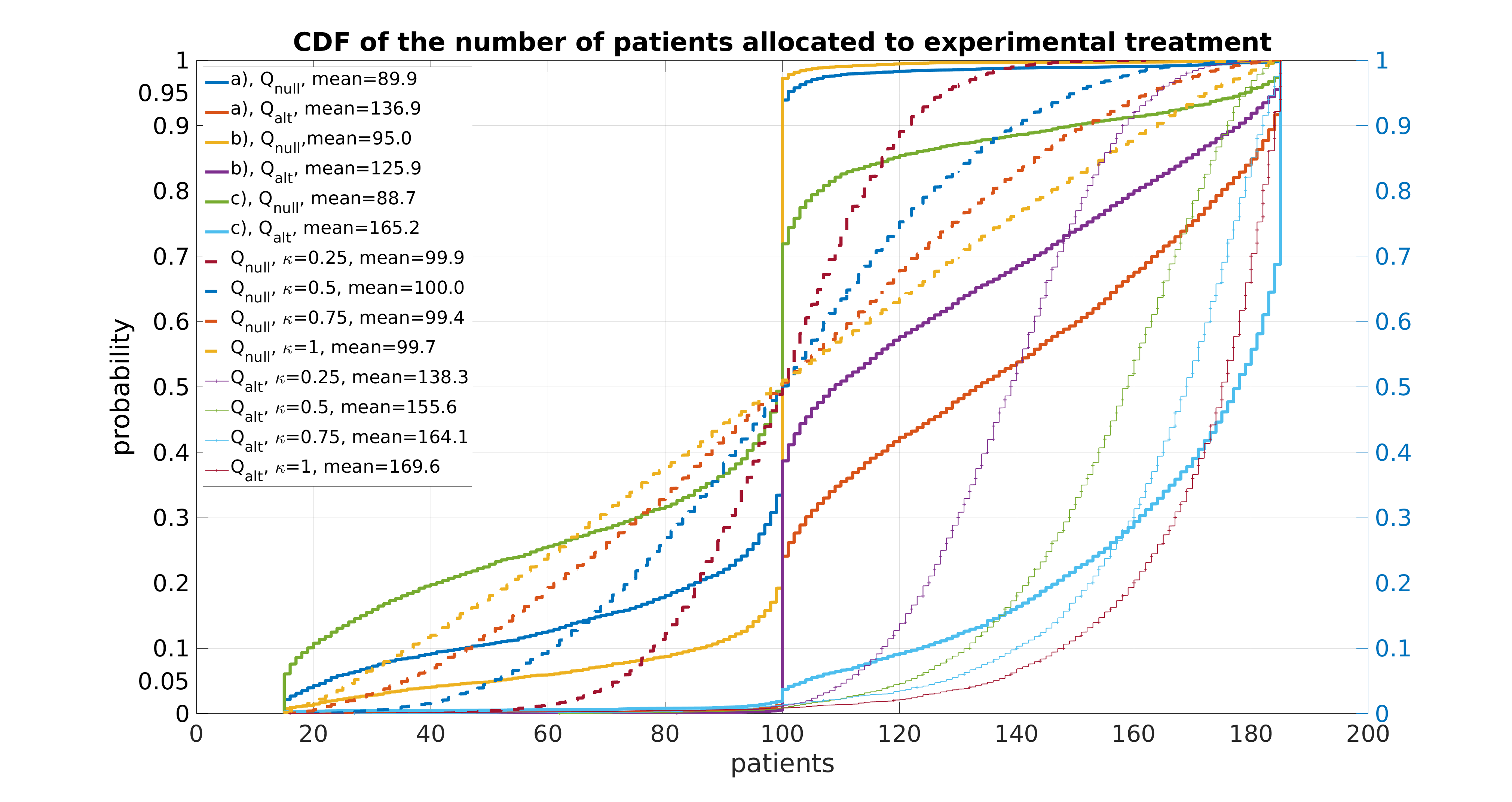}
       \renewcommand\thesubfigure{a}  
     \label{subfig:Ystat:a:null:rule1_200_16-Mar-2021}
     \end{subfigure}
     \vfill 
     \begin{subfigure}[b]{\textwidth}
     \includegraphics[width=\textwidth]{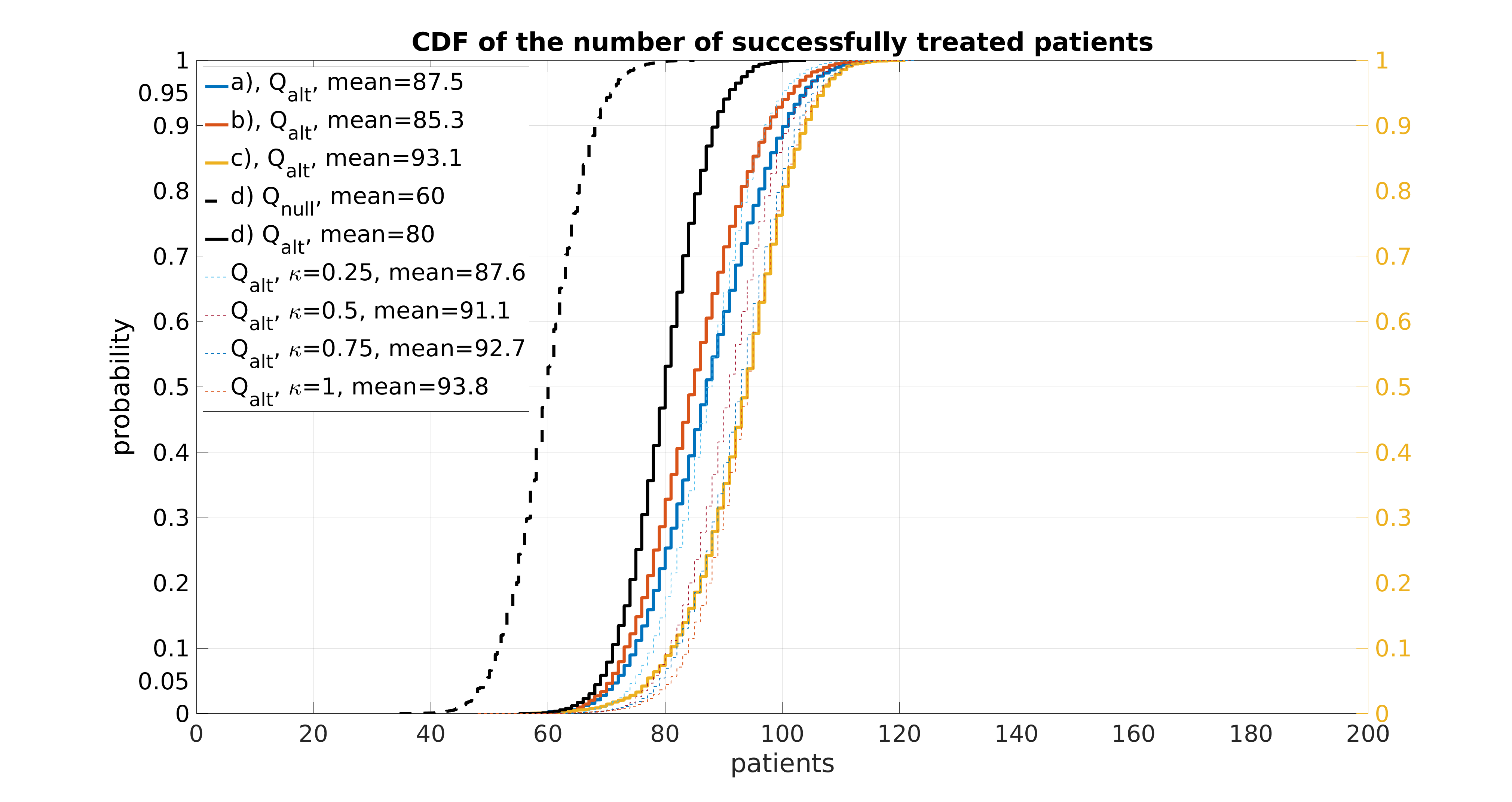}
        \renewcommand\thesubfigure{a} 
 \label{subfig:Ystat:a:alternative:rule1_200_16-Mar-2021}
     \end{subfigure}
\caption{\small Effect of employing a symmetric burn-in period of $n_0=30$ patients,  on the number of patients allocated to the experimental treatment and on the total number of treatment successes. Cumulative distribution functions of $N_1(500)$ (top) and $S(500)$ (bottom) are shown, based on $5000$ simulated data sets,  under $\Q_{null}$ with true parameter values
$\theta_0=\theta_1=0.3$ and $\Q_{alt}$ with values
$\theta_0=0.3,\theta_1=0.5$. Three combinations of the 
design parameters were used:  {(a)} $\varepsilon=0.1$,  $\delta=0.1$, {(b)}   $\varepsilon=0.05$, $\delta=0.1$, {(c)}
$\varepsilon=0.2$, $\delta=0.05$. In addition, {(d)} represents a completely symmetric treatment allocation. For comparison we also plot the corresponding CDF under the alternative hypothesis obtained by using fractional Thompson's rule 
with respective parameters
$\kappa=0.25,0.5,0.75 $ and $1$.   
}
\label{fig:combined_Ystat_RULE1_200_16-Mar-2021}
\end{figure}

 \FloatBarrier
\section{Additional figures and tables to subsection \ref{sub:3.1.3} } 
\label{supplement:B}

\textbf{Effect of trial size on frequentist performance }

\begin{figure}[h!tbp] 
     \begin{subfigure}[b]{1.0\textwidth}
     \includegraphics[height=\macroheight,width=\textwidth]{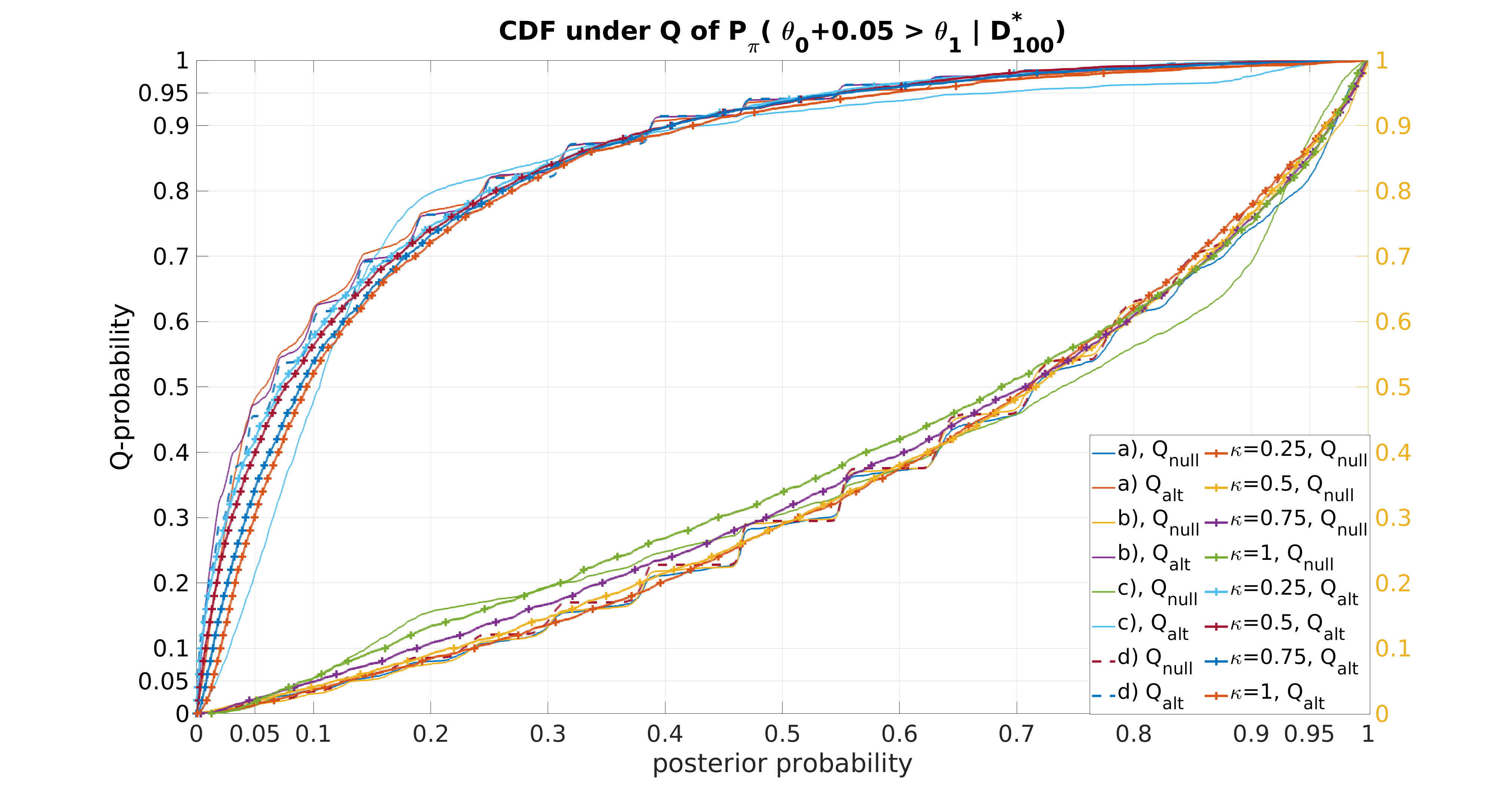}
       \renewcommand\thesubfigure{a}  
         \label{subfig:CDF_of_Pmax_rule1_100:0}
     \end{subfigure} \vfill
    \begin{subfigure}[b]{1.0\textwidth}
     \includegraphics[height=\macroheight,width=\textwidth]{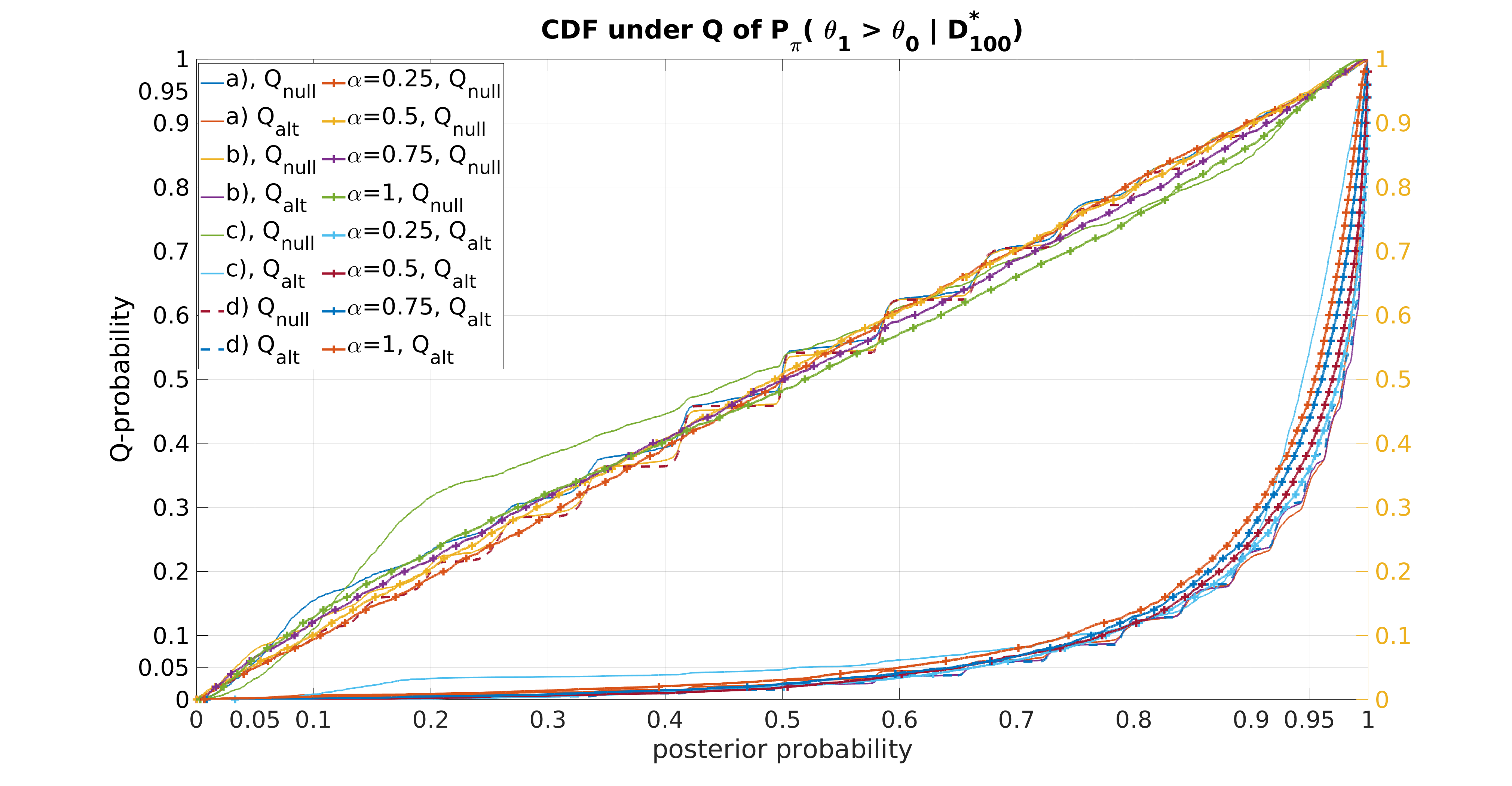}
      \renewcommand\thesubfigure{c}  
\label{subfig:CDF_of_pmax_rule1_100:1}
     \end{subfigure}\captionsetup{singlelinecheck=off}
\caption[short caption]{
\small   Effect of the design parameters $\varepsilon$ and $\delta$ of Rule 1, and $\kappa$ of Thompson's rule, on the CDFs of the posterior probabilities
$\P\bigl( \theta_0 +0.05 \geq \theta_1\big\vert D^*_{100}\bigr)$ (top)
and $\P\bigl( \theta_1 \geq \theta_0\big\vert D^*_{100}\bigr)$ (bottom) in the 2-arm trial of Experiment 1 when applying Rule 1 for treatment allocation and making a final assessment at $i = N_{\max}=100$.  
The results are based on
$5000$ data sets   generated under $\Q_{null}$ and $\Q_{alt}$ when using the following combinations of design parameters:
 {(a)}  
  $\varepsilon=0.1, 
  \delta=0.1$,
 {(b)}
  $\varepsilon=0.05, 
  \delta=0.1$,
  {(c)}
  $\varepsilon=0.2, 
  \delta=0.05$.}
  \label{fig:CDF_of_pmax_rule1_100}
\end{figure}

\begin{figure}[h!tbp] 
     \begin{subfigure}[b]{1.0\textwidth}
     \includegraphics[height=\macroheight,width=\textwidth]{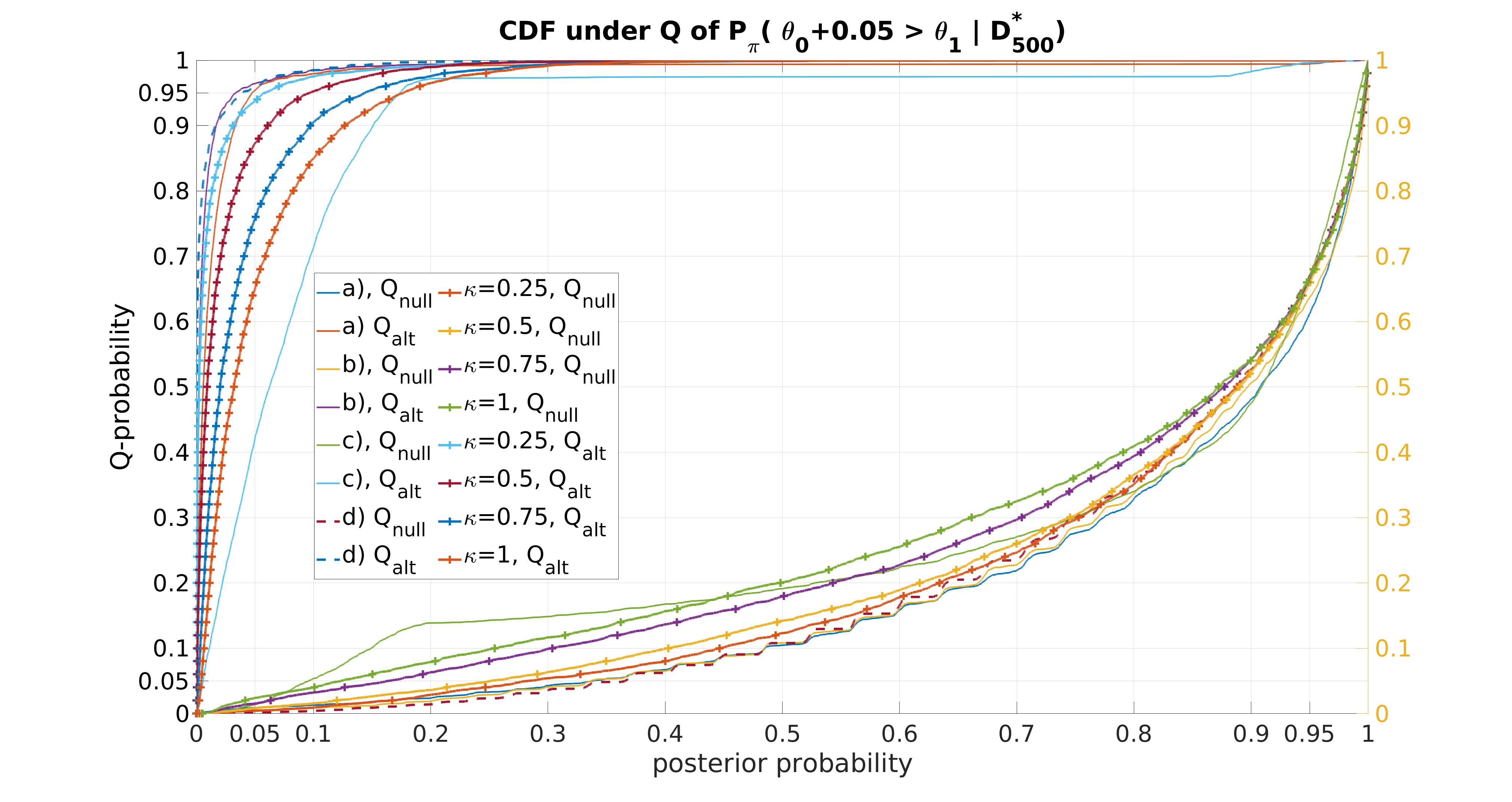}
       \renewcommand\thesubfigure{a}  
         \label{subfig:CDF_of_Pmax_rule1:0}
     \end{subfigure} \vfill
    \begin{subfigure}[b]{1.0\textwidth}
     \includegraphics[height=\macroheight,width=\textwidth]{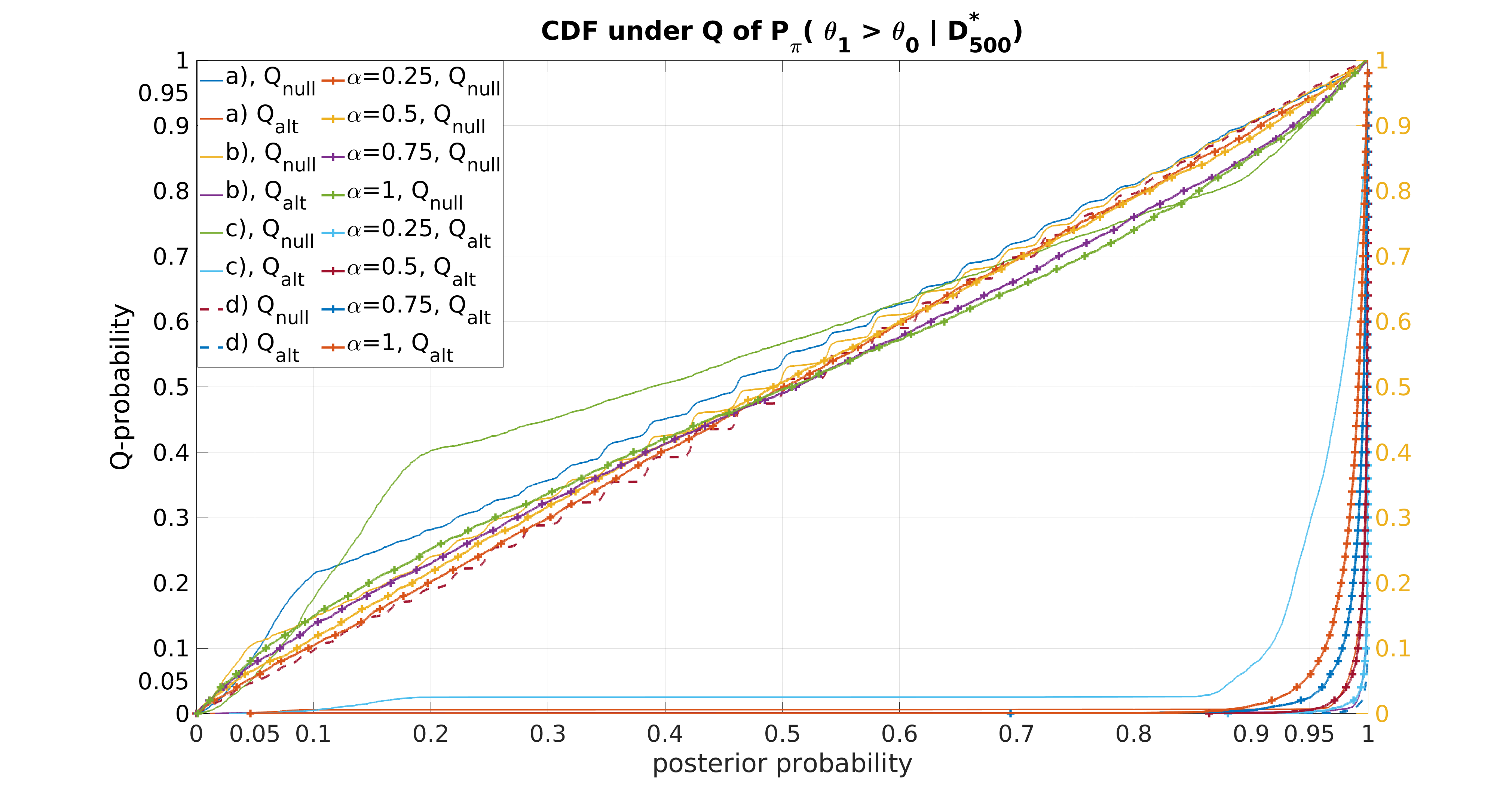}
      \renewcommand\thesubfigure{c}  
\label{subfig:CDF_of_pmax_rule1:1}
     \end{subfigure}\captionsetup{singlelinecheck=off}
\caption[short caption]{
\small   Effect of the design parameters $\varepsilon$ and $\delta$ of Rule 1, and $\kappa$ of Thompson's rule,  on the CDFs of the posterior probabilities
$\P\bigl( \theta_0 +0.05 \geq \theta_1\big\vert D^*_{500}\bigr)$ (top)
and $\P\bigl( \theta_1 \geq \theta_0\big\vert D^*_{500}\bigr)$ (bottom) in the 2-arm trial of Experiment 1 when applying Rule 1 for treatment allocation and making a final assessment at $i = N_{\max}=500$.  
The results are based on
$5000$ data sets  generated under $\Q_{null}$ and $\Q_{alt}$ when using the following combinations of design parameters:
 {(a)}  
  $\varepsilon=0.1, 
  \delta=0.1$,
 {(b)}
  $\varepsilon=0.05, 
  \delta=0.1$,
  {(c)}
  $\varepsilon=0.2, 
  \delta=0.05$. 
 }
 \label{fig:CDF_of_pmax_rule1}
\end{figure}

In subsection \ref{sub:3.1.3} of the main text we studied the performance of different adaptive designs in terms of true and false positive and negative rates,  by considering trial size $N_{max}=200$ in Figure \ref{fig:CDF_of_pmax_rule1_200} and Table \ref{table:1}. Below we present corresponding results for $N_{max}=100$ in Figure \ref{fig:CDF_of_pmax_rule1_100} and Table \ref{table:S1}, and for $N_{max}=500$ in Figure \ref{fig:CDF_of_pmax_rule1} and Table \ref{table:S2}. When combined, these results give us an idea about how such measures depend on the size of the trial.

Figures \ref{fig:CDF_of_pmax_rule1_100} and \ref{fig:CDF_of_pmax_rule1} bear close similarity to   Figure \ref{fig:CDF_of_pmax_rule1_200}. The main differences can be seen in the CDFs arising from data generated under $\Q_{alt}.$ The CDFs of the posterior probabilities $\P_{\pi}( \bftheta_1 \geq \bftheta_0\vert {\mathbf D}_{N_{max}}^{*})$ move to the right  as $N_{max}$ grows from $100$ to $200$ and then to $500,$ thereby  signalling that these probabilities become stochastically larger with growing trial size. A similar movement, somewhat slower and in the opposite direction, is seen in the CDFs of $\P_{\pi}( \bftheta_0+0.05 \geq \bftheta_1\vert {\mathbf D}_{N_{max}}^{*})$ with growing $N_{max}$. 

The following conclusions can now be made from Tables \ref{table:1}, \ref{table:S1} and \ref{table:S2}. Under $\Q_{null}$, the false positive rates are generally somewhat smaller for larger trial sizes, but remain under $0.025$ even in the case of  $N_{max}=100$. The true negative rates are usually larger, by a few percentage points, when the trial size is changed from  $100$ to $200$ and then to $500,$ and the inconclusive rates correspondingly smaller, typically attaining values on either side of ninety percent.   The false negative rates are very small for all considered designs. 

In contrast, as can be expected, the true positive rate (\textit{power}) under $\Q_{alt}$ depends strongly on the size of the trial. As reported in
\ref{sub:3.1.3}, for $N_{max}=200$ it has the moderate level of approximately seventy percent for Rule 1 designs (a), (b) and (d), and almost as high for Thompson's rule with $\kappa = 0.25$. For these same designs and $N_{max}=100,$ the true positive rates are lower, on both sides of 45 percent, but for $N_{max}=500$ already in the range of 95 percent. Again, of interest is to note that, in terms of these frequentist measures,  three adaptive rules perform as well as the symmetric block randomization design (d). For Thompson's rule, larger values of $\kappa$ lead to greater instability in the behavior of the adaptive mechanism and consequent weaker frequentist performance. Of all considered alternatives, the smallest true positive rate is obtained for the design (c) of Rule 1. The false negative rates are very small for all considered designs.

 \FloatBarrier
\begin{table}[H] 
{\small
\begin{center}
\begin{tabular}{ | c | c c c c c c c c | }\hline
 $\varepsilon_0=0.05,\delta_0=0.05$
 & (a) & (b) & (c) &(d) & $\kappa=0.25$ & $\kappa=0.5$ & $\kappa=0.75$ & $\kappa = 1$
 \\  \hline
$\Q_{null}:$ false positive & 
 0.020  &  0.013   & 0.012&    0.016&    0.013   & 0.020&    0.022  &  0.018
\\ \hline 
 $\Q_{null}:$ true negative & 
 0.058 &   0.078   & 0.032 &   0.051&    0.050    &0.054 &   0.066   & 0.066
 \\
    \hline $\Q_{null}:$
 inconclusive &
  0.922  &  0.908  &  0.956  &  0.933  &  0.937  &  0.926  &  0.911  &  0.917
 \\ \hline 
 $\Q_{alt}:$ true positive  &
 0.482  &  0.473   & 0.215 &   0.455&    0.419   & 0.398 &   0.344&    0.303
 \\
  \hline
$\Q_{alt}:$ false negative & 
 0.003  &  0.001   & 0.002         &$\sim 0$ &    $\sim 0$   & 0.001   & 0.001 &   0.002
\\
\hline
$\Q_{alt}:$ inconclusive &
0.516   & 0.525    &0.783&    0.545   & 0.581   & 0.601&    0.655&    0.695
 \\  
 \hline
\end{tabular}
\end{center}
\caption{\label{table:S1}True and false positive and negative rates when applying adaptive treatment allocation with design parameter values $\varepsilon_0 = 0.05$ and $\delta_0 = 0.05$ in a trial of size $N_{max} = 100$.}
}
\end{table}

\begin{table}[H]
{\small
\begin{center}
\begin{tabular}{ | c | c c c c c c c c | }\hline
 $\varepsilon_0=0.05,\delta_0=0.05$
 & (a) & (b) & (c) &(d) & $\kappa=0.25$ & $\kappa=0.5$ & $\kappa=0.75$ & $\kappa = 1$
 \\  \hline
$\Q_{null}:$ false positive &
 0.009  &  0.004   & 0.014    &0.001&    0.005 &   0.008   & 0.015   & 0.024
\\ \hline 
 $\Q_{null}:$ true negative &
 0.092 &   0.108   & 0.059 &    0.049   & 0.057&    0.068   & 0.077 &   0.086
  \\
    \hline $\Q_{null}:$
 inconclusive &
 0.899  &  0.888   & 0.927   & 0.950&    0.939   & 0.924    &0.908 &   0.890
 \\ \hline 
 $\Q_{alt}:$ true positive  &
  0.954  &  0.964 &   0.421 &   0.959 &   0.937 &   0.873 &   0.757 &   0.650
 \\
  \hline
$\Q_{alt}:$ false negative & 
 0.002 &   0.001   & 0.001 &       $\sim 0$
 & $\sim 0$ &
 $\sim 0$ &         $\sim 0$ &    $\sim 0$
\\
\hline
$\Q_{alt}:$ inconclusive &
0.044 &   0.035   & 0.578    &0.041&    0.063   & 0.127    &0.243&    0.350
 \\  
 
 \hline
\end{tabular}
\end{center}
\caption{\label{table:S2}True and false positive and negative rates when applying adaptive treatment allocation with design parameter values $\varepsilon_0 = 0.05$ and $\delta_0 = 0.05$ in a trial of size $N_{max} = 500$.}
}
\end{table}

\textbf{Employing an initial burn-in period} 

In Table \ref{table:1:burnin} we consider the effect of the design modification, where the first $30$ patients are divided evenly, by using a block randomization, to  the two treatments.
Adaptive treatment allocation is then applied after this, either in the form of Rule 1 or Thompson's rule, and the performance measures are evaluated at $N_{max} = 200$ from a simulation experiment of $5000$ repetitions. The numerical values in Table \ref{table:1:burnin} are compared naturally to those in Table \ref{table:1}, where the design was the same except that no burn-in was used. 

Overall, the differences are small. The largest change is in the values of true positive rate (power) for Rule 1 (c), which has increased from $0.303$ in Table \ref{table:1} to $0.443$ due to the stabilizing initial burn-in. Smaller differences can be seen in the false positive rates for Rule 1 (c) and Thompson's rule with $\kappa = 0.75$ and $\kappa =1,$ where burn-in has trimmed down these already rather low rates by small amounts. The conclusion from this experiment is that, in a trial of size $N_{max} = 200$, employing an initial burn-in period has a small to modest stabilizing effect on the frequentist performance of those  adaptive designs in which the adaptive mechanism was strongest.

\begin{table} {\small
\begin{center}
\begin{tabular}{ | c | c c c c c c c c | }\hline
$\varepsilon_0=0.05,\delta=0.05$
 & (a) & (b) & (c) &(d) & $\kappa=0.25$ & $\kappa=0.5$ & $\kappa=0.75$ & $\kappa = 1$
 \\  \hline
$\Q_{null}:$ false positive &
 0.014  &  0.010 &   0.019  &  0.008    &0.011 &   0.015&    0.015   & 0.020 \\
\hline 
 $\Q_{null}:$ true negative 
 &0.077    &0.085&    0.061 &   0.050 &   0.056    &0.057&    0.068 &   0.064 
   \\  \hline $\Q_{null}:$
 inconclusive &
0.909  &  0.905 &   0.920 &    0.942 &    0.934    &0.928 &   0.917   & 0.915
 \\ \hline
 $\Q_{alt}:$ true positive  &
  0.727   & 0.702 &    0.443 &   0.689 &   0.676 &   0.615    &0.533&    0.464
  \\ \hline
$\Q_{alt}:$ false negative   &
$\sim 0$  &       $\sim 0$    &  0.001 &
$\sim 0$    &    $\sim 0$  & $\sim 0 $ &   $\sim 0$ &    0.001
 \\  \hline
$\Q_{alt}:$ inconclusive &
0.272  &  0.298 &    0.556 &    0.311 &   0.324    &0.385&    0.467  &  0.535
 \\
 \hline
\end{tabular}
\end{center}
\caption{\label{table:1:burnin}True and false positive and negative rates when applying adaptive treatment allocation with design parameter values $\varepsilon_0 = 0.05$ and $\delta_0 = 0.05$
and a \textit{burn-in period} of $n_0=30$ patients
in a trial of size $N_{max} = 200$.}
}
\end{table}

\textbf{Remarks on other test variants} 

In the \textbf{first variant}, we consider in Table \ref{table:S3}   the  case $\delta_0 =0$, where the special protection against dropping the control arm in the final test at  $N_{max}$ has been removed. Thus we write 
\textit{false positive rate}  = $\Q_{null}(\P_{\pi}( \bftheta_0 \geq \bftheta_{1}\vert {\mathbf D}_{N_{max}}^{*})\leq \varepsilon_0)$,
\textit{true negative rate} = $ \Q_{null}(\P_{\pi}(\bftheta_{1} \geq \bftheta_0  \vert {\mathbf D}_{N_{max}}^{*}) \leq  \varepsilon_0)$, 
\textit{true positive rate} = $\Q_{alt}(\P_{\pi}( \bftheta_0  \geq \bftheta_{1}\vert {\mathbf D}_{N_{max}}^{*})\leq \varepsilon_0)$ and 
\textit{false negative rate} = $\Q_{alt}(\P_{\pi}(\bftheta_{1} \geq \bftheta_0  \vert {\mathbf D}_{N_{max}}^{*}) \leq  \varepsilon_0).$ 
\textit{Inconclusive rates} are the probabilities
$\Q(\P_{\pi}( \bftheta_0  \geq \bftheta_{1}\vert {\mathbf D}_{N_{max}}^{*}) > \varepsilon_0,  \P_{\pi}(\bftheta_{1} \geq \bftheta_0  \vert {\mathbf D}_{N_{max}}^{*}) > \varepsilon_0)$, for $\Q =\Q_{null}$ and $\Q=\Q_{alt}$. As noted in the main text, this change from the original criteria implies that, compared to the respective values provided in Table \ref{table:1}, all positive rates are now larger, while the negative rates remain intact. Of the former, the rates for Rule 1 (a), (b) and (d), and for Thompson's rule with $\kappa = 0.25$, are again quite similar, with false positive rates varying on both sides of five percent and true positive rates (\textit{power}) reaching levels of almost ninety percent. The frequentist performance of the other designs  is somewhat weaker, deteriorating with increasing instability of the allocation rule. 

\begin{table} {\small
\begin{center}
\begin{tabular}{ | c | c c c c c c c c | }\hline  
 $\varepsilon_0=0.05,\delta_0=0$
 & (a) & (b) & (c) &(d) & $\kappa=0.25$ & $\kappa=0.5$ & $\kappa=0.75$ & $\kappa = 1$
 \\  \hline
$\Q_{null}:$ false positive & 
  0.050  &  0.046 &   0.082&    0.051 &    0.053    &0.056&    0.068 &   0.075
\\ \hline 
 $\Q_{null}:$ true negative & 
  0.074   &  0.086    & 0.040 &    0.052&    0.054  &  0.056 &    0.073 &   0.074
 \\
    \hline $\Q_{null}:$
 inconclusive &
 0.876  &   0.868 &   0.878 &    0.896  &  0.892  &  0.888 &   0.859  &  0.851 
 \\ \hline 
 $\Q_{alt}:$ true positive  & 
 0.897 &   0.896 &   0.622  &    0.891 &   0.886 &   0.857  &  0.794 &   0.739 
  \\
  \hline
$\Q_{alt}:$ false negative & 
0.002  &  0.001 &   0.001 &        $\sim 0$ &        $\sim 0$ &  
$\sim 0$ &    0.001&    0.001 
\\
\hline
$\Q_{alt}:$ inconclusive &
 0.101 &    0.103  &  0.377 &   0.109 &   0.114 &   0.143  &  0.204 &   0.260
 \\  
 \hline
\end{tabular}
\end{center}
\caption{\label{table:S3} True and false positive and negative rates when applying adaptive treatment allocation with design parameter values $\varepsilon_0 = 0.05$ and $\delta_0 = 0$ in a trial of size $N_{max} = 200$. First test variant, see text.}
}
\end{table}
  
In the \textbf{second  variant} of the final test, the experimental arm is dropped if $\P_{\pi}(\bftheta_{1} \geq \bftheta_0 +\delta_0  \vert {\mathbf D}_{N_{max}}^{*}) \leq  \varepsilon_0.$ Therefore, in Table \ref{table:S4} we write 
\textit{false positive rate}  = $\Q_{null}(\P_{\pi}( \bftheta_0+\delta_0 \geq \bftheta_{1}\vert {\mathbf D}_{N_{max}}^{*})\leq \varepsilon_0)$,
\textit{true negative rate} = $ \Q_{null}(\P_{\pi}(\bftheta_{1} \geq \bftheta_0 +\delta_0  \vert {\mathbf D}_{N_{max}}^{*}) \leq  \varepsilon_0)$, 
\textit{true positive rate} = $\Q_{alt}(\P_{\pi}( \bftheta_0+\delta_0 \geq \bftheta_{1}\vert {\mathbf D}_{N_{max}}^{*})\leq \varepsilon_0)$ and 
\textit{false negative rate} = $\Q_{alt}(\P_{\pi}(\bftheta_{1} \geq \bftheta_0 +\delta_0  \vert {\mathbf D}_{N_{max}}^{*}) \leq  \varepsilon_0).$ The probabilities
$\Q(\P_{\pi}( \bftheta_0+\delta_0 \geq \bftheta_{1}\vert {\mathbf D}_{N_{max}}^{*}) > \varepsilon_0,  \P_{\pi}(\bftheta_{1} \geq \bftheta_0 +\delta_0  \vert {\mathbf D}_{N_{max}}^{*}) > \varepsilon_0)$, for $\Q =\Q_{null}$ and $\Q=\Q_{alt}$, are \textit{inconclusive rates}. This change means that the negative rates, both true and false, are now larger than the respective values in Table \ref{table:1}, while the positive rates remain intact.  The true negative rates, which were below ten percent in Table \ref{table:1}, vary in Table \ref{table:S4}  on both sides of twenty percent. The inconclusive rates under $\Q_{null}$ are now lower than in Table \ref{table:S3}, but still rather high, between seventy-five and eighty percent. The false negative rates are slightly higher than in Table \ref{table:1}, but still very low for all allocation rules. The performance of Rule 1 (a), (b) and (d), and of Thompson's rule with $\kappa = 0.25$, is again quite similar.

\begin{table} {\small
\begin{center}
\begin{tabular}{ | c | c c c c c c c c | }\hline
 
 \hline  
 $\varepsilon_0=0.05,\delta_0=0.05$
 & (a) & (b) & (c) &(d) & $\kappa=0.25$ & $\kappa=0.5$ & $\kappa=0.75$ & $\kappa = 1$
 \\  \hline
$\Q_{null}:$ false positive & 
  0.014 &    0.009    & 0.014 &    0.007  &  0.011 &    0.014   & 0.023  &   0.025
\\ \hline 
 $\Q_{null}:$ true negative & 
0.236  &  0.216  &  0.201  &  0.196  &  0.187  &  0.195  &  0.210  &  0.197 \\
    \hline $\Q_{null}:$
 inconclusive &
 0.750 &   0.776 &   0.785 &   0.797 &   0.803 &   0.791 &    0.768 &   0.778
  \\ \hline 
 $\Q_{alt}:$ true positive  & 0.723&    0.711 &   0.303&    0.694 &   0.665   & 0.598  &  0.516 &   0.443 \\
  \hline
$\Q_{alt}:$ false negative & 
0.005  &  0.001 &   0.004&         $\sim 0$  &  $\sim 0$   &  
$\sim 0$  &  0.001 &   0.002
\\
\hline
$\Q_{alt}:$ inconclusive &
0.272  &  0.288&    0.693&    0.306  &  0.335&    0.402&    0.483  &  0.555
 \\  
 \hline
\end{tabular}
\end{center}
\caption{\label{table:S4}True and false positive and negative rates when applying adaptive treatment allocation with design parameter values $\varepsilon_0 = 0.05$ and $\delta_0 = 0.05$ in a trial of size $N_{max} = 200$. Second test variant, see text.}
}
\end{table} 

\end{document}